\documentclass[12pt]{article}
\usepackage{epsfig,amsfonts,amssymb}
\usepackage{hyperref}
\usepackage{comment}
\usepackage{cite}
\input epsf.sty
\usepackage{cite}

\newcommand{\epk}{\epsilon^{\mu\nu}p_{\nu}k^{\rho}k^{\sigma}}
\newcommand{\epkl}{(p\cdot k)\epsilon^{\mu\rho}k^{\sigma}}
\newcommand{\epkll}{\epsilon^{\mu\rho}p_{\rho}k^{\nu}k^{\sigma}}
\newcommand{\epklll}{\epsilon^{\mu\nu}p_{\rho}k^{\rho}k^{\sigma}}

\def\ZZZ{{\hbox{ Z\kern-1.6mm Z}}}
\def\RRR{{\hbox{ R\kern-2.4mm R}}}
\def\CCC{{\hbox{ C\kern-2.0mm C}}}
\def\zzz{{\hbox{z\kern-1mm z}}}

\def\eee{e}

\newcommand{\qeq}{{\hbox{=\kern-2.3mm ? \kern.5mm }}}
\renewcommand{\qeq}{=}

\newcommand{\eps}{\epsilon}

\newcommand{\ve}{\varepsilon}

\newcommand{\BB}{{\cal B}}

\newcommand{\II}{{\cal I}}

\newcommand{\KK}{{\cal K}}

\newcommand{\MM}{{\cal M}}

\newcommand{\OO}{{\cal O}}

\newcommand{\wt}{\widetilde}
\newcommand{\wh}{\widehat}

\newcommand{\RR}{{\cal R}}

\newcommand{\be}{\begin{equation}}
\newcommand{\ee}{\end{equation}}
\newcommand{\ben}{\begin{eqnarray}\displaystyle}
\newcommand{\een}{\end{eqnarray}}

\newcommand{\refb}[1]{(\ref{#1})}
\newcommand{\p}{\partial}
\newcommand{\sectiono}[1]{\section{#1}\setcounter{equation}{0}}

\def\one{{\hbox{ 1\kern-.8mm l}}}
\def\zero{{\hbox{ 0\kern-1.5mm 0}}}

\newcommand{\bea}[1]{\begin{eqnarray}\label{#1} }
\newcommand{\eea}{\end{eqnarray}}

\newcommand{\eqref}{\refb}

\usepackage{bm}
\usepackage[table]{xcolor}

\def\figsoftonefield{

\def\JPicScale{0.6}
\ifx\JPicScale\undefined\def\JPicScale{1}\fi
\unitlength \JPicScale mm

}

\def\figsoftthreefield{

\def\JPicScale{0.6}
\ifx\JPicScale\undefined\def\JPicScale{1}\fi
\unitlength \JPicScale mm
\begin{picture}(135,90)(0,0)
\linethickness{0.3mm}
\put(105.03,48.5){\line(0,1){0.5}}
\multiput(105.02,49.5)(0.01,-0.5){1}{\line(0,-1){0.5}}
\multiput(105,50)(0.02,-0.5){1}{\line(0,-1){0.5}}
\multiput(104.97,50.49)(0.03,-0.5){1}{\line(0,-1){0.5}}
\multiput(104.92,50.99)(0.04,-0.5){1}{\line(0,-1){0.5}}
\multiput(104.87,51.49)(0.06,-0.5){1}{\line(0,-1){0.5}}
\multiput(104.8,51.98)(0.07,-0.49){1}{\line(0,-1){0.49}}
\multiput(104.73,52.47)(0.08,-0.49){1}{\line(0,-1){0.49}}
\multiput(104.64,52.96)(0.09,-0.49){1}{\line(0,-1){0.49}}
\multiput(104.54,53.45)(0.1,-0.49){1}{\line(0,-1){0.49}}
\multiput(104.43,53.94)(0.11,-0.49){1}{\line(0,-1){0.49}}
\multiput(104.31,54.42)(0.12,-0.48){1}{\line(0,-1){0.48}}
\multiput(104.18,54.9)(0.13,-0.48){1}{\line(0,-1){0.48}}
\multiput(104.04,55.38)(0.14,-0.48){1}{\line(0,-1){0.48}}
\multiput(103.89,55.86)(0.15,-0.47){1}{\line(0,-1){0.47}}
\multiput(103.72,56.33)(0.16,-0.47){1}{\line(0,-1){0.47}}
\multiput(103.55,56.79)(0.17,-0.47){1}{\line(0,-1){0.47}}
\multiput(103.37,57.26)(0.09,-0.23){2}{\line(0,-1){0.23}}
\multiput(103.17,57.72)(0.1,-0.23){2}{\line(0,-1){0.23}}
\multiput(102.97,58.17)(0.1,-0.23){2}{\line(0,-1){0.23}}
\multiput(102.76,58.62)(0.11,-0.23){2}{\line(0,-1){0.23}}
\multiput(102.53,59.07)(0.11,-0.22){2}{\line(0,-1){0.22}}
\multiput(102.3,59.51)(0.12,-0.22){2}{\line(0,-1){0.22}}
\multiput(102.06,59.95)(0.12,-0.22){2}{\line(0,-1){0.22}}
\multiput(101.8,60.38)(0.13,-0.21){2}{\line(0,-1){0.21}}
\multiput(101.54,60.8)(0.13,-0.21){2}{\line(0,-1){0.21}}
\multiput(101.27,61.22)(0.14,-0.21){2}{\line(0,-1){0.21}}
\multiput(100.99,61.63)(0.14,-0.21){2}{\line(0,-1){0.21}}
\multiput(100.7,62.04)(0.14,-0.2){2}{\line(0,-1){0.2}}
\multiput(100.4,62.44)(0.15,-0.2){2}{\line(0,-1){0.2}}
\multiput(100.1,62.83)(0.1,-0.13){3}{\line(0,-1){0.13}}
\multiput(99.78,63.21)(0.11,-0.13){3}{\line(0,-1){0.13}}
\multiput(99.46,63.59)(0.11,-0.13){3}{\line(0,-1){0.13}}
\multiput(99.12,63.96)(0.11,-0.12){3}{\line(0,-1){0.12}}
\multiput(98.78,64.33)(0.11,-0.12){3}{\line(0,-1){0.12}}
\multiput(98.43,64.68)(0.12,-0.12){3}{\line(0,-1){0.12}}
\multiput(98.08,65.03)(0.12,-0.12){3}{\line(1,0){0.12}}
\multiput(97.71,65.37)(0.12,-0.11){3}{\line(1,0){0.12}}
\multiput(97.34,65.71)(0.12,-0.11){3}{\line(1,0){0.12}}
\multiput(96.96,66.03)(0.13,-0.11){3}{\line(1,0){0.13}}
\multiput(96.58,66.35)(0.13,-0.11){3}{\line(1,0){0.13}}
\multiput(96.19,66.65)(0.13,-0.1){3}{\line(1,0){0.13}}
\multiput(95.79,66.95)(0.2,-0.15){2}{\line(1,0){0.2}}
\multiput(95.38,67.24)(0.2,-0.14){2}{\line(1,0){0.2}}
\multiput(94.97,67.52)(0.21,-0.14){2}{\line(1,0){0.21}}
\multiput(94.55,67.79)(0.21,-0.14){2}{\line(1,0){0.21}}
\multiput(94.13,68.05)(0.21,-0.13){2}{\line(1,0){0.21}}
\multiput(93.7,68.31)(0.21,-0.13){2}{\line(1,0){0.21}}
\multiput(93.26,68.55)(0.22,-0.12){2}{\line(1,0){0.22}}
\multiput(92.82,68.78)(0.22,-0.12){2}{\line(1,0){0.22}}
\multiput(92.37,69.01)(0.22,-0.11){2}{\line(1,0){0.22}}
\multiput(91.92,69.22)(0.23,-0.11){2}{\line(1,0){0.23}}
\multiput(91.47,69.42)(0.23,-0.1){2}{\line(1,0){0.23}}
\multiput(91.01,69.62)(0.23,-0.1){2}{\line(1,0){0.23}}
\multiput(90.54,69.8)(0.23,-0.09){2}{\line(1,0){0.23}}
\multiput(90.08,69.97)(0.47,-0.17){1}{\line(1,0){0.47}}
\multiput(89.61,70.14)(0.47,-0.16){1}{\line(1,0){0.47}}
\multiput(89.13,70.29)(0.47,-0.15){1}{\line(1,0){0.47}}
\multiput(88.65,70.43)(0.48,-0.14){1}{\line(1,0){0.48}}
\multiput(88.17,70.56)(0.48,-0.13){1}{\line(1,0){0.48}}
\multiput(87.69,70.68)(0.48,-0.12){1}{\line(1,0){0.48}}
\multiput(87.2,70.79)(0.49,-0.11){1}{\line(1,0){0.49}}
\multiput(86.71,70.89)(0.49,-0.1){1}{\line(1,0){0.49}}
\multiput(86.22,70.98)(0.49,-0.09){1}{\line(1,0){0.49}}
\multiput(85.73,71.05)(0.49,-0.08){1}{\line(1,0){0.49}}
\multiput(85.24,71.12)(0.49,-0.07){1}{\line(1,0){0.49}}
\multiput(84.74,71.17)(0.5,-0.06){1}{\line(1,0){0.5}}
\multiput(84.24,71.22)(0.5,-0.04){1}{\line(1,0){0.5}}
\multiput(83.75,71.25)(0.5,-0.03){1}{\line(1,0){0.5}}
\multiput(83.25,71.27)(0.5,-0.02){1}{\line(1,0){0.5}}
\multiput(82.75,71.28)(0.5,-0.01){1}{\line(1,0){0.5}}
\put(82.25,71.28){\line(1,0){0.5}}
\multiput(81.75,71.27)(0.5,0.01){1}{\line(1,0){0.5}}
\multiput(81.25,71.25)(0.5,0.02){1}{\line(1,0){0.5}}
\multiput(80.76,71.22)(0.5,0.03){1}{\line(1,0){0.5}}
\multiput(80.26,71.17)(0.5,0.04){1}{\line(1,0){0.5}}
\multiput(79.76,71.12)(0.5,0.06){1}{\line(1,0){0.5}}
\multiput(79.27,71.05)(0.49,0.07){1}{\line(1,0){0.49}}
\multiput(78.78,70.98)(0.49,0.08){1}{\line(1,0){0.49}}
\multiput(78.29,70.89)(0.49,0.09){1}{\line(1,0){0.49}}
\multiput(77.8,70.79)(0.49,0.1){1}{\line(1,0){0.49}}
\multiput(77.31,70.68)(0.49,0.11){1}{\line(1,0){0.49}}
\multiput(76.83,70.56)(0.48,0.12){1}{\line(1,0){0.48}}
\multiput(76.35,70.43)(0.48,0.13){1}{\line(1,0){0.48}}
\multiput(75.87,70.29)(0.48,0.14){1}{\line(1,0){0.48}}
\multiput(75.39,70.14)(0.47,0.15){1}{\line(1,0){0.47}}
\multiput(74.92,69.97)(0.47,0.16){1}{\line(1,0){0.47}}
\multiput(74.46,69.8)(0.47,0.17){1}{\line(1,0){0.47}}
\multiput(73.99,69.62)(0.23,0.09){2}{\line(1,0){0.23}}
\multiput(73.53,69.42)(0.23,0.1){2}{\line(1,0){0.23}}
\multiput(73.08,69.22)(0.23,0.1){2}{\line(1,0){0.23}}
\multiput(72.63,69.01)(0.23,0.11){2}{\line(1,0){0.23}}
\multiput(72.18,68.78)(0.22,0.11){2}{\line(1,0){0.22}}
\multiput(71.74,68.55)(0.22,0.12){2}{\line(1,0){0.22}}
\multiput(71.3,68.31)(0.22,0.12){2}{\line(1,0){0.22}}
\multiput(70.87,68.05)(0.21,0.13){2}{\line(1,0){0.21}}
\multiput(70.45,67.79)(0.21,0.13){2}{\line(1,0){0.21}}
\multiput(70.03,67.52)(0.21,0.14){2}{\line(1,0){0.21}}
\multiput(69.62,67.24)(0.21,0.14){2}{\line(1,0){0.21}}
\multiput(69.21,66.95)(0.2,0.14){2}{\line(1,0){0.2}}
\multiput(68.81,66.65)(0.2,0.15){2}{\line(1,0){0.2}}
\multiput(68.42,66.35)(0.13,0.1){3}{\line(1,0){0.13}}
\multiput(68.04,66.03)(0.13,0.11){3}{\line(1,0){0.13}}
\multiput(67.66,65.71)(0.13,0.11){3}{\line(1,0){0.13}}
\multiput(67.29,65.37)(0.12,0.11){3}{\line(1,0){0.12}}
\multiput(66.92,65.03)(0.12,0.11){3}{\line(1,0){0.12}}
\multiput(66.57,64.68)(0.12,0.12){3}{\line(1,0){0.12}}
\multiput(66.22,64.33)(0.12,0.12){3}{\line(0,1){0.12}}
\multiput(65.88,63.96)(0.11,0.12){3}{\line(0,1){0.12}}
\multiput(65.54,63.59)(0.11,0.12){3}{\line(0,1){0.12}}
\multiput(65.22,63.21)(0.11,0.13){3}{\line(0,1){0.13}}
\multiput(64.9,62.83)(0.11,0.13){3}{\line(0,1){0.13}}
\multiput(64.6,62.44)(0.1,0.13){3}{\line(0,1){0.13}}
\multiput(64.3,62.04)(0.15,0.2){2}{\line(0,1){0.2}}
\multiput(64.01,61.63)(0.14,0.2){2}{\line(0,1){0.2}}
\multiput(63.73,61.22)(0.14,0.21){2}{\line(0,1){0.21}}
\multiput(63.46,60.8)(0.14,0.21){2}{\line(0,1){0.21}}
\multiput(63.2,60.38)(0.13,0.21){2}{\line(0,1){0.21}}
\multiput(62.94,59.95)(0.13,0.21){2}{\line(0,1){0.21}}
\multiput(62.7,59.51)(0.12,0.22){2}{\line(0,1){0.22}}
\multiput(62.47,59.07)(0.12,0.22){2}{\line(0,1){0.22}}
\multiput(62.24,58.62)(0.11,0.22){2}{\line(0,1){0.22}}
\multiput(62.03,58.17)(0.11,0.23){2}{\line(0,1){0.23}}
\multiput(61.83,57.72)(0.1,0.23){2}{\line(0,1){0.23}}
\multiput(61.63,57.26)(0.1,0.23){2}{\line(0,1){0.23}}
\multiput(61.45,56.79)(0.09,0.23){2}{\line(0,1){0.23}}
\multiput(61.28,56.33)(0.17,0.47){1}{\line(0,1){0.47}}
\multiput(61.11,55.86)(0.16,0.47){1}{\line(0,1){0.47}}
\multiput(60.96,55.38)(0.15,0.47){1}{\line(0,1){0.47}}
\multiput(60.82,54.9)(0.14,0.48){1}{\line(0,1){0.48}}
\multiput(60.69,54.42)(0.13,0.48){1}{\line(0,1){0.48}}
\multiput(60.57,53.94)(0.12,0.48){1}{\line(0,1){0.48}}
\multiput(60.46,53.45)(0.11,0.49){1}{\line(0,1){0.49}}
\multiput(60.36,52.96)(0.1,0.49){1}{\line(0,1){0.49}}
\multiput(60.27,52.47)(0.09,0.49){1}{\line(0,1){0.49}}
\multiput(60.2,51.98)(0.08,0.49){1}{\line(0,1){0.49}}
\multiput(60.13,51.49)(0.07,0.49){1}{\line(0,1){0.49}}
\multiput(60.08,50.99)(0.06,0.5){1}{\line(0,1){0.5}}
\multiput(60.03,50.49)(0.04,0.5){1}{\line(0,1){0.5}}
\multiput(60,50)(0.03,0.5){1}{\line(0,1){0.5}}
\multiput(59.98,49.5)(0.02,0.5){1}{\line(0,1){0.5}}
\multiput(59.97,49)(0.01,0.5){1}{\line(0,1){0.5}}
\put(59.97,48.5){\line(0,1){0.5}}
\multiput(59.97,48.5)(0.01,-0.5){1}{\line(0,-1){0.5}}
\multiput(59.98,48)(0.02,-0.5){1}{\line(0,-1){0.5}}
\multiput(60,47.5)(0.03,-0.5){1}{\line(0,-1){0.5}}
\multiput(60.03,47.01)(0.04,-0.5){1}{\line(0,-1){0.5}}
\multiput(60.08,46.51)(0.06,-0.5){1}{\line(0,-1){0.5}}
\multiput(60.13,46.01)(0.07,-0.49){1}{\line(0,-1){0.49}}
\multiput(60.2,45.52)(0.08,-0.49){1}{\line(0,-1){0.49}}
\multiput(60.27,45.03)(0.09,-0.49){1}{\line(0,-1){0.49}}
\multiput(60.36,44.54)(0.1,-0.49){1}{\line(0,-1){0.49}}
\multiput(60.46,44.05)(0.11,-0.49){1}{\line(0,-1){0.49}}
\multiput(60.57,43.56)(0.12,-0.48){1}{\line(0,-1){0.48}}
\multiput(60.69,43.08)(0.13,-0.48){1}{\line(0,-1){0.48}}
\multiput(60.82,42.6)(0.14,-0.48){1}{\line(0,-1){0.48}}
\multiput(60.96,42.12)(0.15,-0.47){1}{\line(0,-1){0.47}}
\multiput(61.11,41.64)(0.16,-0.47){1}{\line(0,-1){0.47}}
\multiput(61.28,41.17)(0.17,-0.47){1}{\line(0,-1){0.47}}
\multiput(61.45,40.71)(0.09,-0.23){2}{\line(0,-1){0.23}}
\multiput(61.63,40.24)(0.1,-0.23){2}{\line(0,-1){0.23}}
\multiput(61.83,39.78)(0.1,-0.23){2}{\line(0,-1){0.23}}
\multiput(62.03,39.33)(0.11,-0.23){2}{\line(0,-1){0.23}}
\multiput(62.24,38.88)(0.11,-0.22){2}{\line(0,-1){0.22}}
\multiput(62.47,38.43)(0.12,-0.22){2}{\line(0,-1){0.22}}
\multiput(62.7,37.99)(0.12,-0.22){2}{\line(0,-1){0.22}}
\multiput(62.94,37.55)(0.13,-0.21){2}{\line(0,-1){0.21}}
\multiput(63.2,37.12)(0.13,-0.21){2}{\line(0,-1){0.21}}
\multiput(63.46,36.7)(0.14,-0.21){2}{\line(0,-1){0.21}}
\multiput(63.73,36.28)(0.14,-0.21){2}{\line(0,-1){0.21}}
\multiput(64.01,35.87)(0.14,-0.2){2}{\line(0,-1){0.2}}
\multiput(64.3,35.46)(0.15,-0.2){2}{\line(0,-1){0.2}}
\multiput(64.6,35.06)(0.1,-0.13){3}{\line(0,-1){0.13}}
\multiput(64.9,34.67)(0.11,-0.13){3}{\line(0,-1){0.13}}
\multiput(65.22,34.29)(0.11,-0.13){3}{\line(0,-1){0.13}}
\multiput(65.54,33.91)(0.11,-0.12){3}{\line(0,-1){0.12}}
\multiput(65.88,33.54)(0.11,-0.12){3}{\line(0,-1){0.12}}
\multiput(66.22,33.17)(0.12,-0.12){3}{\line(0,-1){0.12}}
\multiput(66.57,32.82)(0.12,-0.12){3}{\line(1,0){0.12}}
\multiput(66.92,32.47)(0.12,-0.11){3}{\line(1,0){0.12}}
\multiput(67.29,32.13)(0.12,-0.11){3}{\line(1,0){0.12}}
\multiput(67.66,31.79)(0.13,-0.11){3}{\line(1,0){0.13}}
\multiput(68.04,31.47)(0.13,-0.11){3}{\line(1,0){0.13}}
\multiput(68.42,31.15)(0.13,-0.1){3}{\line(1,0){0.13}}
\multiput(68.81,30.85)(0.2,-0.15){2}{\line(1,0){0.2}}
\multiput(69.21,30.55)(0.2,-0.14){2}{\line(1,0){0.2}}
\multiput(69.62,30.26)(0.21,-0.14){2}{\line(1,0){0.21}}
\multiput(70.03,29.98)(0.21,-0.14){2}{\line(1,0){0.21}}
\multiput(70.45,29.71)(0.21,-0.13){2}{\line(1,0){0.21}}
\multiput(70.87,29.45)(0.21,-0.13){2}{\line(1,0){0.21}}
\multiput(71.3,29.19)(0.22,-0.12){2}{\line(1,0){0.22}}
\multiput(71.74,28.95)(0.22,-0.12){2}{\line(1,0){0.22}}
\multiput(72.18,28.72)(0.22,-0.11){2}{\line(1,0){0.22}}
\multiput(72.63,28.49)(0.23,-0.11){2}{\line(1,0){0.23}}
\multiput(73.08,28.28)(0.23,-0.1){2}{\line(1,0){0.23}}
\multiput(73.53,28.08)(0.23,-0.1){2}{\line(1,0){0.23}}
\multiput(73.99,27.88)(0.23,-0.09){2}{\line(1,0){0.23}}
\multiput(74.46,27.7)(0.47,-0.17){1}{\line(1,0){0.47}}
\multiput(74.92,27.53)(0.47,-0.16){1}{\line(1,0){0.47}}
\multiput(75.39,27.36)(0.47,-0.15){1}{\line(1,0){0.47}}
\multiput(75.87,27.21)(0.48,-0.14){1}{\line(1,0){0.48}}
\multiput(76.35,27.07)(0.48,-0.13){1}{\line(1,0){0.48}}
\multiput(76.83,26.94)(0.48,-0.12){1}{\line(1,0){0.48}}
\multiput(77.31,26.82)(0.49,-0.11){1}{\line(1,0){0.49}}
\multiput(77.8,26.71)(0.49,-0.1){1}{\line(1,0){0.49}}
\multiput(78.29,26.61)(0.49,-0.09){1}{\line(1,0){0.49}}
\multiput(78.78,26.52)(0.49,-0.08){1}{\line(1,0){0.49}}
\multiput(79.27,26.45)(0.49,-0.07){1}{\line(1,0){0.49}}
\multiput(79.76,26.38)(0.5,-0.06){1}{\line(1,0){0.5}}
\multiput(80.26,26.33)(0.5,-0.04){1}{\line(1,0){0.5}}
\multiput(80.76,26.28)(0.5,-0.03){1}{\line(1,0){0.5}}
\multiput(81.25,26.25)(0.5,-0.02){1}{\line(1,0){0.5}}
\multiput(81.75,26.23)(0.5,-0.01){1}{\line(1,0){0.5}}
\put(82.25,26.22){\line(1,0){0.5}}
\multiput(82.75,26.22)(0.5,0.01){1}{\line(1,0){0.5}}
\multiput(83.25,26.23)(0.5,0.02){1}{\line(1,0){0.5}}
\multiput(83.75,26.25)(0.5,0.03){1}{\line(1,0){0.5}}
\multiput(84.24,26.28)(0.5,0.04){1}{\line(1,0){0.5}}
\multiput(84.74,26.33)(0.5,0.06){1}{\line(1,0){0.5}}
\multiput(85.24,26.38)(0.49,0.07){1}{\line(1,0){0.49}}
\multiput(85.73,26.45)(0.49,0.08){1}{\line(1,0){0.49}}
\multiput(86.22,26.52)(0.49,0.09){1}{\line(1,0){0.49}}
\multiput(86.71,26.61)(0.49,0.1){1}{\line(1,0){0.49}}
\multiput(87.2,26.71)(0.49,0.11){1}{\line(1,0){0.49}}
\multiput(87.69,26.82)(0.48,0.12){1}{\line(1,0){0.48}}
\multiput(88.17,26.94)(0.48,0.13){1}{\line(1,0){0.48}}
\multiput(88.65,27.07)(0.48,0.14){1}{\line(1,0){0.48}}
\multiput(89.13,27.21)(0.47,0.15){1}{\line(1,0){0.47}}
\multiput(89.61,27.36)(0.47,0.16){1}{\line(1,0){0.47}}
\multiput(90.08,27.53)(0.47,0.17){1}{\line(1,0){0.47}}
\multiput(90.54,27.7)(0.23,0.09){2}{\line(1,0){0.23}}
\multiput(91.01,27.88)(0.23,0.1){2}{\line(1,0){0.23}}
\multiput(91.47,28.08)(0.23,0.1){2}{\line(1,0){0.23}}
\multiput(91.92,28.28)(0.23,0.11){2}{\line(1,0){0.23}}
\multiput(92.37,28.49)(0.22,0.11){2}{\line(1,0){0.22}}
\multiput(92.82,28.72)(0.22,0.12){2}{\line(1,0){0.22}}
\multiput(93.26,28.95)(0.22,0.12){2}{\line(1,0){0.22}}
\multiput(93.7,29.19)(0.21,0.13){2}{\line(1,0){0.21}}
\multiput(94.13,29.45)(0.21,0.13){2}{\line(1,0){0.21}}
\multiput(94.55,29.71)(0.21,0.14){2}{\line(1,0){0.21}}
\multiput(94.97,29.98)(0.21,0.14){2}{\line(1,0){0.21}}
\multiput(95.38,30.26)(0.2,0.14){2}{\line(1,0){0.2}}
\multiput(95.79,30.55)(0.2,0.15){2}{\line(1,0){0.2}}
\multiput(96.19,30.85)(0.13,0.1){3}{\line(1,0){0.13}}
\multiput(96.58,31.15)(0.13,0.11){3}{\line(1,0){0.13}}
\multiput(96.96,31.47)(0.13,0.11){3}{\line(1,0){0.13}}
\multiput(97.34,31.79)(0.12,0.11){3}{\line(1,0){0.12}}
\multiput(97.71,32.13)(0.12,0.11){3}{\line(1,0){0.12}}
\multiput(98.08,32.47)(0.12,0.12){3}{\line(1,0){0.12}}
\multiput(98.43,32.82)(0.12,0.12){3}{\line(0,1){0.12}}
\multiput(98.78,33.17)(0.11,0.12){3}{\line(0,1){0.12}}
\multiput(99.12,33.54)(0.11,0.12){3}{\line(0,1){0.12}}
\multiput(99.46,33.91)(0.11,0.13){3}{\line(0,1){0.13}}
\multiput(99.78,34.29)(0.11,0.13){3}{\line(0,1){0.13}}
\multiput(100.1,34.67)(0.1,0.13){3}{\line(0,1){0.13}}
\multiput(100.4,35.06)(0.15,0.2){2}{\line(0,1){0.2}}
\multiput(100.7,35.46)(0.14,0.2){2}{\line(0,1){0.2}}
\multiput(100.99,35.87)(0.14,0.21){2}{\line(0,1){0.21}}
\multiput(101.27,36.28)(0.14,0.21){2}{\line(0,1){0.21}}
\multiput(101.54,36.7)(0.13,0.21){2}{\line(0,1){0.21}}
\multiput(101.8,37.12)(0.13,0.21){2}{\line(0,1){0.21}}
\multiput(102.06,37.55)(0.12,0.22){2}{\line(0,1){0.22}}
\multiput(102.3,37.99)(0.12,0.22){2}{\line(0,1){0.22}}
\multiput(102.53,38.43)(0.11,0.22){2}{\line(0,1){0.22}}
\multiput(102.76,38.88)(0.11,0.23){2}{\line(0,1){0.23}}
\multiput(102.97,39.33)(0.1,0.23){2}{\line(0,1){0.23}}
\multiput(103.17,39.78)(0.1,0.23){2}{\line(0,1){0.23}}
\multiput(103.37,40.24)(0.09,0.23){2}{\line(0,1){0.23}}
\multiput(103.55,40.71)(0.17,0.47){1}{\line(0,1){0.47}}
\multiput(103.72,41.17)(0.16,0.47){1}{\line(0,1){0.47}}
\multiput(103.89,41.64)(0.15,0.47){1}{\line(0,1){0.47}}
\multiput(104.04,42.12)(0.14,0.48){1}{\line(0,1){0.48}}
\multiput(104.18,42.6)(0.13,0.48){1}{\line(0,1){0.48}}
\multiput(104.31,43.08)(0.12,0.48){1}{\line(0,1){0.48}}
\multiput(104.43,43.56)(0.11,0.49){1}{\line(0,1){0.49}}
\multiput(104.54,44.05)(0.1,0.49){1}{\line(0,1){0.49}}
\multiput(104.64,44.54)(0.09,0.49){1}{\line(0,1){0.49}}
\multiput(104.73,45.03)(0.08,0.49){1}{\line(0,1){0.49}}
\multiput(104.8,45.52)(0.07,0.49){1}{\line(0,1){0.49}}
\multiput(104.87,46.01)(0.06,0.5){1}{\line(0,1){0.5}}
\multiput(104.92,46.51)(0.04,0.5){1}{\line(0,1){0.5}}
\multiput(104.97,47.01)(0.03,0.5){1}{\line(0,1){0.5}}
\multiput(105,47.5)(0.02,0.5){1}{\line(0,1){0.5}}
\multiput(105.02,48)(0.01,0.5){1}{\line(0,1){0.5}}

\linethickness{1mm}
\put(30,50){\line(1,0){30}}
\linethickness{1mm}
\multiput(60,90)(0.12,-0.16){125}{\line(0,-1){0.16}}
\linethickness{1mm}
\multiput(104,55)(0.36,0.12){83}{\line(1,0){0.36}}
\linethickness{0.3mm}
\multiput(95,30)(0.12,-0.16){125}{\line(0,-1){0.16}}
\put(35,55){\makebox(0,0)[cc]{$\eps_1,p_1$}}

\put(73,85){\makebox(0,0)[cc]{$\eps_2,p_2$}}

\put(95,80){\makebox(0,0)[cc]{$\cdot$}}

\put(110,70){\makebox(0,0)[cc]{$\cdot$}}

\put(122,55){\makebox(0,0)[cc]{$\eps_N,p_N$}}

\put(110,20){\makebox(0,0)[cc]{$\ve,k$}}

\put(80,50){\makebox(0,0)[cc]{$\wt\Gamma$}}

\end{picture}

}

\def\figthreefield{

\def\JPicScale{0.5}
\ifx\JPicScale\undefined\def\JPicScale{1}\fi
\unitlength \JPicScale mm
\begin{picture}(135,90)(0,0)
\linethickness{0.3mm}
\put(105.03,48.5){\line(0,1){0.5}}
\multiput(105.02,49.5)(0.01,-0.5){1}{\line(0,-1){0.5}}
\multiput(105,50)(0.02,-0.5){1}{\line(0,-1){0.5}}
\multiput(104.97,50.49)(0.03,-0.5){1}{\line(0,-1){0.5}}
\multiput(104.92,50.99)(0.04,-0.5){1}{\line(0,-1){0.5}}
\multiput(104.87,51.49)(0.06,-0.5){1}{\line(0,-1){0.5}}
\multiput(104.8,51.98)(0.07,-0.49){1}{\line(0,-1){0.49}}
\multiput(104.73,52.47)(0.08,-0.49){1}{\line(0,-1){0.49}}
\multiput(104.64,52.96)(0.09,-0.49){1}{\line(0,-1){0.49}}
\multiput(104.54,53.45)(0.1,-0.49){1}{\line(0,-1){0.49}}
\multiput(104.43,53.94)(0.11,-0.49){1}{\line(0,-1){0.49}}
\multiput(104.31,54.42)(0.12,-0.48){1}{\line(0,-1){0.48}}
\multiput(104.18,54.9)(0.13,-0.48){1}{\line(0,-1){0.48}}
\multiput(104.04,55.38)(0.14,-0.48){1}{\line(0,-1){0.48}}
\multiput(103.89,55.86)(0.15,-0.47){1}{\line(0,-1){0.47}}
\multiput(103.72,56.33)(0.16,-0.47){1}{\line(0,-1){0.47}}
\multiput(103.55,56.79)(0.17,-0.47){1}{\line(0,-1){0.47}}
\multiput(103.37,57.26)(0.09,-0.23){2}{\line(0,-1){0.23}}
\multiput(103.17,57.72)(0.1,-0.23){2}{\line(0,-1){0.23}}
\multiput(102.97,58.17)(0.1,-0.23){2}{\line(0,-1){0.23}}
\multiput(102.76,58.62)(0.11,-0.23){2}{\line(0,-1){0.23}}
\multiput(102.53,59.07)(0.11,-0.22){2}{\line(0,-1){0.22}}
\multiput(102.3,59.51)(0.12,-0.22){2}{\line(0,-1){0.22}}
\multiput(102.06,59.95)(0.12,-0.22){2}{\line(0,-1){0.22}}
\multiput(101.8,60.38)(0.13,-0.21){2}{\line(0,-1){0.21}}
\multiput(101.54,60.8)(0.13,-0.21){2}{\line(0,-1){0.21}}
\multiput(101.27,61.22)(0.14,-0.21){2}{\line(0,-1){0.21}}
\multiput(100.99,61.63)(0.14,-0.21){2}{\line(0,-1){0.21}}
\multiput(100.7,62.04)(0.14,-0.2){2}{\line(0,-1){0.2}}
\multiput(100.4,62.44)(0.15,-0.2){2}{\line(0,-1){0.2}}
\multiput(100.1,62.83)(0.1,-0.13){3}{\line(0,-1){0.13}}
\multiput(99.78,63.21)(0.11,-0.13){3}{\line(0,-1){0.13}}
\multiput(99.46,63.59)(0.11,-0.13){3}{\line(0,-1){0.13}}
\multiput(99.12,63.96)(0.11,-0.12){3}{\line(0,-1){0.12}}
\multiput(98.78,64.33)(0.11,-0.12){3}{\line(0,-1){0.12}}
\multiput(98.43,64.68)(0.12,-0.12){3}{\line(0,-1){0.12}}
\multiput(98.08,65.03)(0.12,-0.12){3}{\line(1,0){0.12}}
\multiput(97.71,65.37)(0.12,-0.11){3}{\line(1,0){0.12}}
\multiput(97.34,65.71)(0.12,-0.11){3}{\line(1,0){0.12}}
\multiput(96.96,66.03)(0.13,-0.11){3}{\line(1,0){0.13}}
\multiput(96.58,66.35)(0.13,-0.11){3}{\line(1,0){0.13}}
\multiput(96.19,66.65)(0.13,-0.1){3}{\line(1,0){0.13}}
\multiput(95.79,66.95)(0.2,-0.15){2}{\line(1,0){0.2}}
\multiput(95.38,67.24)(0.2,-0.14){2}{\line(1,0){0.2}}
\multiput(94.97,67.52)(0.21,-0.14){2}{\line(1,0){0.21}}
\multiput(94.55,67.79)(0.21,-0.14){2}{\line(1,0){0.21}}
\multiput(94.13,68.05)(0.21,-0.13){2}{\line(1,0){0.21}}
\multiput(93.7,68.31)(0.21,-0.13){2}{\line(1,0){0.21}}
\multiput(93.26,68.55)(0.22,-0.12){2}{\line(1,0){0.22}}
\multiput(92.82,68.78)(0.22,-0.12){2}{\line(1,0){0.22}}
\multiput(92.37,69.01)(0.22,-0.11){2}{\line(1,0){0.22}}
\multiput(91.92,69.22)(0.23,-0.11){2}{\line(1,0){0.23}}
\multiput(91.47,69.42)(0.23,-0.1){2}{\line(1,0){0.23}}
\multiput(91.01,69.62)(0.23,-0.1){2}{\line(1,0){0.23}}
\multiput(90.54,69.8)(0.23,-0.09){2}{\line(1,0){0.23}}
\multiput(90.08,69.97)(0.47,-0.17){1}{\line(1,0){0.47}}
\multiput(89.61,70.14)(0.47,-0.16){1}{\line(1,0){0.47}}
\multiput(89.13,70.29)(0.47,-0.15){1}{\line(1,0){0.47}}
\multiput(88.65,70.43)(0.48,-0.14){1}{\line(1,0){0.48}}
\multiput(88.17,70.56)(0.48,-0.13){1}{\line(1,0){0.48}}
\multiput(87.69,70.68)(0.48,-0.12){1}{\line(1,0){0.48}}
\multiput(87.2,70.79)(0.49,-0.11){1}{\line(1,0){0.49}}
\multiput(86.71,70.89)(0.49,-0.1){1}{\line(1,0){0.49}}
\multiput(86.22,70.98)(0.49,-0.09){1}{\line(1,0){0.49}}
\multiput(85.73,71.05)(0.49,-0.08){1}{\line(1,0){0.49}}
\multiput(85.24,71.12)(0.49,-0.07){1}{\line(1,0){0.49}}
\multiput(84.74,71.17)(0.5,-0.06){1}{\line(1,0){0.5}}
\multiput(84.24,71.22)(0.5,-0.04){1}{\line(1,0){0.5}}
\multiput(83.75,71.25)(0.5,-0.03){1}{\line(1,0){0.5}}
\multiput(83.25,71.27)(0.5,-0.02){1}{\line(1,0){0.5}}
\multiput(82.75,71.28)(0.5,-0.01){1}{\line(1,0){0.5}}
\put(82.25,71.28){\line(1,0){0.5}}
\multiput(81.75,71.27)(0.5,0.01){1}{\line(1,0){0.5}}
\multiput(81.25,71.25)(0.5,0.02){1}{\line(1,0){0.5}}
\multiput(80.76,71.22)(0.5,0.03){1}{\line(1,0){0.5}}
\multiput(80.26,71.17)(0.5,0.04){1}{\line(1,0){0.5}}
\multiput(79.76,71.12)(0.5,0.06){1}{\line(1,0){0.5}}
\multiput(79.27,71.05)(0.49,0.07){1}{\line(1,0){0.49}}
\multiput(78.78,70.98)(0.49,0.08){1}{\line(1,0){0.49}}
\multiput(78.29,70.89)(0.49,0.09){1}{\line(1,0){0.49}}
\multiput(77.8,70.79)(0.49,0.1){1}{\line(1,0){0.49}}
\multiput(77.31,70.68)(0.49,0.11){1}{\line(1,0){0.49}}
\multiput(76.83,70.56)(0.48,0.12){1}{\line(1,0){0.48}}
\multiput(76.35,70.43)(0.48,0.13){1}{\line(1,0){0.48}}
\multiput(75.87,70.29)(0.48,0.14){1}{\line(1,0){0.48}}
\multiput(75.39,70.14)(0.47,0.15){1}{\line(1,0){0.47}}
\multiput(74.92,69.97)(0.47,0.16){1}{\line(1,0){0.47}}
\multiput(74.46,69.8)(0.47,0.17){1}{\line(1,0){0.47}}
\multiput(73.99,69.62)(0.23,0.09){2}{\line(1,0){0.23}}
\multiput(73.53,69.42)(0.23,0.1){2}{\line(1,0){0.23}}
\multiput(73.08,69.22)(0.23,0.1){2}{\line(1,0){0.23}}
\multiput(72.63,69.01)(0.23,0.11){2}{\line(1,0){0.23}}
\multiput(72.18,68.78)(0.22,0.11){2}{\line(1,0){0.22}}
\multiput(71.74,68.55)(0.22,0.12){2}{\line(1,0){0.22}}
\multiput(71.3,68.31)(0.22,0.12){2}{\line(1,0){0.22}}
\multiput(70.87,68.05)(0.21,0.13){2}{\line(1,0){0.21}}
\multiput(70.45,67.79)(0.21,0.13){2}{\line(1,0){0.21}}
\multiput(70.03,67.52)(0.21,0.14){2}{\line(1,0){0.21}}
\multiput(69.62,67.24)(0.21,0.14){2}{\line(1,0){0.21}}
\multiput(69.21,66.95)(0.2,0.14){2}{\line(1,0){0.2}}
\multiput(68.81,66.65)(0.2,0.15){2}{\line(1,0){0.2}}
\multiput(68.42,66.35)(0.13,0.1){3}{\line(1,0){0.13}}
\multiput(68.04,66.03)(0.13,0.11){3}{\line(1,0){0.13}}
\multiput(67.66,65.71)(0.13,0.11){3}{\line(1,0){0.13}}
\multiput(67.29,65.37)(0.12,0.11){3}{\line(1,0){0.12}}
\multiput(66.92,65.03)(0.12,0.11){3}{\line(1,0){0.12}}
\multiput(66.57,64.68)(0.12,0.12){3}{\line(1,0){0.12}}
\multiput(66.22,64.33)(0.12,0.12){3}{\line(0,1){0.12}}
\multiput(65.88,63.96)(0.11,0.12){3}{\line(0,1){0.12}}
\multiput(65.54,63.59)(0.11,0.12){3}{\line(0,1){0.12}}
\multiput(65.22,63.21)(0.11,0.13){3}{\line(0,1){0.13}}
\multiput(64.9,62.83)(0.11,0.13){3}{\line(0,1){0.13}}
\multiput(64.6,62.44)(0.1,0.13){3}{\line(0,1){0.13}}
\multiput(64.3,62.04)(0.15,0.2){2}{\line(0,1){0.2}}
\multiput(64.01,61.63)(0.14,0.2){2}{\line(0,1){0.2}}
\multiput(63.73,61.22)(0.14,0.21){2}{\line(0,1){0.21}}
\multiput(63.46,60.8)(0.14,0.21){2}{\line(0,1){0.21}}
\multiput(63.2,60.38)(0.13,0.21){2}{\line(0,1){0.21}}
\multiput(62.94,59.95)(0.13,0.21){2}{\line(0,1){0.21}}
\multiput(62.7,59.51)(0.12,0.22){2}{\line(0,1){0.22}}
\multiput(62.47,59.07)(0.12,0.22){2}{\line(0,1){0.22}}
\multiput(62.24,58.62)(0.11,0.22){2}{\line(0,1){0.22}}
\multiput(62.03,58.17)(0.11,0.23){2}{\line(0,1){0.23}}
\multiput(61.83,57.72)(0.1,0.23){2}{\line(0,1){0.23}}
\multiput(61.63,57.26)(0.1,0.23){2}{\line(0,1){0.23}}
\multiput(61.45,56.79)(0.09,0.23){2}{\line(0,1){0.23}}
\multiput(61.28,56.33)(0.17,0.47){1}{\line(0,1){0.47}}
\multiput(61.11,55.86)(0.16,0.47){1}{\line(0,1){0.47}}
\multiput(60.96,55.38)(0.15,0.47){1}{\line(0,1){0.47}}
\multiput(60.82,54.9)(0.14,0.48){1}{\line(0,1){0.48}}
\multiput(60.69,54.42)(0.13,0.48){1}{\line(0,1){0.48}}
\multiput(60.57,53.94)(0.12,0.48){1}{\line(0,1){0.48}}
\multiput(60.46,53.45)(0.11,0.49){1}{\line(0,1){0.49}}
\multiput(60.36,52.96)(0.1,0.49){1}{\line(0,1){0.49}}
\multiput(60.27,52.47)(0.09,0.49){1}{\line(0,1){0.49}}
\multiput(60.2,51.98)(0.08,0.49){1}{\line(0,1){0.49}}
\multiput(60.13,51.49)(0.07,0.49){1}{\line(0,1){0.49}}
\multiput(60.08,50.99)(0.06,0.5){1}{\line(0,1){0.5}}
\multiput(60.03,50.49)(0.04,0.5){1}{\line(0,1){0.5}}
\multiput(60,50)(0.03,0.5){1}{\line(0,1){0.5}}
\multiput(59.98,49.5)(0.02,0.5){1}{\line(0,1){0.5}}
\multiput(59.97,49)(0.01,0.5){1}{\line(0,1){0.5}}
\put(59.97,48.5){\line(0,1){0.5}}
\multiput(59.97,48.5)(0.01,-0.5){1}{\line(0,-1){0.5}}
\multiput(59.98,48)(0.02,-0.5){1}{\line(0,-1){0.5}}
\multiput(60,47.5)(0.03,-0.5){1}{\line(0,-1){0.5}}
\multiput(60.03,47.01)(0.04,-0.5){1}{\line(0,-1){0.5}}
\multiput(60.08,46.51)(0.06,-0.5){1}{\line(0,-1){0.5}}
\multiput(60.13,46.01)(0.07,-0.49){1}{\line(0,-1){0.49}}
\multiput(60.2,45.52)(0.08,-0.49){1}{\line(0,-1){0.49}}
\multiput(60.27,45.03)(0.09,-0.49){1}{\line(0,-1){0.49}}
\multiput(60.36,44.54)(0.1,-0.49){1}{\line(0,-1){0.49}}
\multiput(60.46,44.05)(0.11,-0.49){1}{\line(0,-1){0.49}}
\multiput(60.57,43.56)(0.12,-0.48){1}{\line(0,-1){0.48}}
\multiput(60.69,43.08)(0.13,-0.48){1}{\line(0,-1){0.48}}
\multiput(60.82,42.6)(0.14,-0.48){1}{\line(0,-1){0.48}}
\multiput(60.96,42.12)(0.15,-0.47){1}{\line(0,-1){0.47}}
\multiput(61.11,41.64)(0.16,-0.47){1}{\line(0,-1){0.47}}
\multiput(61.28,41.17)(0.17,-0.47){1}{\line(0,-1){0.47}}
\multiput(61.45,40.71)(0.09,-0.23){2}{\line(0,-1){0.23}}
\multiput(61.63,40.24)(0.1,-0.23){2}{\line(0,-1){0.23}}
\multiput(61.83,39.78)(0.1,-0.23){2}{\line(0,-1){0.23}}
\multiput(62.03,39.33)(0.11,-0.23){2}{\line(0,-1){0.23}}
\multiput(62.24,38.88)(0.11,-0.22){2}{\line(0,-1){0.22}}
\multiput(62.47,38.43)(0.12,-0.22){2}{\line(0,-1){0.22}}
\multiput(62.7,37.99)(0.12,-0.22){2}{\line(0,-1){0.22}}
\multiput(62.94,37.55)(0.13,-0.21){2}{\line(0,-1){0.21}}
\multiput(63.2,37.12)(0.13,-0.21){2}{\line(0,-1){0.21}}
\multiput(63.46,36.7)(0.14,-0.21){2}{\line(0,-1){0.21}}
\multiput(63.73,36.28)(0.14,-0.21){2}{\line(0,-1){0.21}}
\multiput(64.01,35.87)(0.14,-0.2){2}{\line(0,-1){0.2}}
\multiput(64.3,35.46)(0.15,-0.2){2}{\line(0,-1){0.2}}
\multiput(64.6,35.06)(0.1,-0.13){3}{\line(0,-1){0.13}}
\multiput(64.9,34.67)(0.11,-0.13){3}{\line(0,-1){0.13}}
\multiput(65.22,34.29)(0.11,-0.13){3}{\line(0,-1){0.13}}
\multiput(65.54,33.91)(0.11,-0.12){3}{\line(0,-1){0.12}}
\multiput(65.88,33.54)(0.11,-0.12){3}{\line(0,-1){0.12}}
\multiput(66.22,33.17)(0.12,-0.12){3}{\line(0,-1){0.12}}
\multiput(66.57,32.82)(0.12,-0.12){3}{\line(1,0){0.12}}
\multiput(66.92,32.47)(0.12,-0.11){3}{\line(1,0){0.12}}
\multiput(67.29,32.13)(0.12,-0.11){3}{\line(1,0){0.12}}
\multiput(67.66,31.79)(0.13,-0.11){3}{\line(1,0){0.13}}
\multiput(68.04,31.47)(0.13,-0.11){3}{\line(1,0){0.13}}
\multiput(68.42,31.15)(0.13,-0.1){3}{\line(1,0){0.13}}
\multiput(68.81,30.85)(0.2,-0.15){2}{\line(1,0){0.2}}
\multiput(69.21,30.55)(0.2,-0.14){2}{\line(1,0){0.2}}
\multiput(69.62,30.26)(0.21,-0.14){2}{\line(1,0){0.21}}
\multiput(70.03,29.98)(0.21,-0.14){2}{\line(1,0){0.21}}
\multiput(70.45,29.71)(0.21,-0.13){2}{\line(1,0){0.21}}
\multiput(70.87,29.45)(0.21,-0.13){2}{\line(1,0){0.21}}
\multiput(71.3,29.19)(0.22,-0.12){2}{\line(1,0){0.22}}
\multiput(71.74,28.95)(0.22,-0.12){2}{\line(1,0){0.22}}
\multiput(72.18,28.72)(0.22,-0.11){2}{\line(1,0){0.22}}
\multiput(72.63,28.49)(0.23,-0.11){2}{\line(1,0){0.23}}
\multiput(73.08,28.28)(0.23,-0.1){2}{\line(1,0){0.23}}
\multiput(73.53,28.08)(0.23,-0.1){2}{\line(1,0){0.23}}
\multiput(73.99,27.88)(0.23,-0.09){2}{\line(1,0){0.23}}
\multiput(74.46,27.7)(0.47,-0.17){1}{\line(1,0){0.47}}
\multiput(74.92,27.53)(0.47,-0.16){1}{\line(1,0){0.47}}
\multiput(75.39,27.36)(0.47,-0.15){1}{\line(1,0){0.47}}
\multiput(75.87,27.21)(0.48,-0.14){1}{\line(1,0){0.48}}
\multiput(76.35,27.07)(0.48,-0.13){1}{\line(1,0){0.48}}
\multiput(76.83,26.94)(0.48,-0.12){1}{\line(1,0){0.48}}
\multiput(77.31,26.82)(0.49,-0.11){1}{\line(1,0){0.49}}
\multiput(77.8,26.71)(0.49,-0.1){1}{\line(1,0){0.49}}
\multiput(78.29,26.61)(0.49,-0.09){1}{\line(1,0){0.49}}
\multiput(78.78,26.52)(0.49,-0.08){1}{\line(1,0){0.49}}
\multiput(79.27,26.45)(0.49,-0.07){1}{\line(1,0){0.49}}
\multiput(79.76,26.38)(0.5,-0.06){1}{\line(1,0){0.5}}
\multiput(80.26,26.33)(0.5,-0.04){1}{\line(1,0){0.5}}
\multiput(80.76,26.28)(0.5,-0.03){1}{\line(1,0){0.5}}
\multiput(81.25,26.25)(0.5,-0.02){1}{\line(1,0){0.5}}
\multiput(81.75,26.23)(0.5,-0.01){1}{\line(1,0){0.5}}
\put(82.25,26.22){\line(1,0){0.5}}
\multiput(82.75,26.22)(0.5,0.01){1}{\line(1,0){0.5}}
\multiput(83.25,26.23)(0.5,0.02){1}{\line(1,0){0.5}}
\multiput(83.75,26.25)(0.5,0.03){1}{\line(1,0){0.5}}
\multiput(84.24,26.28)(0.5,0.04){1}{\line(1,0){0.5}}
\multiput(84.74,26.33)(0.5,0.06){1}{\line(1,0){0.5}}
\multiput(85.24,26.38)(0.49,0.07){1}{\line(1,0){0.49}}
\multiput(85.73,26.45)(0.49,0.08){1}{\line(1,0){0.49}}
\multiput(86.22,26.52)(0.49,0.09){1}{\line(1,0){0.49}}
\multiput(86.71,26.61)(0.49,0.1){1}{\line(1,0){0.49}}
\multiput(87.2,26.71)(0.49,0.11){1}{\line(1,0){0.49}}
\multiput(87.69,26.82)(0.48,0.12){1}{\line(1,0){0.48}}
\multiput(88.17,26.94)(0.48,0.13){1}{\line(1,0){0.48}}
\multiput(88.65,27.07)(0.48,0.14){1}{\line(1,0){0.48}}
\multiput(89.13,27.21)(0.47,0.15){1}{\line(1,0){0.47}}
\multiput(89.61,27.36)(0.47,0.16){1}{\line(1,0){0.47}}
\multiput(90.08,27.53)(0.47,0.17){1}{\line(1,0){0.47}}
\multiput(90.54,27.7)(0.23,0.09){2}{\line(1,0){0.23}}
\multiput(91.01,27.88)(0.23,0.1){2}{\line(1,0){0.23}}
\multiput(91.47,28.08)(0.23,0.1){2}{\line(1,0){0.23}}
\multiput(91.92,28.28)(0.23,0.11){2}{\line(1,0){0.23}}
\multiput(92.37,28.49)(0.22,0.11){2}{\line(1,0){0.22}}
\multiput(92.82,28.72)(0.22,0.12){2}{\line(1,0){0.22}}
\multiput(93.26,28.95)(0.22,0.12){2}{\line(1,0){0.22}}
\multiput(93.7,29.19)(0.21,0.13){2}{\line(1,0){0.21}}
\multiput(94.13,29.45)(0.21,0.13){2}{\line(1,0){0.21}}
\multiput(94.55,29.71)(0.21,0.14){2}{\line(1,0){0.21}}
\multiput(94.97,29.98)(0.21,0.14){2}{\line(1,0){0.21}}
\multiput(95.38,30.26)(0.2,0.14){2}{\line(1,0){0.2}}
\multiput(95.79,30.55)(0.2,0.15){2}{\line(1,0){0.2}}
\multiput(96.19,30.85)(0.13,0.1){3}{\line(1,0){0.13}}
\multiput(96.58,31.15)(0.13,0.11){3}{\line(1,0){0.13}}
\multiput(96.96,31.47)(0.13,0.11){3}{\line(1,0){0.13}}
\multiput(97.34,31.79)(0.12,0.11){3}{\line(1,0){0.12}}
\multiput(97.71,32.13)(0.12,0.11){3}{\line(1,0){0.12}}
\multiput(98.08,32.47)(0.12,0.12){3}{\line(1,0){0.12}}
\multiput(98.43,32.82)(0.12,0.12){3}{\line(0,1){0.12}}
\multiput(98.78,33.17)(0.11,0.12){3}{\line(0,1){0.12}}
\multiput(99.12,33.54)(0.11,0.12){3}{\line(0,1){0.12}}
\multiput(99.46,33.91)(0.11,0.13){3}{\line(0,1){0.13}}
\multiput(99.78,34.29)(0.11,0.13){3}{\line(0,1){0.13}}
\multiput(100.1,34.67)(0.1,0.13){3}{\line(0,1){0.13}}
\multiput(100.4,35.06)(0.15,0.2){2}{\line(0,1){0.2}}
\multiput(100.7,35.46)(0.14,0.2){2}{\line(0,1){0.2}}
\multiput(100.99,35.87)(0.14,0.21){2}{\line(0,1){0.21}}
\multiput(101.27,36.28)(0.14,0.21){2}{\line(0,1){0.21}}
\multiput(101.54,36.7)(0.13,0.21){2}{\line(0,1){0.21}}
\multiput(101.8,37.12)(0.13,0.21){2}{\line(0,1){0.21}}
\multiput(102.06,37.55)(0.12,0.22){2}{\line(0,1){0.22}}
\multiput(102.3,37.99)(0.12,0.22){2}{\line(0,1){0.22}}
\multiput(102.53,38.43)(0.11,0.22){2}{\line(0,1){0.22}}
\multiput(102.76,38.88)(0.11,0.23){2}{\line(0,1){0.23}}
\multiput(102.97,39.33)(0.1,0.23){2}{\line(0,1){0.23}}
\multiput(103.17,39.78)(0.1,0.23){2}{\line(0,1){0.23}}
\multiput(103.37,40.24)(0.09,0.23){2}{\line(0,1){0.23}}
\multiput(103.55,40.71)(0.17,0.47){1}{\line(0,1){0.47}}
\multiput(103.72,41.17)(0.16,0.47){1}{\line(0,1){0.47}}
\multiput(103.89,41.64)(0.15,0.47){1}{\line(0,1){0.47}}
\multiput(104.04,42.12)(0.14,0.48){1}{\line(0,1){0.48}}
\multiput(104.18,42.6)(0.13,0.48){1}{\line(0,1){0.48}}
\multiput(104.31,43.08)(0.12,0.48){1}{\line(0,1){0.48}}
\multiput(104.43,43.56)(0.11,0.49){1}{\line(0,1){0.49}}
\multiput(104.54,44.05)(0.1,0.49){1}{\line(0,1){0.49}}
\multiput(104.64,44.54)(0.09,0.49){1}{\line(0,1){0.49}}
\multiput(104.73,45.03)(0.08,0.49){1}{\line(0,1){0.49}}
\multiput(104.8,45.52)(0.07,0.49){1}{\line(0,1){0.49}}
\multiput(104.87,46.01)(0.06,0.5){1}{\line(0,1){0.5}}
\multiput(104.92,46.51)(0.04,0.5){1}{\line(0,1){0.5}}
\multiput(104.97,47.01)(0.03,0.5){1}{\line(0,1){0.5}}
\multiput(105,47.5)(0.02,0.5){1}{\line(0,1){0.5}}
\multiput(105.02,48)(0.01,0.5){1}{\line(0,1){0.5}}

\linethickness{1mm}
\put(30,50){\line(1,0){30}}
\linethickness{1mm}
\multiput(60,90)(0.12,-0.16){125}{\line(0,-1){0.16}}
\linethickness{1mm}
\multiput(104,55)(0.36,0.12){83}{\line(1,0){0.36}}
\put(35,55){\makebox(0,0)[cc]{$\eps_1,p_1$}}

\put(78,85){\makebox(0,0)[cc]{$\eps_2,p_2$}}

\put(95,80){\makebox(0,0)[cc]{$\cdot$}}

\put(110,70){\makebox(0,0)[cc]{$\cdot$}}

\put(126,55){\makebox(0,0)[cc]{$\eps_N,p_N$}}

\put(80,50){\makebox(0,0)[cc]{$\Gamma$}}

\end{picture}

}

\def\figsubtleb{

\def\JPicScale{0.5}
\ifx\JPicScale\undefined\def\JPicScale{1}\fi
\unitlength \JPicScale mm
\begin{picture}(140,75.41)(0,0)
\linethickness{0.3mm}
\put(55.4,67.25){\line(0,1){0.5}}
\multiput(55.37,68.24)(0.03,-0.5){1}{\line(0,-1){0.5}}
\multiput(55.31,68.74)(0.06,-0.49){1}{\line(0,-1){0.49}}
\multiput(55.22,69.22)(0.09,-0.49){1}{\line(0,-1){0.49}}
\multiput(55.09,69.71)(0.12,-0.48){1}{\line(0,-1){0.48}}
\multiput(54.94,70.18)(0.15,-0.47){1}{\line(0,-1){0.47}}
\multiput(54.76,70.64)(0.09,-0.23){2}{\line(0,-1){0.23}}
\multiput(54.54,71.09)(0.11,-0.22){2}{\line(0,-1){0.22}}
\multiput(54.3,71.52)(0.12,-0.22){2}{\line(0,-1){0.22}}
\multiput(54.04,71.94)(0.13,-0.21){2}{\line(0,-1){0.21}}
\multiput(53.75,72.35)(0.15,-0.2){2}{\line(0,-1){0.2}}
\multiput(53.43,72.73)(0.11,-0.13){3}{\line(0,-1){0.13}}
\multiput(53.09,73.09)(0.11,-0.12){3}{\line(0,-1){0.12}}
\multiput(52.73,73.43)(0.12,-0.11){3}{\line(1,0){0.12}}
\multiput(52.35,73.75)(0.13,-0.11){3}{\line(1,0){0.13}}
\multiput(51.94,74.04)(0.2,-0.15){2}{\line(1,0){0.2}}
\multiput(51.52,74.3)(0.21,-0.13){2}{\line(1,0){0.21}}
\multiput(51.09,74.54)(0.22,-0.12){2}{\line(1,0){0.22}}
\multiput(50.64,74.76)(0.22,-0.11){2}{\line(1,0){0.22}}
\multiput(50.18,74.94)(0.23,-0.09){2}{\line(1,0){0.23}}
\multiput(49.71,75.09)(0.47,-0.15){1}{\line(1,0){0.47}}
\multiput(49.22,75.22)(0.48,-0.12){1}{\line(1,0){0.48}}
\multiput(48.74,75.31)(0.49,-0.09){1}{\line(1,0){0.49}}
\multiput(48.24,75.37)(0.49,-0.06){1}{\line(1,0){0.49}}
\multiput(47.75,75.4)(0.5,-0.03){1}{\line(1,0){0.5}}
\put(47.25,75.4){\line(1,0){0.5}}
\multiput(46.76,75.37)(0.5,0.03){1}{\line(1,0){0.5}}
\multiput(46.26,75.31)(0.49,0.06){1}{\line(1,0){0.49}}
\multiput(45.78,75.22)(0.49,0.09){1}{\line(1,0){0.49}}
\multiput(45.29,75.09)(0.48,0.12){1}{\line(1,0){0.48}}
\multiput(44.82,74.94)(0.47,0.15){1}{\line(1,0){0.47}}
\multiput(44.36,74.76)(0.23,0.09){2}{\line(1,0){0.23}}
\multiput(43.91,74.54)(0.22,0.11){2}{\line(1,0){0.22}}
\multiput(43.48,74.3)(0.22,0.12){2}{\line(1,0){0.22}}
\multiput(43.06,74.04)(0.21,0.13){2}{\line(1,0){0.21}}
\multiput(42.65,73.75)(0.2,0.15){2}{\line(1,0){0.2}}
\multiput(42.27,73.43)(0.13,0.11){3}{\line(1,0){0.13}}
\multiput(41.91,73.09)(0.12,0.11){3}{\line(1,0){0.12}}
\multiput(41.57,72.73)(0.11,0.12){3}{\line(0,1){0.12}}
\multiput(41.25,72.35)(0.11,0.13){3}{\line(0,1){0.13}}
\multiput(40.96,71.94)(0.15,0.2){2}{\line(0,1){0.2}}
\multiput(40.7,71.52)(0.13,0.21){2}{\line(0,1){0.21}}
\multiput(40.46,71.09)(0.12,0.22){2}{\line(0,1){0.22}}
\multiput(40.24,70.64)(0.11,0.22){2}{\line(0,1){0.22}}
\multiput(40.06,70.18)(0.09,0.23){2}{\line(0,1){0.23}}
\multiput(39.91,69.71)(0.15,0.47){1}{\line(0,1){0.47}}
\multiput(39.78,69.22)(0.12,0.48){1}{\line(0,1){0.48}}
\multiput(39.69,68.74)(0.09,0.49){1}{\line(0,1){0.49}}
\multiput(39.63,68.24)(0.06,0.49){1}{\line(0,1){0.49}}
\multiput(39.6,67.75)(0.03,0.5){1}{\line(0,1){0.5}}
\put(39.6,67.25){\line(0,1){0.5}}
\multiput(39.6,67.25)(0.03,-0.5){1}{\line(0,-1){0.5}}
\multiput(39.63,66.76)(0.06,-0.49){1}{\line(0,-1){0.49}}
\multiput(39.69,66.26)(0.09,-0.49){1}{\line(0,-1){0.49}}
\multiput(39.78,65.78)(0.12,-0.48){1}{\line(0,-1){0.48}}
\multiput(39.91,65.29)(0.15,-0.47){1}{\line(0,-1){0.47}}
\multiput(40.06,64.82)(0.09,-0.23){2}{\line(0,-1){0.23}}
\multiput(40.24,64.36)(0.11,-0.22){2}{\line(0,-1){0.22}}
\multiput(40.46,63.91)(0.12,-0.22){2}{\line(0,-1){0.22}}
\multiput(40.7,63.48)(0.13,-0.21){2}{\line(0,-1){0.21}}
\multiput(40.96,63.06)(0.15,-0.2){2}{\line(0,-1){0.2}}
\multiput(41.25,62.65)(0.11,-0.13){3}{\line(0,-1){0.13}}
\multiput(41.57,62.27)(0.11,-0.12){3}{\line(0,-1){0.12}}
\multiput(41.91,61.91)(0.12,-0.11){3}{\line(1,0){0.12}}
\multiput(42.27,61.57)(0.13,-0.11){3}{\line(1,0){0.13}}
\multiput(42.65,61.25)(0.2,-0.15){2}{\line(1,0){0.2}}
\multiput(43.06,60.96)(0.21,-0.13){2}{\line(1,0){0.21}}
\multiput(43.48,60.7)(0.22,-0.12){2}{\line(1,0){0.22}}
\multiput(43.91,60.46)(0.22,-0.11){2}{\line(1,0){0.22}}
\multiput(44.36,60.24)(0.23,-0.09){2}{\line(1,0){0.23}}
\multiput(44.82,60.06)(0.47,-0.15){1}{\line(1,0){0.47}}
\multiput(45.29,59.91)(0.48,-0.12){1}{\line(1,0){0.48}}
\multiput(45.78,59.78)(0.49,-0.09){1}{\line(1,0){0.49}}
\multiput(46.26,59.69)(0.49,-0.06){1}{\line(1,0){0.49}}
\multiput(46.76,59.63)(0.5,-0.03){1}{\line(1,0){0.5}}
\put(47.25,59.6){\line(1,0){0.5}}
\multiput(47.75,59.6)(0.5,0.03){1}{\line(1,0){0.5}}
\multiput(48.24,59.63)(0.49,0.06){1}{\line(1,0){0.49}}
\multiput(48.74,59.69)(0.49,0.09){1}{\line(1,0){0.49}}
\multiput(49.22,59.78)(0.48,0.12){1}{\line(1,0){0.48}}
\multiput(49.71,59.91)(0.47,0.15){1}{\line(1,0){0.47}}
\multiput(50.18,60.06)(0.23,0.09){2}{\line(1,0){0.23}}
\multiput(50.64,60.24)(0.22,0.11){2}{\line(1,0){0.22}}
\multiput(51.09,60.46)(0.22,0.12){2}{\line(1,0){0.22}}
\multiput(51.52,60.7)(0.21,0.13){2}{\line(1,0){0.21}}
\multiput(51.94,60.96)(0.2,0.15){2}{\line(1,0){0.2}}
\multiput(52.35,61.25)(0.13,0.11){3}{\line(1,0){0.13}}
\multiput(52.73,61.57)(0.12,0.11){3}{\line(1,0){0.12}}
\multiput(53.09,61.91)(0.11,0.12){3}{\line(0,1){0.12}}
\multiput(53.43,62.27)(0.11,0.13){3}{\line(0,1){0.13}}
\multiput(53.75,62.65)(0.15,0.2){2}{\line(0,1){0.2}}
\multiput(54.04,63.06)(0.13,0.21){2}{\line(0,1){0.21}}
\multiput(54.3,63.48)(0.12,0.22){2}{\line(0,1){0.22}}
\multiput(54.54,63.91)(0.11,0.22){2}{\line(0,1){0.22}}
\multiput(54.76,64.36)(0.09,0.23){2}{\line(0,1){0.23}}
\multiput(54.94,64.82)(0.15,0.47){1}{\line(0,1){0.47}}
\multiput(55.09,65.29)(0.12,0.48){1}{\line(0,1){0.48}}
\multiput(55.22,65.78)(0.09,0.49){1}{\line(0,1){0.49}}
\multiput(55.31,66.26)(0.06,0.49){1}{\line(0,1){0.49}}
\multiput(55.37,66.76)(0.03,0.5){1}{\line(0,1){0.5}}

\linethickness{0.3mm}
\put(122.9,54.75){\line(0,1){0.5}}
\multiput(122.87,55.74)(0.03,-0.5){1}{\line(0,-1){0.5}}
\multiput(122.81,56.24)(0.06,-0.49){1}{\line(0,-1){0.49}}
\multiput(122.72,56.72)(0.09,-0.49){1}{\line(0,-1){0.49}}
\multiput(122.59,57.21)(0.12,-0.48){1}{\line(0,-1){0.48}}
\multiput(122.44,57.68)(0.15,-0.47){1}{\line(0,-1){0.47}}
\multiput(122.26,58.14)(0.09,-0.23){2}{\line(0,-1){0.23}}
\multiput(122.04,58.59)(0.11,-0.22){2}{\line(0,-1){0.22}}
\multiput(121.8,59.02)(0.12,-0.22){2}{\line(0,-1){0.22}}
\multiput(121.54,59.44)(0.13,-0.21){2}{\line(0,-1){0.21}}
\multiput(121.25,59.85)(0.15,-0.2){2}{\line(0,-1){0.2}}
\multiput(120.93,60.23)(0.11,-0.13){3}{\line(0,-1){0.13}}
\multiput(120.59,60.59)(0.11,-0.12){3}{\line(0,-1){0.12}}
\multiput(120.23,60.93)(0.12,-0.11){3}{\line(1,0){0.12}}
\multiput(119.85,61.25)(0.13,-0.11){3}{\line(1,0){0.13}}
\multiput(119.44,61.54)(0.2,-0.15){2}{\line(1,0){0.2}}
\multiput(119.02,61.8)(0.21,-0.13){2}{\line(1,0){0.21}}
\multiput(118.59,62.04)(0.22,-0.12){2}{\line(1,0){0.22}}
\multiput(118.14,62.26)(0.22,-0.11){2}{\line(1,0){0.22}}
\multiput(117.68,62.44)(0.23,-0.09){2}{\line(1,0){0.23}}
\multiput(117.21,62.59)(0.47,-0.15){1}{\line(1,0){0.47}}
\multiput(116.72,62.72)(0.48,-0.12){1}{\line(1,0){0.48}}
\multiput(116.24,62.81)(0.49,-0.09){1}{\line(1,0){0.49}}
\multiput(115.74,62.87)(0.49,-0.06){1}{\line(1,0){0.49}}
\multiput(115.25,62.9)(0.5,-0.03){1}{\line(1,0){0.5}}
\put(114.75,62.9){\line(1,0){0.5}}
\multiput(114.26,62.87)(0.5,0.03){1}{\line(1,0){0.5}}
\multiput(113.76,62.81)(0.49,0.06){1}{\line(1,0){0.49}}
\multiput(113.28,62.72)(0.49,0.09){1}{\line(1,0){0.49}}
\multiput(112.79,62.59)(0.48,0.12){1}{\line(1,0){0.48}}
\multiput(112.32,62.44)(0.47,0.15){1}{\line(1,0){0.47}}
\multiput(111.86,62.26)(0.23,0.09){2}{\line(1,0){0.23}}
\multiput(111.41,62.04)(0.22,0.11){2}{\line(1,0){0.22}}
\multiput(110.98,61.8)(0.22,0.12){2}{\line(1,0){0.22}}
\multiput(110.56,61.54)(0.21,0.13){2}{\line(1,0){0.21}}
\multiput(110.15,61.25)(0.2,0.15){2}{\line(1,0){0.2}}
\multiput(109.77,60.93)(0.13,0.11){3}{\line(1,0){0.13}}
\multiput(109.41,60.59)(0.12,0.11){3}{\line(1,0){0.12}}
\multiput(109.07,60.23)(0.11,0.12){3}{\line(0,1){0.12}}
\multiput(108.75,59.85)(0.11,0.13){3}{\line(0,1){0.13}}
\multiput(108.46,59.44)(0.15,0.2){2}{\line(0,1){0.2}}
\multiput(108.2,59.02)(0.13,0.21){2}{\line(0,1){0.21}}
\multiput(107.96,58.59)(0.12,0.22){2}{\line(0,1){0.22}}
\multiput(107.74,58.14)(0.11,0.22){2}{\line(0,1){0.22}}
\multiput(107.56,57.68)(0.09,0.23){2}{\line(0,1){0.23}}
\multiput(107.41,57.21)(0.15,0.47){1}{\line(0,1){0.47}}
\multiput(107.28,56.72)(0.12,0.48){1}{\line(0,1){0.48}}
\multiput(107.19,56.24)(0.09,0.49){1}{\line(0,1){0.49}}
\multiput(107.13,55.74)(0.06,0.49){1}{\line(0,1){0.49}}
\multiput(107.1,55.25)(0.03,0.5){1}{\line(0,1){0.5}}
\put(107.1,54.75){\line(0,1){0.5}}
\multiput(107.1,54.75)(0.03,-0.5){1}{\line(0,-1){0.5}}
\multiput(107.13,54.26)(0.06,-0.49){1}{\line(0,-1){0.49}}
\multiput(107.19,53.76)(0.09,-0.49){1}{\line(0,-1){0.49}}
\multiput(107.28,53.28)(0.12,-0.48){1}{\line(0,-1){0.48}}
\multiput(107.41,52.79)(0.15,-0.47){1}{\line(0,-1){0.47}}
\multiput(107.56,52.32)(0.09,-0.23){2}{\line(0,-1){0.23}}
\multiput(107.74,51.86)(0.11,-0.22){2}{\line(0,-1){0.22}}
\multiput(107.96,51.41)(0.12,-0.22){2}{\line(0,-1){0.22}}
\multiput(108.2,50.98)(0.13,-0.21){2}{\line(0,-1){0.21}}
\multiput(108.46,50.56)(0.15,-0.2){2}{\line(0,-1){0.2}}
\multiput(108.75,50.15)(0.11,-0.13){3}{\line(0,-1){0.13}}
\multiput(109.07,49.77)(0.11,-0.12){3}{\line(0,-1){0.12}}
\multiput(109.41,49.41)(0.12,-0.11){3}{\line(1,0){0.12}}
\multiput(109.77,49.07)(0.13,-0.11){3}{\line(1,0){0.13}}
\multiput(110.15,48.75)(0.2,-0.15){2}{\line(1,0){0.2}}
\multiput(110.56,48.46)(0.21,-0.13){2}{\line(1,0){0.21}}
\multiput(110.98,48.2)(0.22,-0.12){2}{\line(1,0){0.22}}
\multiput(111.41,47.96)(0.22,-0.11){2}{\line(1,0){0.22}}
\multiput(111.86,47.74)(0.23,-0.09){2}{\line(1,0){0.23}}
\multiput(112.32,47.56)(0.47,-0.15){1}{\line(1,0){0.47}}
\multiput(112.79,47.41)(0.48,-0.12){1}{\line(1,0){0.48}}
\multiput(113.28,47.28)(0.49,-0.09){1}{\line(1,0){0.49}}
\multiput(113.76,47.19)(0.49,-0.06){1}{\line(1,0){0.49}}
\multiput(114.26,47.13)(0.5,-0.03){1}{\line(1,0){0.5}}
\put(114.75,47.1){\line(1,0){0.5}}
\multiput(115.25,47.1)(0.5,0.03){1}{\line(1,0){0.5}}
\multiput(115.74,47.13)(0.49,0.06){1}{\line(1,0){0.49}}
\multiput(116.24,47.19)(0.49,0.09){1}{\line(1,0){0.49}}
\multiput(116.72,47.28)(0.48,0.12){1}{\line(1,0){0.48}}
\multiput(117.21,47.41)(0.47,0.15){1}{\line(1,0){0.47}}
\multiput(117.68,47.56)(0.23,0.09){2}{\line(1,0){0.23}}
\multiput(118.14,47.74)(0.22,0.11){2}{\line(1,0){0.22}}
\multiput(118.59,47.96)(0.22,0.12){2}{\line(1,0){0.22}}
\multiput(119.02,48.2)(0.21,0.13){2}{\line(1,0){0.21}}
\multiput(119.44,48.46)(0.2,0.15){2}{\line(1,0){0.2}}
\multiput(119.85,48.75)(0.13,0.11){3}{\line(1,0){0.13}}
\multiput(120.23,49.07)(0.12,0.11){3}{\line(1,0){0.12}}
\multiput(120.59,49.41)(0.11,0.12){3}{\line(0,1){0.12}}
\multiput(120.93,49.77)(0.11,0.13){3}{\line(0,1){0.13}}
\multiput(121.25,50.15)(0.15,0.2){2}{\line(0,1){0.2}}
\multiput(121.54,50.56)(0.13,0.21){2}{\line(0,1){0.21}}
\multiput(121.8,50.98)(0.12,0.22){2}{\line(0,1){0.22}}
\multiput(122.04,51.41)(0.11,0.22){2}{\line(0,1){0.22}}
\multiput(122.26,51.86)(0.09,0.23){2}{\line(0,1){0.23}}
\multiput(122.44,52.32)(0.15,0.47){1}{\line(0,1){0.47}}
\multiput(122.59,52.79)(0.12,0.48){1}{\line(0,1){0.48}}
\multiput(122.72,53.28)(0.09,0.49){1}{\line(0,1){0.49}}
\multiput(122.81,53.76)(0.06,0.49){1}{\line(0,1){0.49}}
\multiput(122.87,54.26)(0.03,0.5){1}{\line(0,1){0.5}}

\linethickness{0.3mm}
\put(57.9,34.75){\line(0,1){0.5}}
\multiput(57.87,35.74)(0.03,-0.5){1}{\line(0,-1){0.5}}
\multiput(57.81,36.24)(0.06,-0.49){1}{\line(0,-1){0.49}}
\multiput(57.72,36.72)(0.09,-0.49){1}{\line(0,-1){0.49}}
\multiput(57.59,37.21)(0.12,-0.48){1}{\line(0,-1){0.48}}
\multiput(57.44,37.68)(0.15,-0.47){1}{\line(0,-1){0.47}}
\multiput(57.26,38.14)(0.09,-0.23){2}{\line(0,-1){0.23}}
\multiput(57.04,38.59)(0.11,-0.22){2}{\line(0,-1){0.22}}
\multiput(56.8,39.02)(0.12,-0.22){2}{\line(0,-1){0.22}}
\multiput(56.54,39.44)(0.13,-0.21){2}{\line(0,-1){0.21}}
\multiput(56.25,39.85)(0.15,-0.2){2}{\line(0,-1){0.2}}
\multiput(55.93,40.23)(0.11,-0.13){3}{\line(0,-1){0.13}}
\multiput(55.59,40.59)(0.11,-0.12){3}{\line(0,-1){0.12}}
\multiput(55.23,40.93)(0.12,-0.11){3}{\line(1,0){0.12}}
\multiput(54.85,41.25)(0.13,-0.11){3}{\line(1,0){0.13}}
\multiput(54.44,41.54)(0.2,-0.15){2}{\line(1,0){0.2}}
\multiput(54.02,41.8)(0.21,-0.13){2}{\line(1,0){0.21}}
\multiput(53.59,42.04)(0.22,-0.12){2}{\line(1,0){0.22}}
\multiput(53.14,42.26)(0.22,-0.11){2}{\line(1,0){0.22}}
\multiput(52.68,42.44)(0.23,-0.09){2}{\line(1,0){0.23}}
\multiput(52.21,42.59)(0.47,-0.15){1}{\line(1,0){0.47}}
\multiput(51.72,42.72)(0.48,-0.12){1}{\line(1,0){0.48}}
\multiput(51.24,42.81)(0.49,-0.09){1}{\line(1,0){0.49}}
\multiput(50.74,42.87)(0.49,-0.06){1}{\line(1,0){0.49}}
\multiput(50.25,42.9)(0.5,-0.03){1}{\line(1,0){0.5}}
\put(49.75,42.9){\line(1,0){0.5}}
\multiput(49.26,42.87)(0.5,0.03){1}{\line(1,0){0.5}}
\multiput(48.76,42.81)(0.49,0.06){1}{\line(1,0){0.49}}
\multiput(48.28,42.72)(0.49,0.09){1}{\line(1,0){0.49}}
\multiput(47.79,42.59)(0.48,0.12){1}{\line(1,0){0.48}}
\multiput(47.32,42.44)(0.47,0.15){1}{\line(1,0){0.47}}
\multiput(46.86,42.26)(0.23,0.09){2}{\line(1,0){0.23}}
\multiput(46.41,42.04)(0.22,0.11){2}{\line(1,0){0.22}}
\multiput(45.98,41.8)(0.22,0.12){2}{\line(1,0){0.22}}
\multiput(45.56,41.54)(0.21,0.13){2}{\line(1,0){0.21}}
\multiput(45.15,41.25)(0.2,0.15){2}{\line(1,0){0.2}}
\multiput(44.77,40.93)(0.13,0.11){3}{\line(1,0){0.13}}
\multiput(44.41,40.59)(0.12,0.11){3}{\line(1,0){0.12}}
\multiput(44.07,40.23)(0.11,0.12){3}{\line(0,1){0.12}}
\multiput(43.75,39.85)(0.11,0.13){3}{\line(0,1){0.13}}
\multiput(43.46,39.44)(0.15,0.2){2}{\line(0,1){0.2}}
\multiput(43.2,39.02)(0.13,0.21){2}{\line(0,1){0.21}}
\multiput(42.96,38.59)(0.12,0.22){2}{\line(0,1){0.22}}
\multiput(42.74,38.14)(0.11,0.22){2}{\line(0,1){0.22}}
\multiput(42.56,37.68)(0.09,0.23){2}{\line(0,1){0.23}}
\multiput(42.41,37.21)(0.15,0.47){1}{\line(0,1){0.47}}
\multiput(42.28,36.72)(0.12,0.48){1}{\line(0,1){0.48}}
\multiput(42.19,36.24)(0.09,0.49){1}{\line(0,1){0.49}}
\multiput(42.13,35.74)(0.06,0.49){1}{\line(0,1){0.49}}
\multiput(42.1,35.25)(0.03,0.5){1}{\line(0,1){0.5}}
\put(42.1,34.75){\line(0,1){0.5}}
\multiput(42.1,34.75)(0.03,-0.5){1}{\line(0,-1){0.5}}
\multiput(42.13,34.26)(0.06,-0.49){1}{\line(0,-1){0.49}}
\multiput(42.19,33.76)(0.09,-0.49){1}{\line(0,-1){0.49}}
\multiput(42.28,33.28)(0.12,-0.48){1}{\line(0,-1){0.48}}
\multiput(42.41,32.79)(0.15,-0.47){1}{\line(0,-1){0.47}}
\multiput(42.56,32.32)(0.09,-0.23){2}{\line(0,-1){0.23}}
\multiput(42.74,31.86)(0.11,-0.22){2}{\line(0,-1){0.22}}
\multiput(42.96,31.41)(0.12,-0.22){2}{\line(0,-1){0.22}}
\multiput(43.2,30.98)(0.13,-0.21){2}{\line(0,-1){0.21}}
\multiput(43.46,30.56)(0.15,-0.2){2}{\line(0,-1){0.2}}
\multiput(43.75,30.15)(0.11,-0.13){3}{\line(0,-1){0.13}}
\multiput(44.07,29.77)(0.11,-0.12){3}{\line(0,-1){0.12}}
\multiput(44.41,29.41)(0.12,-0.11){3}{\line(1,0){0.12}}
\multiput(44.77,29.07)(0.13,-0.11){3}{\line(1,0){0.13}}
\multiput(45.15,28.75)(0.2,-0.15){2}{\line(1,0){0.2}}
\multiput(45.56,28.46)(0.21,-0.13){2}{\line(1,0){0.21}}
\multiput(45.98,28.2)(0.22,-0.12){2}{\line(1,0){0.22}}
\multiput(46.41,27.96)(0.22,-0.11){2}{\line(1,0){0.22}}
\multiput(46.86,27.74)(0.23,-0.09){2}{\line(1,0){0.23}}
\multiput(47.32,27.56)(0.47,-0.15){1}{\line(1,0){0.47}}
\multiput(47.79,27.41)(0.48,-0.12){1}{\line(1,0){0.48}}
\multiput(48.28,27.28)(0.49,-0.09){1}{\line(1,0){0.49}}
\multiput(48.76,27.19)(0.49,-0.06){1}{\line(1,0){0.49}}
\multiput(49.26,27.13)(0.5,-0.03){1}{\line(1,0){0.5}}
\put(49.75,27.1){\line(1,0){0.5}}
\multiput(50.25,27.1)(0.5,0.03){1}{\line(1,0){0.5}}
\multiput(50.74,27.13)(0.49,0.06){1}{\line(1,0){0.49}}
\multiput(51.24,27.19)(0.49,0.09){1}{\line(1,0){0.49}}
\multiput(51.72,27.28)(0.48,0.12){1}{\line(1,0){0.48}}
\multiput(52.21,27.41)(0.47,0.15){1}{\line(1,0){0.47}}
\multiput(52.68,27.56)(0.23,0.09){2}{\line(1,0){0.23}}
\multiput(53.14,27.74)(0.22,0.11){2}{\line(1,0){0.22}}
\multiput(53.59,27.96)(0.22,0.12){2}{\line(1,0){0.22}}
\multiput(54.02,28.2)(0.21,0.13){2}{\line(1,0){0.21}}
\multiput(54.44,28.46)(0.2,0.15){2}{\line(1,0){0.2}}
\multiput(54.85,28.75)(0.13,0.11){3}{\line(1,0){0.13}}
\multiput(55.23,29.07)(0.12,0.11){3}{\line(1,0){0.12}}
\multiput(55.59,29.41)(0.11,0.12){3}{\line(0,1){0.12}}
\multiput(55.93,29.77)(0.11,0.13){3}{\line(0,1){0.13}}
\multiput(56.25,30.15)(0.15,0.2){2}{\line(0,1){0.2}}
\multiput(56.54,30.56)(0.13,0.21){2}{\line(0,1){0.21}}
\multiput(56.8,30.98)(0.12,0.22){2}{\line(0,1){0.22}}
\multiput(57.04,31.41)(0.11,0.22){2}{\line(0,1){0.22}}
\multiput(57.26,31.86)(0.09,0.23){2}{\line(0,1){0.23}}
\multiput(57.44,32.32)(0.15,0.47){1}{\line(0,1){0.47}}
\multiput(57.59,32.79)(0.12,0.48){1}{\line(0,1){0.48}}
\multiput(57.72,33.28)(0.09,0.49){1}{\line(0,1){0.49}}
\multiput(57.81,33.76)(0.06,0.49){1}{\line(0,1){0.49}}
\multiput(57.87,34.26)(0.03,0.5){1}{\line(0,1){0.5}}

\linethickness{1mm}
\put(5,70){\line(1,0){35}}
\linethickness{1mm}
\put(7,35){\line(1,0){35}}
\linethickness{1mm}
\multiput(55,70)(0.66,-0.12){83}{\line(1,0){0.66}}
\linethickness{1mm}
\multiput(55,30)(0.33,0.12){167}{\line(1,0){0.33}}
\linethickness{1mm}
\multiput(120,60)(0.16,0.12){125}{\line(1,0){0.16}}
\linethickness{1mm}
\multiput(120,50)(0.24,-0.12){83}{\line(1,0){0.24}}
\put(130,60){\makebox(0,0)[cc]{$\cdot$}}

\put(130,50){\makebox(0,0)[cc]{$\cdot$}}

\put(20,75){\makebox(0,0)[cc]{$p_j$}}

\put(5,28){\makebox(0,0)[cc]{$p_i$}}

\linethickness{0.3mm}
\put(45,40){\line(0,1){20}}
\put(48,65){\makebox(0,0)[cc]{$\Gamma$}}

\put(48,35){\makebox(0,0)[cc]{$\Gamma$}}

\put(115,55){\makebox(0,0)[cc]{$\Gamma$}}

\linethickness{0.3mm}
\put(38,52){\makebox(0,0)[cc]{$\ell$}}

\put(85,72){\makebox(0,0)[cc]{$p_j-\ell$}}

\put(97,35){\makebox(0,0)[cc]{$p_i+\ell$}}

\end{picture}

}

\def\figirloop{

\def\JPicScale{0.6}
\ifx\JPicScale\undefined\def\JPicScale{1}\fi
\unitlength \JPicScale mm

}

\def\figstringa{

\def\JPicScale{0.7}
\ifx\JPicScale\undefined\def\JPicScale{1}\fi
\unitlength \JPicScale mm
\begin{picture}(110,60)(0,0)
\linethickness{1mm}
\multiput(20,60)(0.18,-0.12){167}{\line(1,0){0.18}}
\linethickness{0.3mm}
\multiput(20,20)(0.18,0.12){167}{\line(1,0){0.18}}
\linethickness{1mm}
\put(50,40){\line(1,0){30}}
\linethickness{1mm}
\multiput(80,40)(0.18,0.12){167}{\line(1,0){0.18}}
\linethickness{1mm}
\multiput(80,40)(0.18,-0.12){167}{\line(1,0){0.18}}
\put(30,60){\makebox(0,0)[cc]{$T$}}

\put(30,20){\makebox(0,0)[cc]{$S$}}

\put(65,35){\makebox(0,0)[cc]{$H$}}

\put(100,60){\makebox(0,0)[cc]{$T$}}

\put(100,20){\makebox(0,0)[cc]{$T$}}

\end{picture}

}

\def\figstringb{

\def\JPicScale{0.7}
\ifx\JPicScale\undefined\def\JPicScale{1}\fi
\unitlength \JPicScale mm
\begin{picture}(110,60)(0,0)
\linethickness{1mm}
\multiput(20,60)(0.18,-0.12){167}{\line(1,0){0.18}}
\linethickness{0.3mm}
\multiput(20,20)(0.18,0.12){167}{\line(1,0){0.18}}
\linethickness{1mm}
\put(50,40){\line(1,0){30}}
\linethickness{1mm}
\multiput(80,40)(0.18,0.12){167}{\line(1,0){0.18}}
\linethickness{1mm}
\multiput(80,40)(0.18,-0.12){167}{\line(1,0){0.18}}
\put(30,60){\makebox(0,0)[cc]{$H$}}

\put(30,20){\makebox(0,0)[cc]{$S$}}

\put(65,35){\makebox(0,0)[cc]{$T$}}

\put(100,60){\makebox(0,0)[cc]{$T$}}

\put(100,20){\makebox(0,0)[cc]{$T$}}

\end{picture}

}

\def\figstringc{

\def\JPicScale{0.7}
\ifx\JPicScale\undefined\def\JPicScale{1}\fi
\unitlength \JPicScale mm
\begin{picture}(110,60)(0,0)
\linethickness{1mm}
\multiput(20,60)(0.18,-0.12){167}{\line(1,0){0.18}}
\linethickness{0.3mm}
\multiput(20,20)(0.18,0.12){167}{\line(1,0){0.18}}
\linethickness{1mm}
\put(50,40){\line(1,0){30}}
\linethickness{1mm}
\multiput(80,40)(0.18,0.12){167}{\line(1,0){0.18}}
\linethickness{1mm}
\multiput(80,40)(0.18,-0.12){167}{\line(1,0){0.18}}
\put(30,60){\makebox(0,0)[cc]{$H$}}

\put(30,20){\makebox(0,0)[cc]{$S$}}

\put(65,35){\makebox(0,0)[cc]{$\phi,H$}}

\put(100,60){\makebox(0,0)[cc]{$T$}}

\put(100,20){\makebox(0,0)[cc]{$T$}}

\end{picture}

}

\begin{document}

\baselineskip 24pt

\begin{center}
{\Large \bf Sub-subleading Soft Graviton Theorem in Generic Theories of Quantum Gravity}

\end{center}

\vskip .6cm
\medskip

\vspace*{4.0ex}

\baselineskip=18pt

\centerline{\large \rm Alok Laddha$^{a}$ and Ashoke Sen$^{b}$}

\vspace*{4.0ex}

\centerline{\large \it ~$^a$Chennai Mathematical Institute, Siruseri, Chennai, India}

\centerline{\large \it ~$^b$Harish-Chandra Research Institute, HBNI}
\centerline{\large \it  Chhatnag Road, Jhusi,
Allahabad 211019, India}


\vspace*{1.0ex}
\centerline{\small E-mail:  aladdha@cmi.ac.in, sen@mri.ernet.in}

\vspace*{5.0ex}

\centerline{\bf Abstract} \bigskip

We analyze  scattering amplitudes with one soft
external graviton and arbitrary number of other finite energy external states 
carrying arbitrary mass and spin to sub-subleading order in the momentum of the
soft graviton. Our result can be expressed as the sum of a universal part that depends
only on the amplitude without the soft graviton and not the other details of the theory and
a non-universal part that depends on the amplitude without the soft graviton, and the
two and three point functions of the theory.  For tree amplitudes our results are
valid in all space-time dimensions while for loop amplitudes, infrared divergences
force us to restrict our analysis to space time dimensions five or more. With this
restriction the results are valid to all orders in perturbation theory. Our results agree
with known results in quantum field theories and string theory.

\bigskip



\vfill \eject

\baselineskip 18pt

\tableofcontents

\sectiono{Introduction and summary} \label{sintro}

Soft graviton theorem has been studied extensively in recent 
years\cite{weinberg1,weinberg2,jackiw1,jackiw2,1103.2981,1404.4091,1404.7749,
1405.1410,1405.2346,
1405.3413,1405.3533,1406.6574,1406.6987,1406.7184,1407.5936,
1407.5982,1408.4179,1410.6406,1412.3699,1503.04816,1504.01364,1507.08882,
1509.07840,1604.00650,1604.03893,1607.02700,1611.02172,1611.03137,1702.02350}, 
primarily due to its connection to asymptotic symmetries\cite{1312.2229,1401.7026,
1411.5745,1506.05789,1509.01406,1605.09094,1608.00685,1612.08294,1701.00496,
1612.05886}
(see \cite{1703.05448} for a recent review). It relates the
scattering amplitude of a set of finite energy particles and a  low momentum (soft)
graviton to an amplitude without the low momentum graviton.
Soft theorems are also known to hold in string theories\cite{ademollo,shapiro,
1406.4172,1406.5155,1411.6661,1502.05258,1505.05854,1507.08829,1511.04921,
1512.00803,1601.03457,1604.03355,1610.03481,1702.03934,1703.00024}. 
Our goal in this paper will be to analyze sub-subleading soft graviton theorem
-- that gives the result for the above mentioned 
scattering amplitude to the sub-subleading
order in the energy of the soft graviton. 

Our main result is that in a generic theory 
the sub-subleading soft graviton amplitude is given by a sum of a standard
set of terms that are universal, independent of the theory, and a non-universal 
term that depends on the theory. The standard terms, reproduced in all but the last line
of
eq.~\refb{etotaln}, can be found {\it e.g.} in
\cite{1404.4091} after appropriate generalizations to arbitrary dimensions.
On the other hand
for scattering of $N$ finite energy particles carrying momenta 
$p_1,\cdots p_N$ and a soft graviton carrying momentum $k$ and
polarization $\ve$, the correction term takes the form
\be \label{esummary}
\{\ve_{\mu\nu}k_{\rho}k_{\sigma}\ -\ \ve_{\mu\sigma}k_{\nu}k_{\rho}\ 
-\ \ve_{\nu\rho}k_{\sigma}k_{\mu}\ +\ \ve_{\rho\sigma}k_{\mu}k_{\nu}\}
\sum_{i=1}^N {1\over p_i\cdot k} \sum_{i'} B_{i,i'}^{\mu\rho\nu\sigma}(p_i)
\, \Gamma_{i\to i'}
\ee
where the sum over $i'$ runs over all on-shell 
states carrying the same mass and momentum as the 
external state $i$ and $\Gamma_{i\to i'}$
denotes the original amplitude without the soft graviton, and the $i$th state
replaced by $i'$. The quantity $B_{i,i'}^{\mu\nu\rho\sigma}(p_i)$ is a function
of the momentum $p_i$ carried by the $i$-th external particle
and depends on the quadratic and cubic terms in the 
one particle irreducible (1PI)  effective action. For a given action 
$B$ can be computed explicitly (see eqs.\refb{etotaln}, \refb{edelalphan}). 

As our analysis is based on general properties
of the 1PI  effective action, our results are valid for any
general coordinate invariant theory of gravity coupled to other fields, including
string theory. For tree amplitudes there is no restriction on the number of space-time
dimensions. However for loop
amplitudes, infrared divergences\cite{1405.1015} 
force us to restrict our analysis to five or more
space-time dimensions. 
A more detailed investigation of soft graviton theorem in generic theories of gravity in
four dimensions is left for future investigation.

If we focus our attention on the theory of massless fields in four dimensions, 
possibly obtained by integrating out other massive fields, then 
Weinberg-Witten theorem excludes the presence of interacting particles of spin 
$>2$. For tree level scattering of massless
particles of spin $\le 2$ we can list all possible three point couplings that can
possibly contribute to the function $B_{i,i'}^{\mu\rho\nu\sigma}(p_i)$ appearing in
\refb{esummary}. These have been listed in eqs.\refb{epossint} and \refb{elistb}.
Their contribution to $B_{i,i'}^{\mu\rho\nu\sigma}(p_i)$ can be evaluated easily.
In the spinor
helicity representation they reproduce the results of \cite{1611.07534}. 
Of course our general result
\refb{esummary} is more general and holds in any space-time 
dimensions and also for massive higher
spin fields. In particular it can also be used to reproduce various results on 
sub-subleading soft graviton amplitudes in string theory\cite{1512.00803,1604.03355} 
involving scattering of 
massless as
well as massive fields.

Our analysis is based on the idea used in \cite{1702.03934,1703.00024} 
in which the coupling of a soft graviton
to the rest of the fields is obtained by covariantizing the gauge fixed 1PI effective
action of the finite energy particles with respect to the soft graviton background.
It is natural to ask if the same technique can be used to extend  the
analysis to next order in soft
momentum. However 
at the end of section \ref{s1} we have
argued that at least this technique is not extendable to the next
order.  

In fact, there maybe a deeper reason as to why for generic configuration of external 
states,  soft theorems do not appear to extend beyond subleading order in gauge 
theories  and sub-subleading order in gravity. It is now becoming increasingly evident 
that soft theorems  are statements about (asymptotic) symmetries of the 
underlying theory\cite{1703.05448}. In the case of QED, 
it was argued in \cite{1605.09677} that if the soft theorems in QED were to extend 
beyond subleading order, the associated asymptotic symmetries will be ill-defined 
in the sense that the corresponding charges will be divergent. One expects similar 
divergences to occur in gravity,  
if one were to extend the emergence of soft theorems from 
asymptotic symmetries beyond 
sub-subleading order\cite{1509.01406,1605.09094,1608.00685}.

For special configurations of external states as in the case of MHV amplitude, it 
was shown in \cite{1405.1410} 
that factorization theorem holds to all orders in graviton 
energy. In view of the discussion above this seems accidental. A more
detailed investigation of such results from the perspective presented in this paper 
is left for future investigation.

The rest of the paper is organized as follows. In section \ref{s1} we analyze
amplitudes with one external soft graviton and arbitrary number of other external
states in any theory of gravity coupled to matter field to sub-subleading order in the
soft momentum. The final result is given in \refb{etotaln}, \refb{edelalphan}. These are the
main results of our paper. In section \ref{scon} we show that our result 
\refb{etotaln}, \refb{edelalphan} depends only on the on-shell data of an amplitude
without the soft graviton, even though individual terms in these equations depend on
the off-shell continuation. Sections \ref{sfield} and \ref{sstring} involve comparing
our general result with known tree level results in quantum field theories and string
theory, and we find perfect agreement.

The usual S-matrix in four space-time dimensions suffers from
infrared divergence in the presence of massless particles. 
Therefore for loop amplitudes we need to restrict our analysis
to five or more space-time dimensions $D$. Even though infrared divergences do not
affect the usual S-matrix elements for $D\ge 5$, they may still alter the behaviour of
an amplitude in the soft limit by producing additional singularities that are not included
in our analysis of section \ref{s1}. In section \ref{sir} we analyze this possibility in detail and
show that no such additional divergences arise. Therefore we can trust the result of
section \ref{s1} for loop amplitudes in $D\ge 5$.

\sectiono{Sub-subleading soft graviton theorem} \label{s1}

\begin{figure}
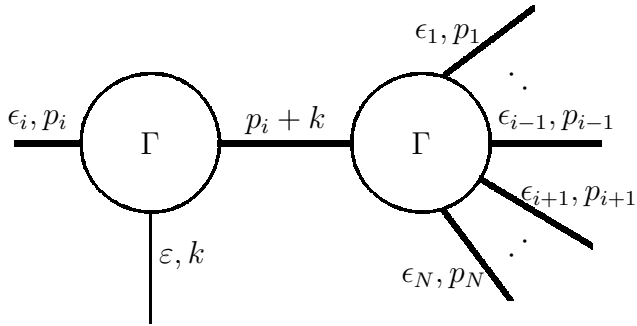

\centerline{\figsoftonefield}
\caption{The leading contribution. \label{f1}}
\end{figure}

\begin{figure}
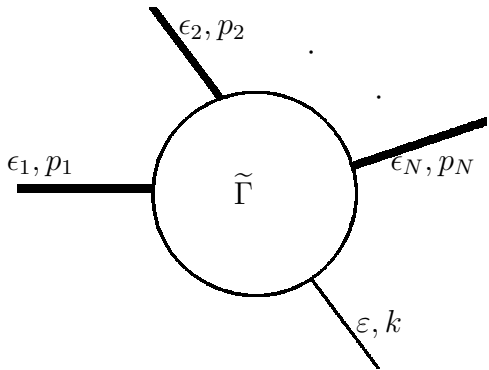

\centerline{\figsoftthreefield}
\caption{The subleading contribution. \label{f2}}
\end{figure}

We consider a general theory of gravity coupled to other matter fields and focus on
a scattering amplitude involving one soft graviton of momentum $k$ and polarization
$\ve$, satisfying the constraints
\be  \label{eepscond}
k^2=0, \qquad
\ve_{\mu\nu}=\ve_{\nu\mu}, \qquad k^\mu \ve_{\mu\nu}=0, \qquad \ve^\mu_{~\mu}=0\, .
\ee
The amplitude is given by a sum of two types of diagrams, shown in Figs.~\ref{f1} and
\ref{f2}. Fig.~\ref{f1} represents sum of all diagrams where the soft graviton is attached to
one of the external finite energy lines via 1PI three point
vertex. Fig.~\ref{f2} contains the sum of the rest of the diagrams. The leading contribution
in the soft limit $k\to 0$ comes from Fig.~\ref{f1} due to the pole associated with the
propagator carrying momentum $p_i+k$. Fig.~\ref{f2} does not have such poles and
therefore begins contributing at the subleading order.

We shall now describe separately the evaluation of these two classes of diagrams. In doing
this we shall follow the strategy of \cite{1702.03934,1703.00024}, i.e.\ first choose a
covariant gauge fixing of the 1PI effective action of finite energy fields (including gravitons),
expanded in a power series in the fields around flat space-time background, 
and then determine the coupling of the soft graviton
to the finite energy fields by replacing the background flat metric by soft graviton
background metric and ordinary derivatives by covariant derivatives computed using the
soft graviton background metric. As in \cite{1703.00024}, the finite energy fields will
be assumed to carry flat tensor indices associated with the tangent space group so that
their covariant derivatives involve the spin connection and not the Christoffel symbol.

\subsection{Evaluation of Fig.~\ref{f2}} \label{sf2}

In this section we shall analyze Fig.~\ref{f2} which begins
contributing at
the subleading order.
Let us denote this by $\wt\Gamma(\ve, k; \eps_1, p_1, \ldots \eps_N, p_N)$,
where $(\ve, k)$ are the polarizations and momentum of the soft graviton and 
$(\eps_i, p_i)$ are the polarizations and momentum of the $i$-th external state. 
All external propagators are amputated in the definition of
$\wt\Gamma$. We shall also assume that all the external
fields are
normalized correctly so that we do not need to keep track of wave-function renormalization
factors in relating the amplitudes to S-matrix elements.
We shall include an explicit momentum conserving delta function in the expression 
for the amplitude and treat the $p_i$'s and $k$ 
as independent variables while taking derivative of the amplitude with respect
to these momenta. We shall not impose any on-shell
condition on $(\eps_i, p_i)$ 
till the end after all the derivatives with respect to momenta are
taken, but the soft graviton will be taken to be on-shell from the beginning. 
Finally we allow the polarization
tensor $\eps_i$ to depend on $p_i$ but no other external momenta and the
polarization $\ve$ of the soft graviton to depend on $k$ but no other
momenta.

\begin{figure}
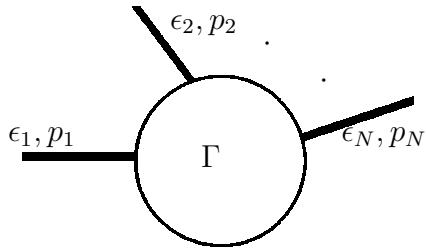

\centerline{\figthreefield}
\caption{The amplitude without the soft graviton. \label{f3}}
\end{figure}

Our goal will be to express $\wt\Gamma$ 
in terms of the amplitude without the soft graviton
shown in Fig.~\ref{f3}. 
This has the form
\be \label{e1}
\eps_{1,\alpha_1} \cdots \eps_{N,\alpha_N} \Gamma^{\alpha_1\ldots \alpha_N}
(p_1,\ldots p_N)  (2\pi)^D \delta^{(D)}(p_1+\cdots p_N)
\ee
where we shall take the index $\alpha$ to run over all the fields $\Phi_\alpha$
present in the theory. We shall assume that all fields carry 
tangent space indices so that the fields $\{\Phi_\alpha\}$ belong to some
large reducible representation of the local Lorentz group.
There is an ambiguity in defining the function 
$\Gamma^{\alpha_1\ldots \alpha_N}
(p_1,\ldots p_N)$ since we can add to it any term that vanishes when 
$\sum_i p_i=0$. We shall not impose any restriction on how we 
resolve this ambiguity except for the (anti-)symmetry of
$\Gamma^{\alpha_1\ldots \alpha_N}
(p_1,\ldots p_N)$ under the exchange $(\alpha_i, p_i)\leftrightarrow (\alpha_j, p_j)$
for any pair $(i,j)$.  We also introduce the shorthand notation
\be \label{e7}
\Gamma_{(i)}^{\alpha_i}(p_i) = \left(\prod_{j=1\atop j\ne i}^N \eps_{j, \alpha_j}\right)
\Gamma^{\alpha_1\ldots \alpha_N}
(p_1,\ldots p_N)  (2\pi)^D \delta^{(D)}(p_1+\cdots p_N)
\ee
where in the argument of $\Gamma_{(i)}$ we have suppressed the momenta
$p_j$ and polarizations $\eps_j$ for $j\ne i$. 

We shall now determine the amplitude shown in Fig.~\ref{f2} from the one in 
Fig~\ref{f3} by noting the following. We can determine the coupling of a soft graviton 
to the finite energy fields by
replacing, in the expression for \refb{e1} written in position space,
all derivatives $\p_\mu$ by covariant derivatives $D_\mu$,
and eventually converting them to 
flat space index  by contracting them with the inverse vielbein $E_a^{~\mu}$.
This procedure can be regarded as the
result of covariantization of the amputated Green's function with respect to
the general coordinate transformation of the background soft graviton field.

To first order in the soft graviton field the inverse vielbein is given by
\be 
E_a^{~\mu} =  \delta_a^{~\mu} - S_a^{~\mu}
\ee
where $S_a^{~\mu}$ is the soft graviton 
\be 
S_{\mu\nu} = \ve_{\mu\nu} e^{ik\cdot x} \, ,
\ee
and all indices are raised and lowered
by the flat metric $\eta$. 
For constructing the covariant derivative we also need the expression for the spin
connection $\omega_\mu^{ab}$ and Christoffel symbol $\Gamma_{\mu\nu}^\rho$.
To first order in the soft graviton field these are given by
\be \label{eomexp}
\omega_\mu^{ab} = \p^b S_\mu^{~a} - \p^a S_\mu^{~b} = i\, e^{ik\cdot x}\, 
\left(k^b \ve_\mu^{~a} - k^a \ve_\mu^{~b}\right) \,,
\ee
and 
\be \label{egamexp}
\Gamma_{\mu\nu}^\rho = \p_\mu S_\nu^{~\rho} + \p_\nu S_\mu^{~\rho}
- \p^\rho S_{\mu\nu}
= i \, e^{ik\cdot x} \, \left\{k_\mu \ve_\nu^{~\rho} + k_\nu 
\ve_\mu^{~\rho}
- k^\rho \ve_{\mu\nu}
\right\} \, .
\ee

The covariant derivative has two kinds of terms. Acting on a field $\Phi_\alpha$
transforming in
some (not necessarily irreducible) representation $R$ of the Lorentz group, it 
has a piece 
\be
{1\over 2} \omega_{\mu}^{ab} (J_{ab})_\alpha^{~\beta} \Phi_\beta
\ee
where $J_{ab}$ is the generator of the Lorentz group in the representation $R$
normalized so that acting on a covariant vector field
\be \label{ejnorm}
(J^{ab})_c^{~d} = \delta^a_{~c} \eta^{bd} - \delta^b_{~c} \eta^{ad}\, .
\ee
The second kind of term arises from the fact that when $D_\mu$ is preceded by
a $D_\nu$ operation, we get a factor of
\be\label{e2}
-\Gamma_{\mu\nu}^\rho D_\rho\, .
\ee
Since $\Gamma_{\mu\nu}^\rho$ already contains a factor of soft graviton 
field snd since we shall work to first order in the soft graviton field, we can
replace $D_\rho$ by $\p_\rho$ in \refb{e2}. This leads to the simple rule that 
for every pair of derivatives we get a factors of $-\Gamma_{\mu\nu}^\rho \p_\rho$.

Since in momentum space a derivative is replaced by $ip_\mu$, the above
considerations give the following expression for the amplitude in 
Fig~\ref{f2} in terms of the amplitude in Fig.~\ref{f3} to order $k^\rho$:
\ben \label{e3}
 (2\pi)^{D} \, \delta^{(D)}(p_1+\cdots p_N+k)  \, 
\prod_{j=1}^{N} \eps_{j, \alpha_j} \hskip 1in \nonumber \\
\sum_{i=1}^N \left[-\delta_{\beta_i}^{~\alpha_i} \ve_{\mu}^{~\nu} p_{i\nu} {\p
\over \p p_{i\mu}} + k^b \ve_\mu^{~a}
(J_{ab})_{\beta_i}^{~\alpha_i} {\p
\over \p p_{i\mu}} - {1\over 2} \delta_{\beta_i}^{~\alpha_i}
\left\{k_\mu \ve_\nu^{~\rho} + k_\nu 
\ve_\mu^{~\rho}
- k^\rho \ve_{\mu\nu}
\right\} \, p_{i\rho} \, {\p^2 \over \p p_{i\mu} \p p_{i\nu}}
\right] \nonumber \\
\hskip 2in\Gamma^{\alpha_1\cdots \alpha_{i-1}\beta_i\alpha_{i+1}\cdots \alpha_N}
(p_1,\ldots , p_N)\, .
 \hskip 1in 
\een
In this the first term inside the square bracket is the effect of multiplication by
inverse vielbein to convert the space-time indices carried by the momenta 
to tangent space indices. The
second term represents the effect of the spin connection term in the covariant
derivative and the third factor represents the effect of the Christoffel symbol
term in the covariant derivative. The shift by $k$ of the argument of the delta function
represents the effect of the multiplicative factor of $e^{ik\cdot x}$ from the soft graviton field
in the position space representation of the amplitude. 

We shall now show that we can bring the momentum conserving delta function
in \refb{e3} to the right of the derivatives so that the 
derivatives also act on the delta function.
We begin with 
second and the third
terms in the square bracket. Their contribution to \refb{e3} may be expressed as
\ben\label{e3a}
&& \prod_{j=1}^{N} \eps_{j, \alpha_j} 
\sum_{i=1}^N \left[k^b \ve_\mu^{~a}
(J_{ab})_{\beta_i}^{~\alpha_i} {\p
\over \p p_{i\mu}} - {1\over 2} \delta_{\beta_i}^{~\alpha_i}
\left\{k_\mu \ve_\nu^{~\rho} + k_\nu 
\ve_\mu^{~\rho}
- k^\rho \ve_{\mu\nu}
\right\} \, p_{i\rho} \, {\p^2 \over \p p_{i\mu} \p p_{i\nu}}
\right] \nonumber \\
&& \hskip .5in \left\{
\Gamma^{\alpha_1\cdots \alpha_{i-1}\beta_i\alpha_{i+1}\cdots \alpha_N}
(p_1,\ldots , p_N)  (2\pi)^{D} \, \delta^{(D)}(p_1+\cdots p_N+k) 
\right\} + J_1 + J_2\, ,
\een
where 
\ben \label{eJ1}
J_1 &=& (2\pi)^{D} \, \left\{\prod_{j=1}^{N} \eps_{j, \alpha_j} \right\} \, 
\Gamma^{\alpha_1\ldots , \alpha_N}
(p_1,\ldots , p_N) {1\over 2}
\left\{k_\mu \ve_\nu^{~\rho} + k_\nu 
\ve_\mu^{~\rho}
- k^\rho \ve_{\mu\nu}
\right\} \, \nonumber \\ &&
\sum_{i=1}^N\, p_{i\rho} \, {\p^2 \over \p p_{i\mu} \p p_{i\nu}}
\delta^{(D)}(p_1+\cdots p_N+k) 
\een
and 
\ben \label{e4pre}
J_2 &=& -(2\pi)^{D} \, \left\{\prod_{j=1}^{N} \eps_{j, \alpha_j} \right\} \, \sum_{i=1}^N
{\p
\over \p p_{i\mu}} \delta^{(D)}(p_1+\cdots p_N+k)
\nonumber \\ &&
\left[ k^b \ve_\mu^{~a}
(J_{ab})_{\beta_i}^{~\alpha_i} -  \delta_{\beta_i}^{~\alpha_i}
\left\{k_\mu \ve_\nu^{~\rho} + k_\nu 
\ve_\mu^{~\rho}
- k^\rho \ve_{\mu\nu} \right\} p_{i\rho} {\p\over \p p_{i\nu}}\right]
\Gamma^{\alpha_1\ldots \alpha_{i-1}\beta_i \alpha_{i+1},
\ldots , \alpha_N}
(p_1,\ldots , p_N)  \, . \nonumber \\
\een
$J_1$ cancels the term where 
both derivatives in the last term in the square bracket in \refb{e3a}
act on the delta function, whereas 
$J_2$ cancels the terms in \refb{e3a} where one momentum derivative acts on the
delta function.

We shall first analyze $J_1$. In \refb{eJ1}
we can replace ${\p^2 / \p p_{i\mu} \p p_{i\nu}}$ by 
$\p^2 / \p k_\mu \p k_\nu$
using the fact that the argument of the delta function contains sum of all the
$p_i$'s and $k$. We can now 
bring the $\sum_{i=1}^N p_{i\rho}$ factor inside the
derivative and finally replace it by $-k_\rho$ using the delta function. This gives
 \ben \label{evalJ1}
J_1 &=& (2\pi)^{D} \, \left\{\prod_{j=1}^{N} \eps_{j, \alpha_j} \right\} \, 
\Gamma^{\alpha_1\ldots , \alpha_N}
(p_1,\ldots , p_N) {1\over 2}
\left\{k_\mu \ve_\nu^{~\rho} + k_\nu 
\ve_\mu^{~\rho}
- k^\rho \ve_{\mu\nu}
\right\}  \, \nonumber \\ &&
\, {\p^2 \over \p k_\mu \p k_\nu} \left[ - k_\rho
\delta^{(D)}(p_1+\cdots p_N+k) \right] \nonumber \\
&=& (2\pi)^{D} \, \left\{\prod_{j=1}^{N} \eps_{j, \alpha_j} \right\} \, 
\Gamma^{\alpha_1\ldots , \alpha_N}
(p_1,\ldots , p_N) {1\over 2}
\left\{k_\mu \ve_\nu^{~\rho} + k_\nu 
\ve_\mu^{~\rho}
- k^\rho \ve_{\mu\nu}
\right\}  \, \nonumber \\ &&
\, \left[ - k_\rho
{\p^2 \over \p k_\mu \p k_\nu} 
- \delta_\rho^{~\mu} {\p \over \p k_\nu}
- \delta_\rho^{~\nu} {\p \over \p k_\mu} 
\right]\delta^{(D)}(p_1+\cdots p_N+k)
\nonumber \\ &=& 0
\een
where in the last step we have used \refb{eepscond}.

On the other hand in the expression \refb{e4pre}
for $J_2$ we can replace each of the $\p / \p p_{i\mu}$
operator acting on the momentum conserving delta function by $\p / \p k_\mu$ and express it as
\ben \label{e4}
J_2 &=& -(2\pi)^{D} \, \left\{\prod_{j=1}^{N} \eps_{j, \alpha_j} \right\} \, 
{\p
\over \p k_\mu} \delta^{(D)}(p_1+\cdots p_N+k)
\nonumber \\ 
&& \hskip -.3in \times
\sum_{i=1}^N \left[ k^b \ve_\mu^{~a}
(J_{ab})_{\beta_i}^{~\alpha_i} -  \delta_{\beta_i}^{~\alpha_i}
\left\{k_\mu \ve_\nu^{~\rho} + k_\nu 
\ve_\mu^{~\rho}
- k^\rho \ve_{\mu\nu} \right\}  p_{i\rho}  {\p\over \p p_{i\nu}}\right]
\Gamma^{\alpha_1\ldots \alpha_{i-1}\beta_i \alpha_{i+1},
\ldots , \alpha_N}
(p_1,\ldots , p_N) \, .\nonumber \\
\een
Now Lorentz covariance of $\Gamma^{\alpha_1\ldots \alpha_N}(p_1,\ldots,
p_N)$ implies
\be
\sum_{i=1}^N  \left[ (J^{ab})_{\beta_i}^{~\alpha_i} 
\Gamma^{\alpha_1\ldots \alpha_{i-1}\beta_i \alpha_{i+1},
\ldots , \alpha_N}
(p_1,\ldots , p_N)  - \{ p_i^a {\p \over \p p_{ib}} - 
 p_i^b {\p \over \p p_{ia}} \} \Gamma^{\alpha_1\ldots \alpha_N}
 (p_1,\ldots p_N)
\right] = 0\, .
\ee
Using this \refb{e4} may be expressed as
\ben \label{e10}
J_2 &=& -(2\pi)^{D} \, \left\{\prod_{j=1}^{N} \eps_{j, \alpha_j} \right\} \, 
{\p
\over \p k_\mu} \delta^{(D)}(p_1+\cdots p_N+k) \nonumber \\ &&
\sum_{i=1}^N \left[k_b \ve_{\mu a} \{ p_i^a {\p \over \p p_{ib}} - 
 p_i^b {\p \over \p p_{ia}} \}  - \left\{k_\mu \ve_\nu^{~\rho} + k_\nu 
\ve_\mu^{~\rho}
- k^\rho \ve_{\mu\nu} \right\}  p_{i\rho} {\p\over \p p_{i\nu}}
\right]
\Gamma^{\alpha_1\ldots \alpha_N}(p_1,\ldots,
p_N) \nonumber \\
&=& (2\pi)^{D} \, \left\{\prod_{j=1}^{N} \eps_{j, \alpha_j} \right\} \, 
{\p
\over \p k_\mu} \delta^{(D)}(p_1+\cdots p_N+k) k_\mu \ve_\nu^{~\rho}
\sum_{i=1}^N  p_{i\rho} {\p\over \p p_{i\nu}} 
\Gamma^{\alpha_1\ldots \alpha_N}(p_1,\ldots,
p_N)\, . \nonumber \\
\een

We now turn to the contribution from the first term inside 
the square bracket in \refb{e3}. By expanding the delta function in Taylor series expansion in $k$ and
keeping terms up to order $k^\mu$, we get
\be \label{e5}
-(2\pi)^{D} \, \left\{\prod_{j=1}^{N} \eps_{j, \alpha_j} \right\} \, 
\delta^{(D)}(p_1+\cdots p_N) \ve_\nu^{~\rho}
\sum_{i=1}^N  p_{i\rho} {\p\over \p p_{i\nu}} 
\Gamma^{\alpha_1\ldots \alpha_N}(p_1,\ldots,
p_N) + J_3
\ee
where
\be \label{edefJ3}
J_3 = -(2\pi)^{D} \, \left\{\prod_{j=1}^{N} \eps_{j, \alpha_j} \right\} \, 
{\p
\over \p k_\mu} \delta^{(D)}(p_1+\cdots p_N+k) \, k_\mu \ve_\nu^{~\rho}
\sum_{i=1}^N  p_{i\rho} {\p\over \p p_{i\nu}} 
\Gamma^{\alpha_1\ldots \alpha_N}(p_1,\ldots,
p_N)\, .
\ee
Note that in a Taylor series expansion we would normally set $k$ in the argument of
$\delta^{(D)}$ to zero after taking the derivative. However to this order in the expansion
in powers of $k$, it does not make any difference.
Using the relation
\ben
&& \ve_\nu^{~\rho} \sum_{i=1}^N  p_{i\rho} {\p\over \p p_{i\nu}} 
\delta^{(D)}(p_1+\cdots p_N)  
= \ve_\nu^{~\rho} \sum_{i=1}^N  p_{i\rho} {\p\over \p p_{1\nu}} 
\delta^{(D)}(p_1+\cdots p_N)  \nonumber \\
&=&   {\p\over \p p_{1\nu}}  \sum_{i=1}^N  p_{i\rho} \,
\delta^{(D)}(p_1+\cdots p_N) - \ve_\nu^{~\nu} \delta^{(D)}(p_1+\cdots p_N) =0
\een
we can express \refb{e5} as
\be \label{e3b}
-\left\{\prod_{j=1}^{N} \eps_{j, \alpha_j} \right\} \, 
\ve_\nu^{~\rho}
\sum_{i=1}^N  p_{i\rho} {\p\over \p p_{i\nu}} 
\{\Gamma^{\alpha_1\ldots \alpha_N}(p_1,\ldots,
p_N) (2\pi)^{D} \, \delta^{(D)}(p_1+\cdots p_N) \} + J_3 \, .
\ee

The  total contribution from Fig.~\ref{f2},
given in \refb{e3}, is given by the sum of \refb{e3a} and \refb{e3b}. 
We have seen from \refb{evalJ1} that $J_1$ vanishes. On the other hand \refb{e10} and
\refb{edefJ3} shows that $J_2+J_3$ vanishes. 
Furthermore, since we need the terms up to order $k$, we can set $k=0$ in the argument
of the delta function in \refb{e3a}. This gives the net contribution to Fig.~\ref{f2}
to order $k$ as
\ben\label{e6}
I_0&=& \prod_{j=1}^{N} \eps_{j, \alpha_j} 
\sum_{i=1}^N \left[- \ve_\nu^{~\rho} p_{i\rho} {\p\over \p p_{i\nu}}  
\delta_{\beta_i}^{~\alpha_i}
+ k^b \ve_\mu^{~a}
(J_{ab})_{\beta_i}^{~\alpha_i} {\p
\over \p p_{i\mu}} \right. \nonumber \\ &&
\hskip 1in \left. - {1\over 2} \delta_{\beta_i}^{~\alpha_i}
\left\{k_\mu \ve_\nu^{~\rho} + k_\nu 
\ve_\mu^{~\rho}
- k^\rho \ve_{\mu\nu}
\right\} \, p_{i\rho} \, {\p^2 \over \p p_{i\mu} \p p_{i\nu}}
\right] \nonumber \\
&& \hskip 1in \left\{
\Gamma^{\alpha_1\cdots \alpha_{i-1}\beta_i\alpha_{i+1}\cdots \alpha_N}
(p_1,\ldots , p_N)  (2\pi)^{D} \, \delta^{(D)}(p_1+\cdots p_N) 
\right\} \, .
\een
Using \refb{e7} and the fact that $\eps_j$ is independent of $p_i$ for $j\ne i$,
\refb{e6} may be rewritten as
\ben \label{e8}
I_0&=&  \sum_{i=1}^N \eps_{i,\alpha_i} 
\left[- \ve_\nu^{~\rho} p_{i\rho} {\p\over \p p_{i\nu}}  
\delta_{\beta_i}^{~\alpha_i}
+ k^b \ve_\mu^{~a}
(J_{ab})_{\beta_i}^{~\alpha_i} {\p
\over \p p_{i\mu}} \right. \nonumber \\ &&
\hskip 1in \left. - {1\over 2} \delta_{\beta_i}^{~\alpha_i}
\left\{k_\mu \ve_\nu^{~\rho} + k_\nu 
\ve_\mu^{~\rho}
- k^\rho \ve_{\mu\nu}
\right\} \, p_{i\rho} \, {\p^2 \over \p p_{i\mu} \p p_{i\nu}}
\right] \Gamma_{(i)}^{\beta_i}(p_i) \, .
\een

\subsection{Evaluation of Fig.~\ref{f1}} \label{sf1}

We now turn to the evaluation of the contribution from 
Fig.~\ref{f1}. 
To evaluate this we begin by writing the quadratic term in the 1PI effective 
action:
\be \label{es2n}
S^{(2)}
= {1\over 2} \int {d^D q_1\over (2\pi)^D} \, {d^D q_2\over (2\pi)^D} \,  \Phi_\alpha(q_1) 
\KK^{\alpha\beta}(q_2) \Phi_\beta(q_2)\, (2\pi)^D \delta^{(D)}(q_1+q_2)
\, ,
\ee
where we take
\be \label{ekabn}
\KK^{\alpha\beta}(q) = \KK^{\beta\alpha}(-q)\, .
\ee
For grassmann odd fields there will be an additional minus sign on the right hand side
of \refb{ekabn}, but the final result is unaffected by this. For this reason we shall proceed
by taking the fields to be grassmann even.
The full propagator computed from this action has the form
\be 
\Xi(q) \, (q^2 + M^2)^{-1}\, ,
\ee
where
\be \label{edefXin}
\Xi(q) = i(q^2 + M^2) \, \KK^{-1}(q)\, .
\ee
At this stage $M$ is taken to be an arbitrary mass parameter. 
Lorentz covariance of $\KK$ and $\Xi$ implies the relations
\be \label{e4.9n}
\KK^{\alpha\gamma}(q) (J^{ab})_\gamma^{~\beta} + \KK^{\gamma\beta}(q) (J^{ab})_\gamma^{~\alpha}
= q^a {\p \KK^{\alpha\beta}(q)\over \p q_b} -  q^b {\p \KK^{\alpha\beta}(q)\over \p q_a}
\, ,
\ee
\be \label{e4.9xin}
-\, \Xi_{\alpha\gamma}(q) (J^{ab})_\beta^{~\gamma} - \Xi_{\gamma\beta}(q) (J^{ab})_\alpha^{~\gamma}
= q^a {\p \Xi_{\alpha\beta}(q)\over \p q_b} -  q^b {\p \Xi_{\alpha\beta}(q)\over \p q_a}
\, .
\ee
For computing the
propagator carrying momentum $p_i+k$ in Fig.~\ref{f1} we shall take $M$ to be the
mass $M_i$ of the $i$-th incoming particle and call the corresponding $\Xi(q)$ as
$\Xi^i(q)$.  In that case the polarization vector
$\eps_{i,\alpha}$ and the momenta $p_i$ of the $i$-th external state will satisfy 
the on-shell condition 
\be \label{epoln}
\eps_{i,\alpha} \KK^{\alpha\beta}(p_i) = 0\, , \quad
p_i^2+M_i^2=0\, .
\ee

We shall now determine the coupling of the soft graviton to a pair of finite energy
particles by covariantizing the  action \refb{es2n} 
with respect to the background soft 
graviton field.
We shall assume that while covariantizing, we replace ordinary derivatives
by covariant derivatives and symmetrize under arbitrary permutations of these
derivatives. This may differ from the 
actual action by terms proportional to the Riemann
tensor of the soft graviton. 
The effect of such additional couplings will be taken care of separately.

We now list the effect of coupling the action \refb{es2n} to the soft graviton field carrying
momentum $k$ and polarization $\ve$, up to sub-subleading order in the soft momentum
$k$:
\begin{enumerate}
\item Since the soft graviton carries momentum $k$, the $\delta^{(D)}(q_1+q_2)$
is replaced by $\delta^{(D)}(q_1+q_2+k)$. 
\item For every derivative $\p_\mu$ acting on 
$\Phi_\beta$ or its derivatives, 
we get a term $-\ve_\mu^{~\nu} \p_\nu$ from having 
to convert the
space-time index associated with $\p_\mu$ 
to tangent space index by the replacement $\p_\mu \to
E_\mu^{~\nu}\p_\nu$. This is done at the very last step after all the other steps
mentioned below have been performed.
Once this replacement is made, the indices can be 
contracted using the flat metric $\eta$.
\item For every derivative $\p_\mu$ acting on 
$\Phi_\beta$ we get a term ${1\over 2}
\omega_\mu^{ab} (J_{ab})_{\beta}^{~\gamma}
\Phi_\gamma$ from having to replace ordinary derivatives by covariant derivatives.
\item For every pair of derivatives $\p_\mu$ and $\p_\nu$ acting on
$\Phi_\beta$, we get an additional 
term $-\Gamma^\rho_{\mu\nu}\p_\rho\Phi_\beta$ due to the
fact that $D_\mu$ acting on $D_\nu$ generates a term $-\Gamma_{\mu\nu}^\rho
D_\rho$. This factor is independent of the relative order of $D_\mu$ and $D_\nu$.
\item For every pair of derivatives $\p_\mu$ and $\p_\nu$ acting on $\Phi_\beta$
we get an additional term  ${1\over 2}
\p_{(\mu} \omega_{\nu)}^{ab} (J_{ab})_{\beta}^{~\gamma}
\Phi_\gamma$ due to the left-most derivative acting on the spin connection.
\item For every triplet of derivatives $\p_\mu$, $\p_\nu$ and $\p_\rho$ acting
on $\Phi_\beta$, we get an additional term
$-\p_{(\rho} \Gamma_{\mu\nu)}^{\sigma} \p_\sigma\Phi_\beta$ 
due to the left-most derivative acting
on the Christoffel symbol.
\end{enumerate}
Of these the first four effects also appeared in our analysis of Fig.~\ref{f2} in section
\ref{sf2}. The last two effects generate two powers of soft momentum and do not affect the 
evaluation of Fig.~\ref{f2} which begins to contribute only at the subleading order.
Using \refb{eomexp}, \refb{egamexp} and 
the fact that in momentum space $\p_\mu\Phi_\alpha$ is represented by 
$iq_\mu \Phi_\alpha(q)$, 
we get the following action describing the coupling of the soft graviton to
the $\Phi$ field
\ben \label{e2.14n}
S^{(3)} &=& {1\over 2} \int {d^D q_1\over (2\pi)^D} \, {d^D q_2\over (2\pi)^D} \,  (2\pi)^D \delta^{(D)}(q_1+q_2+k)
\nonumber \\ && \times \Phi_\alpha(q_1) 
\Bigg[ - \ve_{\mu \nu} q_2^\nu  {\p\over \p q_{2\mu}}\KK^{\alpha\beta} (q_2)
+ {1\over 2} (k_b \, \ve_{a\mu} - k_a \, \ve_{b\mu}) {\p\over \p q_{2\mu}}\KK^{\alpha\gamma}(q_2) 
\left(J^{ab}\right)_{\gamma}^{~\beta}  \nonumber \\ &&
- {1\over 2} {\p^2 \KK^{\alpha\beta}(q_2)\over \p q_{2\mu} \p q_{2\nu}}
q_{2\rho} \left(k_\mu \ve_\nu^{~\rho} + 
k_\nu \ve_\mu^{~\rho} - k^\rho \ve_{\mu\nu}\right) +{1\over 4} {\p^2 \KK^{\alpha\gamma}(q_2)\over \p q_{2\mu} \p q_{2\nu}}
k_\mu \left(k^b \ve_\nu^{~a} -  k^a \ve_\nu^{~b}\right) (J_{ab})_\gamma^{~\beta}
\nonumber \\
&& -{1\over 6} {\p^3 \KK^{\alpha\beta}(q_2)\over \p q_{2\mu} \p q_{2\nu} \p
q_{2\rho}} q_{2\sigma} k_\rho \left(k_\mu \ve_\nu^{~\sigma} + 
k_\nu \ve_\mu^{~\sigma} - k^\sigma \ve_{\mu\nu}\right)
\Bigg]\Phi_\beta(q_2) \, . 
\een
To \refb{e2.14n} we could add an additional coupling of the fields
$\Phi_\alpha$ to the Riemann tensor constructed from the soft graviton:
\be \label{eriemannn}
\bar S^{(3)} \equiv
{1\over 2} \int {d^D q_1\over (2\pi)^D} \, {d^D q_2\over (2\pi)^D} \,  (2\pi)^D \delta^{(D)}(q_1+q_2+k) 
\, \RR_{\mu\rho\nu\sigma} \, \Phi_\alpha(q_1) \,
\BB^{\alpha\beta;\mu\rho\nu\sigma}(q_2) \, \Phi_\beta(q_2) \, .
\ee
where
 \begin{equation} \label{edefR}
\begin{array}{lll}
{\cal R}^{\mu\rho\nu\sigma}\ \equiv \ \ve^{\mu\nu}k^{\rho}k^{\sigma}\ -\ \ve^{\mu\sigma}k^{\nu}k^{\rho}\ 
-\ \ve^{\nu\rho}k^{\sigma}k^{\mu}\ +\ \ve^{\rho\sigma}k^{\mu}k^{\nu}
\end{array}
\end{equation}
is the linearized Riemann tensor of the soft graviton written in the momentum space.
For the Riemann tensor we are using the convention
 \be
 \RR^\mu_{~\rho\nu\sigma} = \p_\nu \Gamma^\mu_{\sigma\rho} - 
\p_\sigma \Gamma^\mu_{\nu\rho} + \hbox{ $\Gamma$ $\Gamma$ terms}\, .
 \ee 
 \refb{edefR} includes an extra minus sign from having to convert $\p_\rho$ to $i\, k_\rho$
 when we go from position space description to momentum space description.
\refb{eriemannn} represents a non-minimal coupling of the soft graviton to the fields $\Phi_\alpha$
that is not obtained from covariantization of the kinetic term. 
We can choose $\BB^{\alpha\beta;\mu\rho\nu\sigma}(q_2)
= \BB^{\beta\alpha;\mu\rho\nu\sigma}(-q_2-k)$.

We now turn to the evaluation of Fig.~\ref{f1}.
The propagator
gives a factor of $\Xi^i_{\alpha\beta}(-p_i) \, \{(p_i+k)^2 + M_i^2\}^{-1}=
(2p_i\cdot k)^{-1} \, \Xi^i_{\alpha\beta}(-p_i)$ where now
\be 
\Xi^i(p) \equiv \Xi(p)|_{M=M_i} = i \, (p^2 + M_i^2) \, \KK^{-1}(p)\, .
\ee
Therefore the contribution to Fig.~\ref{f1} is given by
\be \label{e4.13pren}
(2p_i\cdot k)^{-1} \eps_{i,\alpha}\, \left\{
\Gamma^{(3)\alpha\beta}(\ve, k; p_i, -p_i-k)
+ \bar\Gamma^{(3)\alpha\beta}(\ve, k; p_i, -p_i-k) \right\} \,
\Xi^i_{\beta\delta} (-p_i-k) \, 
\Gamma_{(i)}^{\delta}(p_i+k)\, ,
\ee
where $\Gamma^{(3)}$ and $\bar\Gamma^{(3)}$ are the contributions of 
$S^{(3)}$ and $\bar S^{(3)}$ to the 1PI 
three point vertices of two finite energy external
states carrying labels $\alpha$ and $\beta$ and momenta $p_i$ and $-p_i-k$
and one external soft graviton carrying momentum $k$ and polarization $\ve$.
We have from \refb{e2.14n}, \refb{eriemannn}
\ben \label{e4.8n}
&& \Gamma^{(3)\alpha\beta}(\ve, k; p, -p-k) \nonumber \\
&=& {i\over 2} \Bigg[- \ve_{\mu \nu} (p+k)^\nu  {\p\over \p p_{\mu}}\KK^{\alpha\beta} (-p-k)
- \ve_{\mu \nu} p^\nu  {\p\over \p p_{\mu}}\KK^{\beta\alpha} (p)
\nonumber \\ &&
+ {1\over 2} (k_a \, \ve_{b\mu} - k_b \, \ve_{a\mu}) {\p\over \p p_{\mu}}\KK^{\alpha\gamma}(-p-k) 
\left(J^{ab}\right)_{\gamma}^{~\beta} 
- {1\over 2} (k_a \, \ve_{b\mu} - k_b \, \ve_{a\mu}) {\p\over \p p_{\mu}}\KK^{\beta\gamma}(p) 
\left(J^{ab}\right)_{\gamma}^{~\alpha}  \nonumber \\ &&
+{1\over 4} {\p^2 \KK^{\alpha\gamma}(-p-k)\over \p p_{\mu} \p p_{\nu}}
k_\mu \left(k^b \ve_\nu^{~a} -  k^a \ve_\nu^{~b}\right) (J_{ab})_\gamma^{~\beta}
+{1\over 4} {\p^2 \KK^{\beta\gamma}(p)\over \p p_{\mu} \p p_{\nu}}
k_\mu \left(k^b \ve_\nu^{~a} -  k^a \ve_\nu^{~b}\right) (J_{ab})_\gamma^{~\alpha}
\nonumber \\ &&
-{1\over 2} {\p^2 \KK^{\alpha\beta}(-p-k)\over \p p_{\mu} \p p_{\nu}}
(-p_{\rho}-k_\rho) \left(k_\mu \ve_\nu^{~\rho} + 
k_\nu \ve_\mu^{~\rho} - k^\rho \ve_{\mu\nu}\right) 
\nonumber \\ &&
-{1\over 2} {\p^2 \KK^{\beta\alpha}(p)\over \p p_{\mu} \p p_{\nu}}
p_{\rho} \left(k_\mu \ve_\nu^{~\rho} + 
k_\nu \ve_\mu^{~\rho} - k^\rho \ve_{\mu\nu}\right)  \nonumber \\ &&
-{1\over 6} {\p^3 \KK^{\alpha\beta}(-p-k)\over \p p_{\mu} \p p_{\nu} \p
p_{\rho}} (p_{\sigma}+ k_\sigma) k_\rho \left(k_\mu \ve_\nu^{~\sigma} + 
k_\nu \ve_\mu^{~\sigma} - k^\sigma \ve_{\mu\nu}\right)
\nonumber \\ &&
-{1\over 6} {\p^3 \KK^{\beta\alpha}(p)\over \p p_{\mu} \p p_{\nu} \p
p_{\rho}} p_{\sigma} k_\rho \left(k_\mu \ve_\nu^{~\sigma} + 
k_\nu \ve_\mu^{~\sigma} - k^\sigma \ve_{\mu\nu}\right)
\Bigg] \, ,
\een
and
\be \label{eaddvern}
\bar\Gamma^{(3)\alpha\beta}(\ve, k; p, -p-k)
=i\, \RR_{\mu\rho\nu\sigma} \, \BB^{\alpha\beta;\mu\rho\nu\sigma}(-p)\, ,
\ee
to order $k^2$ in Taylor series expansion in powers of the soft momentum $k$.

The contribution of \refb{eaddvern} to \refb{e4.13pren} is easy to evaluate. Since we already
have two factors of soft momentum in the vertex, we can set $k=0$ in the argument of $\Xi^i$ and
$\Gamma_{(i)}$.
Therefore this contribution is given by
\be \label{ei4n}
I_1 = {i\over 2} 
\sum_{i=1}^N (p_i\cdot k)^{-1} \, \RR_{\mu\rho\nu\sigma} \, 
\eps_{i,\alpha} \,  
\BB^{\alpha\beta;\mu\rho\nu\sigma}(-p_i) \Xi^i_{\beta\delta}(-p_i) 
\Gamma_{(i)}^\delta(p_i) \, .
\ee

In order to evaluate the contribution from the $\Gamma^{(3)}$ part of the vertex 
to \refb{e4.13pren} we
follow the following strategy:
\begin{enumerate}
\item First we replace all factor of $\KK^{rs}(p_i)$ by $\KK^{sr}(-p_i)$ using \refb{ekabn}.
\item In each product of $\KK$, $\Xi^i$ and $J^{ab}$ factors, we first 
use \refb{e4.9n}, \refb{e4.9xin} to
move the $J^{ab}$
factors to the extreme right so that its index is contracted with 
that of $\Gamma_{(i)}$. 
For this we have to rewrite \refb{e4.9n}, \refb{e4.9xin} as
\ben \label{emove}
&& \KK^{\gamma\beta}(q) (J^{ab})_\gamma^{~\alpha}
= - \KK^{\alpha\gamma}(q) (J^{ab})_\gamma^{~\beta} 
+ q^a {\p \KK^{\alpha\beta}(q)\over \p q_b} -  q^b {\p \KK^{\alpha\beta}(q)\over \p q_a}
\, , \nonumber \\
&&  \Xi^i_{\gamma\beta}(q) (J^{ab})_\alpha^{~\gamma}
= -\, \Xi^i_{\alpha\gamma}(q) (J^{ab})_\beta^{~\gamma} -
q^a {\p \Xi^i_{\alpha\beta}(q)\over \p q_b} + 
q^b {\p \Xi^i_{\alpha\beta}(q)\over \p q_a}
\, .
\een
\item We now expand $\KK^{\alpha\beta}(-p_i-k)$, $\Xi^i_{\beta\gamma}
(-p_i-k)$ and $\Gamma_{(i)}^\alpha(p_i+k)$ 
in Taylor series expansion
in powers of soft momenta to appropriate order relevant for computing the
sub-subleading contribution to the amplitude.
\item The expression that results after this has products of (derivatives of) $\KK$,
$\Xi^i$ and $\Gamma_{(i)}$. We now use the derivatives of the 
relation $\KK(q) \, \Xi^i(q) = i (q^2 + M_i^2)$ to transfer the derivatives from $\KK$ to $\Xi^i$
to the maximal possible extent. This requires for example using the relations
\ben \label{eexchange}
{\p \KK(-p)\over \p p_\mu} \Xi^i(-p) &=& -  \KK(-p) {\p \Xi^i(-p)\over \p p_\mu} 
+ 2\, i\, p^\mu\, ,
\nonumber \\ 
{\p^2 \KK(-p)\over \p p_\mu \p p_\nu} \Xi^i(-p) &\hskip -.2in=& \hskip -.2in - {\p \KK(-p)\over \p p_\mu}
{\p \Xi^i(-p)\over \p p_\nu} - {\p \KK(-p)\over \p p_\nu}
{\p \Xi^i(-p)\over \p p_\mu} - \KK(-p) {\p^2 \Xi^i(-p)\over \p p_\mu \p p_\nu}
+2 \, i\, \eta^{\mu\nu}\, , 
\nonumber \\
{\p^3 \KK(-p)\over \p p_\mu \p p_\nu \p p_\rho} \Xi^i(-p) &=& - {\p^2 \KK(-p)\over \p p_\mu
\p p_\nu}
{\p \Xi^i(-p)\over \p p_\rho} - {\p^2 \KK(-p)\over \p p_\mu
\p p_\rho}
{\p \Xi^i(-p)\over \p p_\nu}  - {\p^2 \KK(-p)\over \p p_\nu
\p p_\rho}
{\p \Xi^i(-p)\over \p p_\mu} \nonumber \\ &&
- {\p \KK(-p)\over \p p_\mu}
{\p^2 \Xi^i(-p)\over \p p_\nu \p p_\rho} - {\p \KK(-p)\over \p p_\nu}
{\p^2 \Xi^i(-p)\over \p p_\mu \p p_\rho} - {\p \KK(-p)\over \p p_\rho}
{\p^2 \Xi^i(-p)\over \p p_\mu \p p_\nu} \nonumber \\
&& - \KK(-p) \, {\p^3 \Xi^i(-p) \over \p p_\mu \p p_\nu \p p_\rho}\, . 
\een
\item In the final step we use \refb{epoln} to set to zero terms involving $\KK$ without
derivatives since they are always contracted with $\eps_{i,\alpha}$.
\end{enumerate}
The final result for the contribution of \refb{e4.8n} to \refb{e4.13pren}  is given by
\ben \label{ei2new}
I_2 &=& 
{1\over 2} \sum_{i=1}^N (p_i\cdot k)^{-1} \, 
\eps_{i,\alpha}\,  
\, (k_a \, \ve_{b\mu} - k_b \, \ve_{a\mu})  \, 
\Bigg[p_i^\mu \, (J^{ab})_\delta^{~\alpha} 
\Gamma_{(i)}^{\delta}(p_i)+  
p_i^\mu \, k_\rho (J^{ab})_\delta^{~\alpha} 
{\p \Gamma_{(i)}^{\delta}(p_i)\over \p p_{i\rho}} \Bigg]
\nonumber \\
&& -{i\over 4} \sum_{i=1}^N  
(2 p_i\cdot k)^{-1} \,  \, (k_\rho\, k_a \, \ve_{b\mu} - k_\rho\, 
k_b \, \ve_{a\mu} - k_\mu\, k_a \, \ve_{b\rho} + k_\mu\, 
k_b \, \ve_{a\rho} ) \, \eps_{i,\alpha}\, 
\nonumber \\
&& \hskip 1in {\p \KK^{\alpha\gamma}(-p_i) \over \p p_{i\mu}} {\p
\Xi^i(-p_i)_{\gamma\beta}\over \p p_{i\rho}}
(J^{ab})_\delta^{~\beta} \Gamma_{(i)}^{\delta}(p_i)
\nonumber \\ && 
+\sum_{i=1}^N (p_i\cdot k)^{-1} \, \eps_{i,\alpha} \, 
\ve^{\mu\nu} \, p_{i\mu} \, p_{i\nu}\, 
\left\{ \Gamma_{(i)}^\alpha(p_i) + k_\rho {\p \Gamma_{(i)}^\alpha(p_i)
\over \p p_{i\rho}}
+ {1\over 2}\, k_\rho\, k_\sigma\,  {\p^2 \Gamma_{(i)}^\alpha(p_i) 
 \over \p p_{i\rho} \p p_{i\sigma}}
\right\} \nonumber \\
&& + {i\over 4} 
\left\{\ve_{\mu\nu} \, k_\rho \, k_\sigma - \ve_{\rho\nu} \, k_\mu \, k_\sigma
-\ve_{\mu\sigma} \, k_\rho \, k_\nu + \ve_{\rho\sigma} \, k_\mu \, k_\nu\right\}
\nonumber \\ && \hskip -.2in
 \sum_{i=1}^N (p_i\cdot k)^{-1}  \, \eps_{i,\alpha} \, \left\{
 {2\over 3} p_i^\nu {\p K^{\alpha\beta}(-p_i)\over \p p_{i\mu}} 
 \, {\p^2 \Xi^i_{\beta\delta}(-p_i)\over \p p_{i\rho} p_{i\sigma}}  
 -{1\over 3} {\p^2 K^{\alpha\beta}(-p_i)\over \p p_{i\mu}\p p_{i\nu}}
 \, p_i^\rho \, {\p \Xi^i_{\beta\delta}(-p_i)\over \p p_{i\sigma}} 
 \right\}\, 
 \Gamma_{(i)}^\delta(p_i) \, . \nonumber \\ 
\een

\subsection{Final result}

Using \refb{e8}, \refb{ei4n} and \refb{ei2new} we now get the total amplitude to
sub-subleading order
\ben \label{etotaln}
I &=& I_0+I_1+I_2 \nonumber \\
&=& 
\sum_{i=1}^N (p_i\cdot k)^{-1}  \eps_{i,\alpha_i} 
\ve_{\mu \nu}\,  p_i^\mu \,  p_i^\nu \Gamma_{(i)}^{\alpha_i}(p_i)
\nonumber \\
&& +  \sum_{i=1}^N (p_i\cdot k)^{-1}  \eps_{i,\alpha}  
\ve_{\mu b}\,  p_i^\mu  k_a \left[ \left\{ p_{i}^{b} {\p\over \p p_{ia}}
- p_{i}^a {\p\over \p p_{ib}} \right\} \delta_{\beta}^{~\alpha}
+ (J^{ab})_{\beta}^{~\alpha}\right] \Gamma_{(i)}^{\beta}(p_i)
\nonumber \\
&& + {1\over 2} 
\sum_{i=1}^N (p_i\cdot k)^{-1}  \eps_{i,\alpha}  \ve_{ac} k_b k_d
\left[ \left\{ p_{i}^{b} {\p\over \p p_{ia}}
- p_{i}^a {\p\over \p p_{ib}} \right\} \delta_{\beta}^{~\alpha}
+ (J^{ab})_{\beta}^{~\alpha}\right] \nonumber \\ && \hskip 1in
\left[ \left\{ p_{i}^{d} {\p\over \p p_{ic}}
- p_{i}^c {\p\over \p p_{id}} \right\} \delta_{\gamma}^{~\beta}
+ (J^{cd})_{\gamma}^{~\beta}\right] \Gamma_{(i)}^{\gamma}(p_i)
\nonumber \\ &&
+ {1\over 2} \, 
 \sum_{i=1}^N (p_i\cdot k)^{-1}  \, \eps_{i,\alpha} \, 
 \Delta^{\alpha}_{~\beta}(-p_i, k) \,  \Gamma_{(i)}^\beta(p_i) \, ,
 \een
 where
 \ben  \label{edelalphan}
 \Delta^{\alpha}_{~\delta}(-p_i, k) &=& \left\{\ve_{\mu\nu} \, k_\rho \, k_\sigma - \ve_{\rho\nu} \, k_\mu \, k_\sigma
-\ve_{\mu\sigma} \, k_\rho \, k_\nu + \ve_{\rho\sigma} \, k_\mu \, k_\nu\right\}\nonumber \\
&& \times \, 
 \Bigg\{
 {1\over 3} \, i\, p^\nu {\p \KK^{\alpha\beta}(-p_i)\over \p p_{i\mu}} 
 \, {\p^2 \Xi^i_{\beta\delta}(-p_i)\over \p p_{i\rho} p_{i\sigma}}  
 -{1\over 6}\, i\,  {\p^2 \KK^{\alpha\beta}(-p_i)\over \p p_{i\mu}\p p_{i\nu}}
 \, p_i^\rho \, {\p \Xi^i_{\beta\delta}(-p_i)\over \p p_{i\sigma}} 
\nonumber \\ && \hskip .5in 
+ {i\over 4}  {\p \KK^{\alpha\gamma}(-p_i) \over \p p_{i\mu}} {\p
\Xi^i_{\gamma\beta}(-p_i)\over \p p_{i\rho}}  (J^{\nu\sigma})_{\delta}^{~\beta} 
- {1\over 4} (J^{\mu\rho})_{\beta}^{~\alpha} 
(J^{\nu\sigma})_{\delta}^{~\beta} \nonumber \\
&&  \hskip .5in  
+ i\, \BB^{\alpha\beta;\mu\rho\nu\sigma}(-p_i) \, \Xi^i_{\beta\delta}(-p_i) 
\Bigg\} \, .
 \een
Eqs.~\refb{etotaln}, \refb{edelalphan} are our main results.
  
We end by making a few comments:
\begin{enumerate}
\item If the indices $\alpha$ and $\delta$ in \refb{edelalphan} label scalar fields, then
the tensor inside the curly bracket must be constructed from 
the vector $p_i$ and the invariant
tensor $\eta$. Contraction of $\eta$ with $\RR_{\mu\nu\rho\sigma}$ vanishes as a
result of \refb{eepscond}. Therefore the only possibility is the tensor $p_i^\mu p_i^\rho 
p_i^\nu
p_i^\sigma$. The contraction of this with $\RR_{\mu\nu\rho\sigma}$ vanishes due
to antisymmetry of $\RR$ in the first two indices and last two indices. Therefore
\refb{edelalphan} shows that 
for scalars there are no corrections to the sub-subleading soft graviton theorem. This
is in agreement with known results.
\item \refb{edelalphan} represents correction to the universal part of the sub-subleading factor. 
The first three terms on the right hand side show that unlike the leading and subleading soft factors, 
sub-subleading soft factors are sensitive to the 
(infrared-finite) loop corrections to the propogator.  Even at tree level the contribution from these terms
may be non-zero for higher spin fields -- we shall discuss the case of Rarita-Schwinger fields in
section \ref{sminfer}.
The fourth 
term represents an additional contribution due to spin-angular momentum of the finite energy 
particles and may
give non-vanishing contribution even at tree-level. We shall discuss its contribution for a 
graviton line in section \ref{se-m}. The fifth and the final term shows that the 
sub-subleading factor depends on corrections to 
the three point function involving a soft graviton and a pair of finite energy particles, as
given in  eq.(\ref{eriemannn}).

\item The line of argument followed here cannot be
used to extend the analysis to higher order in the soft momentum. This is due to the
fact that the  contribution from Fig.~\ref{f2} can have terms in which 
the linearized Riemann tensor of the soft graviton 
given in \refb{edefR} is contracted with
an arbitrary function of the finite
external momenta $p_i$ -- bearing no relation to the amplitude without the soft
graviton. As a result terms of this type do not have factorized form
and prevent us from extending the soft graviton theorem.
\end{enumerate}

 \sectiono{Consistency check} \label{scon}

 The right hand side of \refb{etotaln} apparently depends on off-shell data through its
 dependence of $\Gamma_{(i)}^\delta$. This arises from the following sources.
 A scattering amplitude of $n$ finite energy particles is given by the amplitude
 $\Gamma^{\alpha_1\cdots \alpha_n}(p_1,\cdots p_n)$ after setting the external
 momenta $p_i$ on-shell, i.e. satisfy $p_i^2+M_i^2=0$, and then contracting them with
 physical external polarization $\eps_{i,\alpha}$ satisfying \refb{epoln}. Therefore if we
 add to $\Gamma^{\alpha_1\cdots \alpha_n}(p_1,\cdots p_n)$ (or equivalently to
 $\Gamma_{(i)}^\alpha$) a term proportional to $p_i^2+M_i^2$ then the scattering
 amplitude of the finite energy particles do not change. On the other hand individual
 terms on the right hand side of \refb{etotaln} do get modified due to the derivative
 operation with respect to $p_{i\mu}$. Acting on a term proportional to $p_i^2+M_i^2$ 
 this gives a terms proportional to $p_i^\mu$, which do not vanish on-shell. Similarly
 if we add to $\Gamma_{(i)}^\alpha(p_i)$ a term  proportional to $\KK^{\alpha\beta}(-p_i)
 \MM_{(i)\beta}$ for any $\MM_{(i)\beta}$, 
 then the amplitudes involving finite energy external states do not
 get affected due to the on-shell condition \refb{epoln}. However the individual terms on
 the right hand side of \refb{etotaln} change under this transformation. Our goal will be
 to show that when we add all the contributions, the right hand side of \refb{etotaln} 
 actually remains invariant under these deformations of $\Gamma_{(i)}^\alpha$.
 
 First let us consider the effect of adding a term proportional to $p_i^2+M_i^2$ to
 $\Gamma_{(i)}^\alpha$. Using the fact that
 \be 
 \left\{ p_{i}^{b} {\p\over \p p_{ia}}
- p_{i}^a {\p\over \p p_{ib}}\right\} (p_i^2+M_i^2)=0
\ee
it is easy to check that the change of the right hand side of \refb{etotaln} vanishes
after setting $p_i^2+M_i^2=0$.
 
 Next let us consider the effect of shifting $\Gamma_{(i)}^\alpha$ by a term of the form
 $\KK^{\alpha\beta}(-p_i) \MM_{(i)\beta}$. It is easy to see that the 
 first term on the right hand side of
 \refb{etotaln} does not change under this deformation as long as $\eps_{i,\alpha}$
 satisfies \refb{epoln}. For the terms in the second and the third lines on the right hand
 side of \refb{etotaln}, we can use \refb{e4.9n} to bring $\KK$ to the left so that it
 is contracted with $\eps_{i,\alpha}$. The result then vanishes by \refb{epoln}. Therefore
 we need to focus on the contribution from the 
 last term on the right hand side of \refb{etotaln} given by
 \be
 {1\over 2} \,
 \sum_{i=1}^N (p_i\cdot k)^{-1}  \, \eps_{i,\alpha} \, 
 \Delta^{\alpha}_{~\delta}(-p_i, k) \,  \KK^{\delta\gamma} \MM_{(i)\gamma} (p_i)\, .
 \ee
 $\Delta^{\alpha}_{~\delta}(-p_i, k)$ has been given in
 \refb{edelalphan}. The contribution from the last term in \refb{edelalphan}
 is proportional to $\Xi^i_{\beta\delta}(-p_i) \KK^{\delta\gamma} (-p_i)
 = i\, (p_i^2+M_i^2) \, \delta_\beta^{~\gamma}$ and vanishes using the on-shell condition. 
 The contribution from the rest of the terms may be manipulated as follows.
 \begin{enumerate}
 \item First we move all the $J$'s to the right using \refb{emove} so that the index 
 of $J$ is
 contracted with that of $\MM$. 
 \item The resulting expression has products of (derivatives of) $\Xi^i$ and $\KK$ contracted
 with each other. We now transfer the derivatives from the left-most $\KK$ to the right to
 the extent possible using \refb{eexchange} and its analog with $\KK$ and $\Xi^i$ 
 exchanged:
 \ben 
{\p \Xi^i(-p)\over \p p_\mu} \KK(-p) &=& -  \Xi^i(-p) {\p \KK(-p)\over \p p_\mu} 
+ 2\, i\, p^\mu\, ,
\\ 
{\p^2 \Xi^i(-p)\over \p p_\mu \p p_\nu} \KK(-p) &\hskip -.2in=& \hskip -.2in - {\p \Xi^i(-p)\over \p p_\mu}
{\p \KK(-p)\over \p p_\nu} - {\p \Xi^i(-p)\over \p p_\nu}
{\p \KK(-p)\over \p p_\mu} - \Xi^i(-p) {\p^2 \KK(-p)\over \p p_\mu \p p_\nu}
+2 \, i\, \eta^{\mu\nu}\, .\nonumber 
\een
 \item In the final step we set the terms in which the left-most $\KK$ has no derivatives
 to zero using \refb{epoln}.
 \end{enumerate}
 The net result of this analysis yields
\be
-\frac{i}{12}\,  \sum_{i=1}^N (p_i\cdot k)^{-1}  \, 
p_{i\rho}\, \epsilon_{i,\alpha}\, R^{\mu\rho\nu\sigma}\left[ 
{\partial{\cal K} \over \p p_i^\mu} \,
{\partial \Xi^i \over \p p_i^\nu} \, {\partial {\cal K}\over \p p_i^\sigma} 
+ {\partial{\cal K} \over \p p_i^\sigma} \,
{\partial \Xi^i \over \p p_i^\mu} \, {\partial {\cal K}\over \p p_i^\nu} 
+ {\partial{\cal K} \over \p p_i^\nu} \,
{\partial \Xi^i \over \p p_i^\sigma} \, {\partial {\cal K}\over \p p_i^\mu} 
\right]^{\alpha\gamma}\, \MM_{(i)\gamma} (p_i) =0
\ee
where in the last line we have used the algebraic Bianchi identity of the 
Riemann tensor.

This shows that \refb{etotaln} is insensitive to the off-shell information in 
$\Gamma_{(i)}^\alpha$, leading to the form given in \refb{esummary}. We shall now
show that $\Delta^\alpha_{~\delta}$ appearing in \refb{edelalphan} 
depends only on the on-shell
three point function involving the external soft graviton. We shall do this using
factorization property of the full amplitude -- namely that if we adjust the {\it direction}
of $k$ so that $p_i\cdot k\to 0$, the amplitude \refb{etotaln} must factorize into a
product of the on-shell three point function involving external states with momenta
$p_i$, $k$ and $-p_i-k$ and the on-shell $N$-point function involving external states
carrying momenta $p_1,\cdots , p_{i-1}, p_i+k, p_{i+1},\cdots , p_N$. It then follows from
\refb{etotaln} that $\Delta^\alpha_{~\delta}$ in the limit $p_i\cdot k\to 0$ is determined
in terms of the on-shell three point amplitude. Our goal will be to show that the knowledge
of $\Delta^\alpha_{~\delta}$ in the $p_i\cdot k\to 0$ limit is enough to determine
$\Delta^\alpha_{~\delta}$ for general direction of $k$.

To proceed, let us suppress 
the indices $\alpha,\delta$ from $\Delta^\alpha_{~\delta}$, and express \refb{edelalphan}
as
\be \label{eadd1}
\Delta =
\left\{\ve_{\mu\nu} \, k_\rho \, k_\sigma - \ve_{\rho\nu} \, k_\mu \, k_\sigma
-\ve_{\mu\sigma} \, k_\rho \, k_\nu + \ve_{\rho\sigma} \, k_\mu \, k_\nu\right\}
B^{\mu\rho\nu\sigma}\, .
\ee
It is understood that $B$ carries the indices $\alpha,\delta$. $B$ depends on $p_i$
but not on $\ve$ or $k$ to this order in expansion in powers of $k$. Without loss
of generality we can assume that $B^{\mu\rho\nu\sigma}$ has the symmetries of
the Riemann tensor.
In this case the question of whether $\Delta$ is determined from on-shell
three point function reduces to whether it is possible to add some terms to 
$B^{\mu\rho\nu\sigma}$ so that the contribution from this term to \refb{eadd1} vanishes
for $p_i\cdot k=0$ but not in general. In order to make use of the $p_i\cdot k=0$ constraint, the
additional terms in
$B^{\mu\rho\nu\sigma}$ must be proportional to $p_i$. Let us make the most general ansatz
for this ambiguity consistent with the symmetries of $B^{\mu\rho\nu\sigma}$:
\be \label{eshift1}
p_i^\mu A^{\rho\nu\sigma} - p_i^\rho A^{\mu\nu\sigma}
+ p_i^\nu A^{\sigma\mu\rho} - p_i^\sigma A^{\nu\mu\rho}\, ,
\ee
where $A^{\rho\nu\sigma}$ is antisymmetric under $\nu\leftrightarrow\sigma$. Substituting this
into \refb{eadd1} we see that under this shift $\Delta$ changes by
\be
4 (\ve_{\mu\nu} p_i^\mu k_\rho k_\sigma - \ve_{\mu\sigma} p_i^\mu k_\rho k_\nu)\,
A^{\rho\nu\sigma}
\ee
up to terms proportional to $p_i\cdot k$. Since this does not vanish identically for
$p_i\cdot k=0$, we see
that different values of $A$ are
still distinguishable near the pole at $p_i\cdot k=0$.  
This can be rectified by taking $A^{\rho\nu\sigma}$ to be either
proportional to $p_i^\rho B^{\nu\sigma}$ for any anti-symmetric tensor $B$, or 
proportional to $(\eta^{\rho\nu} C^\sigma-\eta^{\rho\sigma}C^\nu)$ for any
vector $C^\nu$, or
by taking it to
be totally anti-symmetric in $\nu,\rho,\sigma$. It is easy to see that
in the first case \refb{eshift1} vanishes identically, while in the last two cases
\refb{eshift1} does not generate any change in \refb{eadd1}.
Therefore we conclude that there is no
ambiguity in determining $\Delta$ from its value near the pole at $p_i\cdot k=0$,
and therefore in terms of on-shell three point function.

\sectiono{Comparison with tree level results for massless fields} \label{sfield}

In this section we shall compare our final result, given in \refb{etotaln}, 
\refb{edelalphan} with some known results in the theory of massless fields
at tree level.

\subsection{Einstein-Maxwell theory} \label{se-m}

 For Einstein-Maxwell theory without any higher derivative terms, the sub-subleading
 soft graviton theorem is known to include only the contribution from the first 
 three lines on the
 right hand side of \refb{etotaln}\cite{1404.4091}. 
 Therefore $\Delta^\alpha_{~\beta}$ 
 given in \refb{edelalphan} must vanish
 for these theories. We shall now verify this explicitly.
 
 First let us consider the case where the $i$-th external finite energy state is a photon.
 We shall choose the Feynman gauge.
 In this case the indices $\alpha$, $\delta$ can be taken to be 
 covariant vector indices $m,n$, and 
 $\KK^{mn}(q)$ is simply 
 $-q^2 \, \eta^{mn}$.   Therefore we have $\Xi^i_{mn}(q) =- i\, \eta_{mn}$ and the first
 three terms on the right hand side of \refb{edelalphan} involving derivatives of $\Xi^i$ must
 vanish. To compute the fourth term we recall that in this case
the components of $J^{ab}$ are given by \refb{ejnorm}. 
This gives
 \be \label{e4.1}
 (J^{\mu\rho})_{p}^{~m} 
(J^{\nu\sigma})_{n}^{~p} = \eta^{\mu\sigma} \eta^{\rho m} \delta^\nu_{~n} 
- \eta^{\rho\sigma} \eta^{\mu m} \delta^\nu_{~n} -
\eta^{\mu\nu} \eta^{\rho m} \delta^\sigma_{~n} + \eta^{\rho\nu} \eta^{\mu m} 
\delta^\sigma_{~n} \, .
 \ee
This has to be contracted with $\RR_{\mu\rho\nu\sigma}$ given in \refb{edefR}. 
Using \refb{eepscond} one can easily verify that all the terms vanish. This shows that the
contribution to \refb{etotaln} from the fourth term on the right hand side of 
\refb{edelalphan} also vanishes.

It remains to analyze the contribution from the last term in \refb{edelalphan}. To calculate
$\BB$ in this case we need to start with the Einstein-Maxwell action in Feynman gauge
and compare with \refb{e2.14n}. Now the part of the Einstein-Maxwell action involving
the gauge field, together with the gauge fixing term, is given by
\ben \label{ephoton}
&& - \int d^D x\,  \sqrt{-\det g} \, \left[ {1\over 4} (D_\mu A_\nu - D_\nu A_\mu) 
 (D^\mu A^\nu - D^\nu A^\mu) - {1\over 2} D_\mu A^\mu D_\nu A^\nu \right]
\nonumber \\ &=&  {1\over 2} \int d^D x\, \sqrt{-\det g} \,   \eta^{mn} \, 
A_m \, D^\rho D_\rho 
\, A_n
\, ,
 \een
 where we have used the fact that $\RR^\mu_{~\nu \mu \sigma}$ vanishes as a
 consequence of \refb{eepscond}. The right hand side of \refb{ephoton} is the 
 covariantization of the free Maxwell action in Feynman gauge for which $\KK^{mn}
 = -q^2 \eta^{mn}$ and therefore the terms linear in the soft graviton field
 computed from \refb{ephoton} coincides with \refb{e2.14n}. Therefore
 in this case $\BB^{\alpha\beta;\mu\nu\rho\sigma}$ vanishes. This in turn shows 
 that the entire contribution to
 \refb{etotaln} from the $\Delta^\alpha_{~\beta}$ term vanishes.
 
 Next we turn to the case where the $i$-th external state is a finite energy graviton. 
 We shall
 use de Donder gauge.
 In this case each of  the indices $\alpha$, $\delta$ can be taken to be a pair of covariant
 vector indices $(mn)$, and we have $\KK^{mn,pq}(q) = - q^2\, \eta^{mp} \eta^{nq}$.\footnote{We
 omit the symmetrization under $m\leftrightarrow n$ and $p\leftrightarrow q$, and removal
 of the trace part, since they are
 taken care of by the symmetry and tracelessness 
 of $h_{mn}$.} In this gauge we have
 $\Xi^i_{mn,pq}(q)=-i\eta_{mp}\eta_{nq}$ and again the first three terms on the right hand side of
 \refb{edelalphan} vanishes. On the other hand we have
  \be
 {(J^{\mu\rho})_{mn}}^{pq} =  \delta^\mu_{~m} \, \eta^{\rho \, p} \, \delta_n^{~q}
 - \delta^\rho_{~m} \, \eta^{\mu \, p} \, \delta_n^{~q} + 
 \delta^\mu_{~n} \, \eta^{\rho \, q} \, \delta_m^{~p} -
 \delta^\rho_{~n} \, \eta^{\mu \, q} \, \delta_m^{~p}\, .
 \ee
 This gives 
 \be 
\eps_{i,pq} \, \RR_{\mu\rho\nu\sigma} \, {(J^{\mu\rho})_{mn}}^{pq} {(J^{\nu\sigma})_{rs}}^{mn} 
 = 8\, \eps_{i, pq} {{{\RR_r}^p}_s}^q
 \ee
 where we have again used the fact that ${\RR^\mu}_{\nu\mu\sigma}=0$.
Therefore the contribution to \refb{etotaln} from the fourth term in \refb{edelalphan} is given by
\be \label{eca1}
- \, \sum_{i=1}^N (p_i\cdot k)^{-1} \,  \eps_{i, pq} \, {{{\RR_r}^p}_s}^q \, \Gamma_{(i)}^{rs}(p_i)\, .
\ee

It remains to calculate the contribution from the last term in \refb{edelalphan}.  For this we need to
determine $\BB$. This can be
calculated in two different ways. The first approach will be to begin with Einstein action in de Donder gauge
and then expand it in powers of the fluctuations $h_{mn}$ to quadratic order 
{\it around a soft graviton background}.
This is then brought to the form $(1/2) \int \sqrt{-\det g} \, h^{mn} \, D^\rho D_\rho \,
h_{mn} + \cdots$
where the
$\cdots$ term, proportional to the Riemann tensor of the soft graviton, determines the action 
$\bar S^{(3)}$ in \refb{eriemannn} and therefore $\BB^{\alpha\beta;\mu\rho\nu\sigma}$
(see {\it e.g.} eq.(7.5.23) of \cite{wald}). The other
possibility is to expand the Einstein action in
the de Donder gauge in powers of the fluctuation $H_{mn}$ around the
{\it flat background} to cubic order\cite{9411092}, 
split $H_{mn}$ as the sum of a soft and a finite energy parts, and then determine
the coupling between a single soft graviton and a pair of finite energy gravitons. Comparing this with the
action \refb{e2.14n} one can determine the missing part $\bar S^{(3)}$. Both approaches yield
\be 
\bar S^{(3)} =  \int d^D x \, \sqrt{-\det g} \, \RR^{mpnq} \, h_{mn} h_{pq}\, .
\ee
Comparing this with \refb{eriemannn} we get
\be
\RR_{\mu\rho\nu\sigma}\BB^{mn,pq;\mu\rho\nu\sigma} = 2\, \RR^{mpnq} \, .
\ee
Using the fact that $\Xi^i_{pq,rs}=-i\eta_{pr} \eta_{qs}$, the contribution from the last term in
\refb{edelalphan} to \refb{etotaln} is seen to be
\be \label{eca2}
\sum_{i=1}^N (p_i\cdot k)^{-1} \,  \eps_{i, pq} \, {{{\RR_r}^p}_s}^q \, \Gamma_{(i)}^{rs}(p_i)\, .
\ee
This cancels \refb{eca1}. Therefore we see that even for external finite energy gravitons the
sub-subleading soft graviton theorem in the Einstein-Maxwell theory
is given by the first four lines on the
right hand side of \refb{etotaln}.

\subsection{Fermions with minimal coupling to gravity} \label{sminfer}

We shall now generalize the analysis of section \ref{se-m} to the case of fermion fields
minimally coupled to gravity. 
We shall work with real fermions by taking the   real and imaginary parts of
a complex field as independent fields -- 
this effectively doubles the dimension of the $\gamma$
matrices but makes them purely imaginary. 
First let us consider the
case of Dirac field. 
Denoting the spinor indices by $r,s$, we have
\be \label{eDirac}
\KK^{rs}(-p) = \left\{ \gamma^0\, (p_\mu \gamma^\mu -M)\right\}_{rs}, \qquad 
\Xi_{rs}(-p) = -i \, \left\{ (p_\mu \gamma^\mu +M) \gamma^0\right\}_{rs}\, ,
\ee
where the $\gamma^\mu$'s satisfy
\be \label{egampr}
\{\gamma^\mu, \gamma^\nu\}=-2\, \eta^{\mu\nu}, \quad 
(\gamma^\mu)^* =
-\gamma^\mu,  \quad
(\gamma^0)^T=-\gamma^0, \quad (\gamma^i)^T = 
\gamma^i \quad \hbox{for $1\le i\le 
(D-1)$}\, .
\ee
In this case the terms in 
\refb{edelalphan} involving two derivatives of $\KK$ or $\Xi$ vanish. Also for minimal
coupling to gravity, $\BB^{\alpha\beta;\mu\nu\rho\sigma}$ vanishes. This leaves us with
the terms in the second line of \refb{edelalphan}. 
Now for spin 1/2 fermions
$(J^{\mu\rho})_r^{~s}$, where $r,s$ represent spinor indices, is given by 
\be \label{espinJ}
(J_S^{\mu\rho})_r^{~s} = -{1\over 2} (\gamma^{\mu\rho})_{rs}, \qquad
\gamma^{\mu\rho} \equiv {1\over 2} \, 
(\gamma^\mu\gamma^\rho- \gamma^\rho\gamma^\mu)\, .
\ee
The sign and normalization of $J_S^{\mu\rho}$ defined in \refb{espinJ} can be shown to
be consistent with that used in \refb{ejnorm} 
by comparing the algebra of the $J^{\mu\rho}$'s in
the spinor and the vector representation.
On the other hand
\refb{eDirac} gives
\be
{\p \KK^{rs}(-p)\over \p p_\mu} =  (\gamma^0 \gamma^\mu)_{rs}, \qquad
{\p \Xi_{rs}(-p)\over \p p_\rho} = -i(\gamma^\rho \gamma^0)_{rs}\, ,
\ee
and therefore
\be \label{e4.13}
{\p \KK^{rt}(-p)\over \p p_{[\mu}}  {\p \Xi_{tu}(-p)\over \p p_{\rho]}} 
= -i \, (\gamma^0 \gamma^{\mu\rho} \gamma^0)_{ru}
= i\, (\gamma^{\mu\rho})_{ur}\,  ,
\ee
where in the last step we have used \refb{egampr}.
Using this we see that sum of the terms  in the second line of \refb{edelalphan}
is given by
\be\label{exa1}
\Delta^r_{~s} = \RR_{\mu\rho\nu\sigma} \,  \left\{
{1\over 8}  \, \gamma^{\nu\sigma} \, \gamma^{\mu\rho}- 
{1\over 16} \, \gamma^{\nu\sigma} \,\gamma^{\mu\rho} \right\}_{sr} \, .
\ee
In arriving at \refb{exa1} we have used the fact that in order to interpret the product of
$J$'s given in \refb{edelalphan} as matrix multiplication as in \refb{exa1} we have to
transpose the matrices costing a sign. This does not change the sign of the second
term but gives an additional minus sign in the first term.
We
now use the identity
\be \label{exa2}
\gamma^{\nu\sigma} \gamma^{\mu\rho} = \gamma^{\nu\sigma\mu\rho}
- (\eta^{\mu\nu} \, \gamma^{\rho\sigma} - \eta^{\rho\nu} \, \gamma^{\mu\sigma}
- \eta^{\mu\sigma} \, \gamma^{\rho\nu} +\eta^{\rho\sigma} \, \gamma^{\mu\nu})
- (\eta^{\mu\nu} \, \eta^{\rho\sigma} - \eta^{\rho\nu} \, \eta^{\mu\sigma}
- \eta^{\mu\sigma} \, \eta^{\rho\nu} +\eta^{\rho\sigma} \, \eta^{\mu\nu})\, ,
\ee
where $\gamma^{\nu\sigma\mu\rho}$ is the totally anti-symmetrized version of
$\gamma^\nu\gamma^\sigma\gamma^\mu\gamma^\rho$.
Using \refb{eepscond}
and the algebraic Bianchi identity of $\RR_{\mu\nu\rho\sigma}$, we can see that 
individual terms in \refb{exa1} vanish.
Therefore $\Delta^r_{~s}$ vanishes and the sub-subleading soft graviton
amplitude is given by the terms in the first four lines on the right hand side of
\refb{etotaln}.

For the massless Rarita-Schwinger field $\psi_{a,r}$,with $a,b,c,d$ denoting vector
indices and $r,s,t,u$ labelling spinor indices, 
we can fix harmonic gauge so that
$\KK$ and $\Xi$ take simple form
\be \label{eRar}
(\KK)^{a,r;b,s} = p_\mu (\gamma^0\gamma^\mu)_{rs} \, \eta^{ab}, \qquad
(\Xi)_{a,r;b,s} = -i\, p_\mu (\gamma^\mu\gamma^0)_{rs} \, \eta_{ab}\, .
\ee
Also we have 
\be
(J^{\mu\rho})_{a,r}^{~~~b,s} = (J_V^{\mu\rho})_a^{~b} \, \delta_r^{~s} + 
\delta_a^{~b} \, (J_S^{\mu\rho})_r^{~s}\, ,
\ee
where $J_V$ and $J_S$ denote the representation of $J$ in vector and spinor
representations, given respectively in \refb{ejnorm} and \refb{espinJ}.
Using \refb{eRar} we again see that the contribution from the first line on the right
hand side of \refb{edelalphan} vanishes.  For minimal coupling to gravity, the contribution
from the third line also vanishes. In the second line of \refb{edelalphan}, noting that the first term is
proportional to $(J_S^{\mu\rho}) J^{\nu\sigma}$ due to \refb{e4.13}, 
we see that there are three kind of
contributions from the first term, proportional to $J_S J_S$, $J_S J_V$ and 
$J_V J_V$. The second term in the second line of \refb{edelalphan} 
is proportional to $(J_S+J_V) (J_S+J_V)$.
The terms
proportional to $J_SJ_S$ have the same structure as \refb{exa1} and vanish using
\refb{exa2}.
The terms proportional to $J_V J_V$ have
the same structure as \refb{e4.1} and vanish after contraction with 
$\RR_{\mu\rho\nu\sigma}$. Therefore we are left with the term proportional to
$J_V J_S$ and $J_S J_V$. Their contribution is given by
\be
\Delta^{a,r}_{~~~b,s}=\RR_{\mu\rho\nu\sigma}\,  \left\{
-{1\over 4} (\gamma^{\mu\rho})_{sr} \, 
(J_V^{\nu\sigma})_b^{~a} + {1\over 8}  (\gamma^{\mu\rho})_{sr} \, 
(J_V^{\nu\sigma})_b^{~a} + {1\over 8}  (\gamma^{\nu\sigma})_{sr} \, 
(J_V^{\mu\rho})_b^{~a} \right\} = 0\, ,
\ee
where in the last step we have used the
symmetry of $\RR_{\mu\rho\nu\sigma}$ under 
$\mu,\rho\leftrightarrow \nu,\sigma$.
Therefore even for massless Rarita Schwinger field minimally coupled to gravity,
the contribution to the sub-subleading 
soft graviton theorem is given by the terms in the first four lines on the right hand side
of \refb{etotaln}.

\subsection{Four dimensional quantum field theories with higher derivative corrections}

Ref.~\cite{1611.07534} discussed soft graviton theorem for
massless fields in four dimensions in the presence of higher derivative corrections.
In this section we shall compare our results with the results of \cite{1611.07534}. 
The relevant bosonic fields here include massless scalar $\phi$, massless gauge
field $A_\mu$ and massless graviton. In the fermionic sector we can have 
massless spin 3/2 and spin 1/2 fields.

First let us consider the case of massless bosonic fields only. 
We shall choose harmonic gauge so that $\KK^{\alpha\beta}(q)$ is given by $-q^2\delta^{\alpha\beta}$ and
$\Xi^i_{\alpha\beta}=-i\, \delta_{\alpha\beta}$. 
In this case the contributions from the derivatives of $\Xi^i$ in 
\refb{edelalphan} vanish. Furthermore as seen in section \ref{se-m}, 
the contribution from the $J^{\mu\rho} J^{\nu\sigma}$ term 
vanishes for scalar and the gauge fields, while for gravity this
term cancels a term arising out of expansion of the Einstein-Hilbert action around a soft background.
Therefore the contribution to \refb{edelalphan} comes only from the interaction terms 
involving non-minimal coupling of gravity to other fields.
It is easy to classify the 
possible terms that could contribute. They are\footnote{We shall
not consider theories with superreormalizable couplings {\it e.g.} a three point
coupling without derivative between the massless scalars.}
\ben \label{epossint}
&& \int d^4 x \, \sqrt{-\det g} \, \phi \, R_{\mu\nu\rho\sigma} \, R^{\mu\nu\rho\sigma}, \quad 
\int d^4 x \, \sqrt{-\det g}\,  R_{\mu\nu\rho\sigma}\, F^{\mu\nu} \, F^{\rho\sigma}, \nonumber \\
&& \int d^4 x \, \sqrt{-\det g} \, \phi \, R_{\mu\nu\rho\sigma} \, \wt R^{\mu\nu\rho\sigma}, \quad 
\int d^4 x \, \sqrt{-\det g} \, R_{\mu\nu\rho\sigma} \, F^{\mu\nu} \, \wt F^{\rho\sigma} \, ,
\een
where $R_{\mu\nu\rho\sigma}$ is the Riemann tensor, $F_{\mu\nu} = \p_\mu A_\nu - \p_\nu A_\mu$ is
the gauge field strength and $\wt R$, $\wt F$ denote Hodge duals:
\be
\wt R_{\mu\nu\rho\sigma} = \left(\sqrt{-\det g}\right)^{-1}\, \eps_{\mu\nu\mu'\nu'} \, 
{R^{\mu'\nu'}}_{\rho\sigma}, 
\quad \wt F_{\mu\nu} = \left(\sqrt{-\det g}\right)^{-1}\, 
\eps_{\mu\nu\mu'\nu'} \, F^{\mu'\nu'}\, .
\ee
One could also consider a term with three Riemann tensors appropriately contracted, but when we
take one of the external states to be soft and another on-shell, the vertex contains more than two
powers of soft momentum and therefore does not contribute to the amplitude at the sub-subleading order. In higher dimensions the term with two Riemann tensors with their
indices contracted gives rise to a three graviton vertex but in four dimensions this is
equivalent to the sum of Gauss-Bonnet term which is a total derivative and terms involving
Ricci tensor that vanish on-shell. Therefore this does not contribute in the soft limit.

The three point vertices listed in \refb{epossint} affect the sub-subleading contribution by modifying the
three point vertex in Fig.~\ref{f1}. Two of the external states of this vertex, including the soft graviton,
are on-shell while the third one, representing the internal line, is nearly on-shell. Since we are to
evaluate the leading contribution from this vertex in the soft limit, we can regard the internal line also as on-shell by 
decomposing the numerator factor $\Xi^i$ from the internal propagators into a sum over 
physical and unphysical polarizations and using the fact that in the final amplitude the contribution from
the unphysical polarizations will cancel. Therefore the computation reduces to the problem of computing the
contribution of \refb{epossint} to an on-shell three point amplitude.

A further simplification in four dimensions comes from the fact that in four dimensions by 
appropriate choice of gauge the polarization tensor of 
a massless graviton can be taken to be the square of that of a massless photon carrying the
same momentum. By making this choice we
write 
\be
\ve_{\mu\nu} = \ve_\mu \ve_\nu, \quad \eee_{\mu\nu} = \eee_\mu \eee_\nu\, ,
\ee
for the polarizations of soft and hard gravitons respectively. Then in the momentum space,
to linearized order the Riemann tensors associated with the soft and the finite
energy graviton fields
take the form
\be \label{esoftlin}
R^{(s)}_{\mu\rho\nu\sigma} = \{\ve_{\mu\nu}k_{\rho}k_{\sigma}\ -\ \ve_{\mu\sigma}k_{\nu}k_{\rho}\ 
-\ \ve_{\nu\rho}k_{\sigma}k_{\mu}\ +\ \ve_{\rho\sigma}k_{\mu}k_{\nu}\}
= (\ve_\mu k_\rho - \ve_\rho k_\mu) (\ve_\nu k_\sigma - \ve_\sigma k_\nu)
\ee
\be 
R^{(h)}_{\mu\rho\nu\sigma} = \{\eee_{\mu\nu}p_{\rho}p_{\sigma}\ -\ \eee_{\mu\sigma}p_{\nu}p_{\rho}\ 
-\ \eee_{\nu\rho}p_{\sigma}p_{\mu}\ +\ \eee_{\rho\sigma}p_{\mu}p_{\nu}\}
= (\eee_\mu p_\rho - \eee_\rho p_\mu) (\eee_\nu p_\sigma - \eee_\sigma p_\nu)
\ee
respectively. Here $p$ denotes the momentum carried by the finite energy graviton.
Using this we see that the contribution to the
three point vertex from the $\phi \, R_{\mu\nu\rho\sigma}\, R^{\mu\nu\rho\sigma}$ term
is proportional to
\be \label{eampli}
\{(\ve_\mu k_\rho - \ve_\rho k_\mu) (\eee^\mu p^\rho - \eee^\rho p^\mu)\}^2\, .
\ee

Now in flat space-time background, a polarization vector $\ve$ carried by a massless 
particle of
momentum $k$ is defined to have helicity $\pm$ if
\be \label{edual}
\eps_{\mu\nu\rho\sigma} \, (k^{\rho} \ve^{\sigma} - k^\sigma \ve^\rho) = \pm \, 2 \, i\, (k_\mu \, \ve_\nu 
-k_\nu \ve_\mu)\, .
\ee
Using this it is easy to see that 
\be
(\ve_\mu k_\rho - \ve_\rho k_\mu) (\eee^\mu p^\rho - \eee^\rho p^\mu)=0 \, ,
\ee
unless $\ve$ and $\eee$ carry same helicity. For example if $\ve$ has positive
helicity and $\eee$ has negative helicity then we can write
\be
(\ve_\mu k_\rho - \ve_\rho k_\mu) (\eee^\mu p^\rho - \eee^\rho p^\mu)
= {1\over 2\, i} \eps_{\mu\rho \mu'\rho'} (\ve^{\mu'} k^{\rho'} - \ve^{\rho'} k^{\mu'}) (\eee^\mu p^\rho - \eee^\rho p^\mu)
= - (\ve^{\mu'} k^{\rho'} - \ve^{\rho'} k^{\mu'}) (\eee_{\mu'} p_{\rho'} - \eee_{\rho'} p_{\mu'})
\, .
\ee
Since the two sides of this equation are negatives of each other the result vanishes. Therefore we shall
take $\ve$ and $\eee$ to have the same helicity. This analysis also shows that once we have chosen the
helicity of the soft graviton, the contribution
from the $\phi \, R_{\mu\nu\rho\sigma}\, \wt R^{\mu\nu\rho\sigma}$ term differs from the one given in
\refb{eampli} by a factor of $\pm2\, i$. 
Therefore we shall not analyze its contribution separately.

For the $R_{\mu\rho\nu\sigma} F^{\mu\rho} F^{\nu\sigma}$ term, the three point vertex receives
a contribution proportional to 
\be \label{eampli2}
\{(\ve_\mu k_\rho - \ve_\rho k_\mu) (\eee^\mu p^\rho - \eee^\rho p^\mu)\}
\{(\ve_\nu k_\sigma - \ve_\sigma k_\nu) (\bar\eee^\nu p^\sigma - \bar\eee^\sigma p^\nu)\}\, ,
\ee
where $\eee$ and $\bar \eee$ represent the polarizations of the external and the internal photons.
The previous argument now shows that this vanishes unless the helicities of $\eee$ and $\bar \eee$
agree with that of $\ve$. Since for soft external graviton, the momenta of the two photons connected
to the vertex are nearly 
equal and opposite, this shows that $\bar\eee$ is equal to $\eee$ (up to gauge 
transformation). Therefore \refb{eampli2} reduces to \refb{eampli}.

In order to compare this with the result of \cite{1611.07534} we need to convert \refb{eampli} to the 
spinor helicity 
notation (see {\it e.g.} \cite{1308.1697,1310.5353} for a review). 
We label each of the null vectors $p$ and $k$ by a pair of two component spinors 
\be \label{epkrep}
p\to (\mu_\alpha,\tilde \mu_{\dot{\alpha}}), \qquad 
k\to (\lambda_\alpha,\tilde \lambda_{\dot{\alpha}})\, ,
\ee 
via the relation
\be \label{espinrep}
p_\mu (\gamma^\mu)_{\alpha\dot{\alpha}} = \mu_\alpha \, \tilde \mu_{\dot{\alpha}}, \qquad
k_\mu (\gamma^\mu)_{\alpha\dot{\alpha}} = \lambda_\alpha \, 
\tilde \lambda_{\dot{\alpha}}, 
\ee
and introduce the notation
\be
[\tilde a\, \tilde b] 
= \eps^{\dot{\alpha}\dot{\beta}} \tilde a_{\dot{\alpha}} 
\tilde b_{\dot{\beta}}, \quad \langle a\, b\rangle = 
\eps^{\alpha\beta} a_\alpha b_\beta\, ,
\ee
where $\eps=i\sigma_2$, $\sigma_i$'s being Pauli matrices. In this notation we have
\be \label{epdotk}
p\cdot k = -{1\over 2} [\tilde\lambda  \, \tilde\mu] \langle \lambda\, \mu\rangle\, .
\ee
For describing polarization vectors $\ve^\mu$ and $\eee^\mu$ we introduce an auxiliary pair of
spinors $(x_\alpha, \tilde x_{\dot{\alpha}})$ for the soft particle and another pair of
spinors $(y_\alpha, \tilde y_{\dot{\alpha}})$ for the finite energy particle. In terms of these spinors
we can label the normalized positive helicity polarization vectors $\ve$ and $\eee$ 
as\footnote{The spinors $\tilde x$ and $\tilde y$ are necessary for describing negative helicity
polarization vectors.}
\be \label{eepsrep}
\ve \to \sqrt 2\, (x_\alpha, \tilde \lambda_{\dot{\alpha}})/ \langle \lambda\, x\rangle, \qquad
\eee\to \sqrt 2\, (y_\alpha, \tilde \mu_{\dot{\alpha}}) / \langle \mu\, y\rangle\, .
\ee 
Now we can easily
generalize \refb{epdotk} as
\be 
\ve \cdot p =-{1\over \sqrt 2} {[\tilde \mu\, \tilde \lambda] \langle\mu \, x\rangle \over  \langle \lambda\, x\rangle}, 
\quad \eee \cdot k =  -{1\over \sqrt 
2} {[\tilde \lambda\, \tilde \mu] \langle\lambda \, y\rangle \over  \langle \mu\, y\rangle}, 
\quad \ve \cdot \eee = - 
{ [\tilde \lambda\, \tilde \mu] \langle x \, y\rangle \over  \langle \lambda\, x\rangle
\langle \mu\, y\rangle} \, .
\ee
We can simplify our analysis by making the gauge choice $y=\lambda$. In that case $\eee\cdot k$ vanishes
and we have
\be \label{esoso1}
\{(\ve_\mu k_\rho - \ve_\rho k_\mu) (\eee^\mu p^\rho - \eee^\rho p^\mu)\} = 
[\tilde \lambda\, \tilde \mu]^2\, .
\ee
Therefore for the three point vertex induced from any of the terms listed in 
\refb{epossint},  the soft factor associated with the amplitude in Fig.~\ref{f1} is proportional to
\be \label{esoso}
{1\over 2 p\cdot k} \{(\ve_\mu k_\rho - \ve_\rho k_\mu) 
(\eee^\mu p^\rho - \eee^\rho p^\mu)\}^2 =
-  {[\tilde \lambda\, \tilde \mu]^3\over \langle \lambda\, \mu\rangle }\, .
\ee
This agrees with the result of \cite{1611.07534}.

Finally we consider the inclusion of spin $3/2$ and spin 1/2 Dirac spinors $\psi_\rho$ and $\chi$. The
terms in the action that can lead to the coupling of a soft graviton to a pair of  finite energy nearly on-shell
fermions are of the form
\be \label{elistb}
\int \, d^4 x\, \sqrt{-\det g} \, 
R^{\mu\rho\nu\sigma} \bar\psi_\mu \gamma_{\nu\sigma} \p_\rho \chi, \quad \int \, d^4 x\, \sqrt{-\det g} \, 
R^{\mu\rho\nu\sigma} \bar\psi_\mu \gamma_{\nu\sigma} \gamma^5\, \p_\rho \chi\, .
\ee
For given helicity of $\chi$ the contribution from the two terms are proportional to each other; so let us
focus on the first term.
Using \refb{esoftlin} this leads to the following coupling between the soft graviton of momentum $k$
and the finite energy (nearly)
on-shell fermions of momentum $p$:
\be \label{eamam1}
(\ve_\mu k_\rho - \ve_\rho k_\mu) (\ve_\nu k_\sigma - \ve_\sigma k_\nu)\, 
p^\rho\, \bar\psi^\mu \gamma^{\nu\sigma} \chi \, .
\ee
Using \refb{edual} and the corresponding result for the spinors in flat space:
\be 
\eps_{\mu\nu\rho\sigma} \, \gamma^{\rho\sigma} = 2\, i\, \gamma_{\mu\nu} \, \gamma^5 = 
2\, i\,  \gamma^5 \, \gamma_{\mu\nu}\, ,
\ee
it is easy to see that the amplitude \refb{eamam1}
vanishes unless the $\bar\psi$ and $\chi$ fields carry the same
helicity as the soft graviton. For positive helicity of the soft graviton this means that 
\be
\bar \psi_\rho \, \gamma^5 = \bar\psi^\rho, \quad \gamma^5 \chi = \chi, \quad
\eps^{\mu\rho\nu\sigma} (p_\rho\, \bar\psi_\mu - p_\mu\, \bar\psi_\rho) =
2\, i \, (p^\sigma\, \bar\psi^\nu - p^\nu\, \bar\psi^\sigma)
\ee
Therefore $\bar\psi_\rho$ can be taken to be proportional to the positive helicity 
polarization vector $\eee_\rho$ and that in spinor space both $\bar\psi_\rho$ and 
$\chi$ carry dotted index and can be taken to be proportional to $\tilde \mu_{\dot\alpha}$
introduced in \refb{epkrep}.  
Up to overall normalization, the soft factor is then given by
\be \label{esoso2}
{1\over 2 p\cdot k} (\ve_\mu k_\rho - \ve_\rho k_\mu) (\ve_\nu k_\sigma - \ve_\sigma k_\nu)
\, p^\rho \, \eee^\mu\, (\gamma^{\sigma\nu})^{\dot{\alpha}\dot{\beta}} \, \tilde\mu_{\dot{\alpha}}
\tilde\mu_{\dot{\beta}}\, .
\ee
Now we have, using \refb{espinrep}, \refb{eepsrep}
\be 
\ve_\nu k_\sigma (\gamma^{\sigma\nu})^{\dot{\alpha}\dot{\beta}} \, \tilde\mu_{\dot{\alpha}}
\tilde\mu_{\dot{\beta}}
\propto { \langle \lambda\, x\rangle [\tilde\lambda\, \tilde \mu]^2 \over \langle \lambda\, x\rangle 
} =  [\tilde\lambda\, \tilde \mu]^2\, .
\ee
Using this, and \refb{epdotk}, \refb{esoso1} we see that \refb{esoso2} reduces to
\be
{ [\tilde\lambda\, \tilde \mu]^3\over \langle\lambda\, \mu\rangle}
\ee
up to normalization factor. This is identical to \refb{esoso}, in agreement with \cite{1611.07534}.

For specific helicity configurations, the nature of soft theorems can be completely governed by the 
non-universal terms. An example of this is as follows. Consider a tree level 4-graviton amplitude 
${\cal M}_4(p_{1}^{+},\dots, p_{4}^{+})$ in which all the gravitons have positive helicity. As is well known \cite{1308.1697}, in pure gravity this amplitude vanishes. However suppose we compute this 
amplitude in the theory where gravity is non-minimally coupled to a massless scalar via 
$\int\sqrt{-\textrm{det}g} \ \phi\ R_{\mu\nu\rho\sigma}R^{\mu\nu\rho\sigma}$. In this case the amplitude 
${\cal M}_4(p_{1}^{+},\dots, p_{4}^{+})$ will not be zero due to the additional vertices involving the 
scalar, leading to a scalar exchange diagram.
We can also see that in the limit $p_{4}\rightarrow\ 0$ we get
\be\label{sp-case}
{\cal M}_4(p_{1},\dots, p_{4})\ =\left[\tilde{S}^{(2)}_{1}{\cal M}_{3}(p_{1}, p_2^+,
p_{3}^{+}) + 
\tilde{S}^{(2)}_{2}{\cal M}_{3}(p_{1}^+, p_2,
p_{3}^{+}) +
\tilde{S}^{(2)}_{3}{\cal M}_{3}(p_{1}^+, p_2^+,
p_{3}) \right]
+ O(E_{n}^{2})
\ee
where  $\tilde{S}^{(2)}_{i}$ is the sub-subleading factor given in eq.(\ref{esoso}) with $(e,p)$ replaced by
$(e_i, p_i)$ and the $i$-th 3 
point amplitude on the right hand side of eq.(\ref{sp-case}) is an amplitude involving $2$ gravitons and a 
scalar with momentum $p_{i}$. In the above equation, there is no universal soft factor due to the 
fact that universal soft factors (to sub-subleading order) are precisely governed by the pure gravity 
three point vertices. These factors will dress a 3 graviton amplitude which is computed via 
Einstein Hilbert Lagrangian and such an amplitude vanishes as all the gravitons have the same 
helicity.

\sectiono{Comparison with results from tree level string theory} \label{sstring}

In this section we shall compare our results with the results of
\cite{1512.00803,1604.03355} which computed bosonic
string tree amplitudes with external graviton
and other states in the soft limit. 

\subsection{Two tachyon two graviton amplitude} 

Ref.~\cite{1512.00803} computed the
scattering amplitude involving a pair of external gravitons and a pair of external tachyons
in the limit when one of the graviton momentum becomes soft. 
At the sub-subleading order the result of \cite{1512.00803} contained an extra term
besides the ones given by the first four lines on the right hand side of \refb{etotaln}.
If we denote by $k$ and
$\ve$ the momentum and polarization of the soft graviton, by $p_1$ and $\eee$ the
momentum and polarization of the finite energy graviton, and by $p_2$ and $p_3$ the
momenta of the tachyons then, up to an overall normalization,
the extra term obtained in \cite{1512.00803} (after correcting a 
typographical error and the overall sign) can be written as
\ben \label{ecc1}
&& -{\alpha'\over 4} \, \bigg\{- k\cdot p_- \, p_1^\mu\, \ve_{\mu\nu} \, \eee^{\nu\rho} \, p_{-\rho}
+ k\cdot p_1 \, p_-^\mu\, \ve_{\mu\nu} \, \eee^{\nu\rho} \, p_{-\rho}
+{1\over p_1\cdot k} \, k \cdot p_- \, k_\mu \, \eee^{\mu\nu} \, p_{-\nu}\, 
p_{1\rho} \, \ve^{\rho\sigma} \, p_{1\sigma} \nonumber \\ && \hskip 1in
- k_\mu \, \eee^{\mu\nu} \, p_{-\nu} \, p_{1\rho}\, \ve^{\rho\sigma} \, p_{-\sigma}
\bigg\}, 
\een
where
\be
p_- = p_2-p_3\, .
\ee
Since $p_2$ and $p_3$ satisfy the on-shell condition $p_2^2=p_3^2 = - m_T^2$ where
$m_T^2$ is the tachyon mass$^2$, we have, using momentum conservation,
\be 
p_1\cdot p_- = - (p_2+p_3+k)\cdot (p_2-p_3) = \OO(k)\, .
\ee
Using this we can express \refb{ecc1} as 
(up to term suppressed by additional powers of
soft momentum)
\be \label{ecc2}
-{\alpha'\over 8} \, {1\over 
p_1\cdot k} \, R^{(s)}_{\mu\rho\nu\sigma} \, R^{(h)\mu\rho\tau\sigma} \, 
p_-^\nu \, p_{-\tau} 
\ee
where $R^{(s)}$ and $R^{(h)}$ are the linearized
Riemann tensors for the soft and finite energy external
gravitons respectively:
\be \label{eRs}
R^{(s)}_{\mu\rho\nu\sigma} = \{\ve_{\mu\nu}k_{\rho}k_{\sigma}\ -\ \ve_{\mu\sigma}k_{\nu}k_{\rho}\ 
-\ \ve_{\nu\rho}k_{\sigma}k_{\mu}\ +\ \ve_{\rho\sigma}k_{\mu}k_{\nu}\}\, ,
\ee
and
\be  \label{eRh}
R^{(h)}_{\mu\rho\nu\sigma} 
= \{\eee_{\mu\nu}p_{1\rho}p_{1\sigma}\ -\ \eee_{\mu\sigma}p_{1\nu}p_{1\rho}\ 
-\ \eee_{\nu\rho}p_{1\sigma}p_{1\mu}\ +\ \eee_{\rho\sigma}p_{1\mu}p_{1\nu}\}\, .
\ee
\refb{ecc2} may be written in a more suggestive form by noting that the three
point coupling between the finite energy graviton of momentum $p_1$ and
polarization $\eee$ and a pair of tachyons of momenta $p_2$ and $p_3$ has the
form $\eee_{\mu\nu} \Gamma^{\mu\nu}$ where\cite{1512.00803}
\be
\Gamma^{\mu\nu} 
= p_-^\mu p_-^\nu / 4\, .
\ee 
The three point coupling between the two tachyons and a dilaton is given by the same
formula if we choose $e_{\mu\nu}\propto \eta_{\mu\nu}$.
Therefore we can express \refb{ecc2} as
\be \label{ecc2a}
-{\alpha'\over 2} \, {1\over 
p_1\cdot k} \, R^{(s)}_{\mu\rho\nu\sigma} \, R^{(h)\mu\rho\tau\sigma} \, 
\Gamma^\nu_{~\tau}\, .
\ee

\begin{figure}
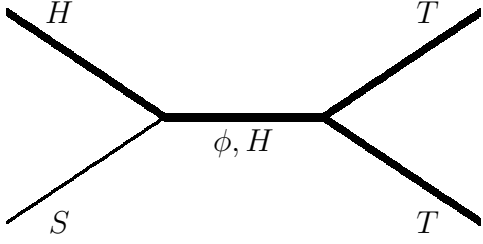

\begin{center}
\figstringc 
\end{center}

\vskip -.6in

\caption{Possible sources of correction to the sub-subleading soft graviton theorem for
two graviton, two tachyon scattering in string theory. $S$ denotes the soft graviton, 
$T$ denotes external tachyon, $H$ denotes finite energy external or internal graviton and
$\phi$ denotes a finite energy internal dilaton. These represent additional contribution to Fig.~\ref{f1} 
besides the one arising from the three point vertex \refb{e4.8n} representing minimal coupling.
\label{fstr2}
}
\end{figure}

Eq.\refb{ecc2a}, being proportional to $(p_1\cdot k)^{-1} 
R^{(s)}_{\mu\rho\nu\sigma}$, clearly has the structure of the
corrections given in the last term on the 
right hand side of \refb{etotaln}. 
We shall now explore their origin is some more detail.
In the Siegel gauge $K^{\alpha\beta}(q)$ is proportional
to $q^2$ and therefore $\Xi^i_{\alpha\beta}$ is independent of $q$. Therefore the
contribution from the terms involving derivatives of $\Xi^i$ in \refb{edelalphan} 
vanish.  Also the quadratic term in $J$ vanishes for the tachyon and for the
graviton it cancels against a
term from the Einstein-Hilbert action as in section \ref{se-m}.  Therefore the correction term
\refb{ecc2a} can only come from a higher derivative three point coupling involving one soft
graviton and a pair of finite energy particles.
If in \refb{ecc2a} we decompose
\be \label{edecomp}
\Gamma^\nu_{~\tau} = {1\over D}\, \Gamma^\rho_{~\rho}\,  \delta^\nu_\tau + 
\{ \Gamma^\nu_{~\tau} - {1\over D}\, \Gamma^\rho_{~\rho}\,  \delta^\nu_\tau \} 
\ee
then the contribution from the first term to \refb{ecc2a} 
gives the dilaton mediated coupling 
in Fig.~\ref{fstr2} where we choose the internal line to be the dilaton $\phi$.
This  requires a three point coupling proportional to
\be 
\int d^D x\, \phi \, R^{(s)}_{\mu\nu\rho\sigma} \, R^{(h)\mu\nu\rho\sigma} \, ,
\ee
which comes via the correction to the effective action
of the form
\be \label{e5.8}
\int d^D x \, \sqrt{-\det g} \, \phi\, R_{\mu\nu\rho\sigma} \, R^{\mu\nu\rho\sigma}\, .
\ee
This is known to be present in the bosonic and heterotic string theory. 
Contribution from the second term in \refb{edecomp} to 
\refb{ecc2a} can be identified as the graviton
mediated amplitude where we pick the intermediate state  in Fig.~\ref{fstr2} to be
a finite energy graviton $H$. This requires a higher derivative three point 
coupling involving one soft and two finite energy gravitons of the form
\be \label{esofthard}
\int\, d^D x \, R^{(s)}_{\mu\rho\nu\sigma} \, 
R^{(h)\mu\rho\tau\sigma} \, {h_\tau}^\nu\, .
\ee
This can come from the following term in the original action
\be \label{efull}
\int d^D x \, \sqrt{-\det g} \, R_{\mu\nu\rho\sigma} \, R^{\mu\nu\rho\sigma}\, .
\ee
It is easy to verify that in the soft limit, the coupling of a soft graviton to
a pair of finite energy gravitons computed from \refb{esofthard} and
\refb{efull} are the same (up to overall normalization).
For $D=4$ \refb{efull} does not contribute to the three point function since it is equivalent
to the Gauss-Bonnet action on-shell. However in higher dimensions the contribution from
this term does not vanish.

\subsection{Scattering of gravitons and dilatons}

Ref.~\cite{1604.03355} computed the scattering amplitude in the bosonic string
theory for massless external states, and found corrections to the soft graviton
theorem at sub-subleading order. If the soft particle carries polarization 
$\ve$ and momentum $k$, and the finite energy particles carry momenta 
$p_1,\cdots p_N$ and polarizations $\eee_{1}^{\mu\nu},\cdots \eee_{N}^{\mu\nu}$,
then the correction to the sub-subleading 
soft graviton theorem was found to be given by:
\be \label{ema0}
{\alpha'\over 2} \ve^{\mu\nu}  \sum_{i=1}^N 
\left\{k_\sigma p_{i\nu} \eta_{\rho\mu} + k_\rho p_{i\mu} 
\eta_{\sigma\nu} - \eta_{\rho\mu} \eta_{\sigma\nu} p_i\cdot k -{1\over p_i\cdot k} 
k_\rho k_\sigma p_{i\mu} p_{i\nu}
\right\}  \Pi_i^{\{\rho,\sigma\}} \Gamma\, ,
\ee
where $\Gamma$ is the amplitude without the soft graviton and
the operation $\Pi_i^{\{\rho,\sigma\}}$ is defined as follows. If we label the
polarization $\eee_i^{\rho\sigma}$ as $\eee_i^\rho \bar\eee_i^\sigma$ then
\be 
\Pi_i^{\{\rho,\sigma\}}  = {1\over 2} \left[\eee_i^\rho {\p\over\p \eee_{i,\sigma} }+ \bar\eee_i^\rho 
{\p\over\p \bar\eee_{i,\sigma}} +
\eee_i^\sigma {\p\over\p \eee_{i,\rho} }+ \bar\eee_i^\sigma
{\p\over\p \bar\eee_{i,\rho}}
\right]\, .
\ee
In string theory $\eee_i^{\rho\sigma}$ may be symmetric or anti-symmetric under the
exchange  $\rho\leftrightarrow\sigma$. If we restrict to the symmetric case, representing
graviton or dilaton state, then
\be
\Pi_i^{\{\rho,\sigma\}}  \Gamma = \eee_{i~\tau}^{~\rho} \, \Gamma_{(i)}^{\sigma\tau}
+ \eee_{i~\tau}^{~\sigma} \, \Gamma_{(i)}^{\rho\tau}
\ee
where $\Gamma_{(i)}^{\sigma\tau}$ is defined such that $\eee_{i,\rho\sigma}
\Gamma_{(i)}^{\rho\sigma} = \Gamma$. This allows us to express \refb{ema0} as
\be \label{ema1}
{\alpha'} \, \ve^{\mu\nu}  \sum_{i=1}^N 
\left\{k_\sigma p_{i\nu} \eta_{\rho\mu} + k_\rho p_{i\mu} 
\eta_{\sigma\nu} - \eta_{\rho\mu} \eta_{\sigma\nu} p_i\cdot k -{1\over p_i\cdot k} 
k_\rho k_\sigma p_{i\mu} p_{i\nu}
\right\}  \eee_{i~~\tau}^{~\rho} \Gamma_{(i)}^{\sigma\tau}\, .
\ee 
Now using the gauge invariance of $\Gamma$:
\be
p_{i,\rho} \, \Gamma_{(i)}^{\rho\tau} = 0, \qquad p_{i,\tau} \, \Gamma_{(i)}^{\rho\tau} = 0\, ,
\ee
one can express \refb{ema1} as
\be \label{ema2}
-{\alpha'\over 2} \, R^{(s)}_{\mu\rho\nu\sigma} \, \sum_{i=1}^N 
{1\over p_i\cdot k} \, R^{(i)\mu\rho\tau\sigma}
\Gamma_{(i)\tau}^{~~\nu}\, ,
\ee
where $R^{(s)}$ has been defined in \refb{eRs}, and $R^{(i)}$ is
given  by
\be  \label{eRi}
R^{(i)}_{\mu\rho\nu\sigma} 
= \{\eee_{i,\mu\nu}p_{i\rho}p_{i\sigma}\ -\ \eee_{i,\mu\sigma}p_{i\nu}p_{i\rho}\ 
-\ \eee_{i,\nu\rho}p_{i\sigma}p_{i\mu}\ +\ \eee_{i,\rho\sigma}p_{i\mu}p_{i\nu}\}\, .
\ee

Eq.~\refb{ema2} has a structure 
identical to the one obtained in \refb{ecc2a}. As in that case,
decomposing ${\Gamma_\tau}^\nu$ as
\be \label{e5.18}
{\Gamma_\tau}^\nu = {1\over D} \, {\delta_\tau}^\nu \, {\Gamma_\rho}^\rho+
\left\{ {\Gamma_\tau}^\nu -  {1\over D} \, {\delta_\tau}^\nu \, {\Gamma_\rho}^\rho
\right\}
\ee
we can interpret the contribution to \refb{ema2} from the first term in \refb{e5.18}
as due to
an intermediate finite energy dilaton and the contribution to \refb{ema2} 
from the rest of the terms in \refb{e5.18} as due to
an intermediate finite energy graviton. The relevant three point interactions
arise from \refb{e5.8} and \refb{efull}.

\subsection{Amplitude for two tachyons, one graviton and one massive particle}

We shall now consider the four point scattering in bosonic string theory
of a pair of tachyons carrying momenta $p_1$ and $p_2$,
a rank 4 symmetric tensor field at the first massive level carrying momentum $p_3$ and
polarization $\eps_3$ and a soft graviton carrying momentum $k$ and polarization $\ve$.
The full amplitude 
can be read out from eq.~(70) of \cite{1512.00803} with the following replacement:
\be
k_3\to k, \quad a_{3\mu} \, \tilde a_{3\nu} \to \ve_{\mu\nu}, \quad p_4\to p_3, \quad
H_{\mu\nu} \wt H_{\rho\sigma} \to \eps_{3,\mu\nu\rho\sigma}\, .
\ee 
With this we find that the leading and subleading soft graviton amplitudes
agree with the expected result given in the first two lines on the right hand side of 
\refb{etotaln} if we take the amplitude without the soft graviton to be
\be \label{egiven}
-{1\over 16} \, 
\eps_{3,\mu\nu\rho\sigma} \, p_-^\mu \, p_-^\nu\, p_-^\rho \, p_-^\sigma\, , \qquad
p_-\equiv p_1 - p_2\, .
\ee
Given \refb{egiven}, sub-subleading contribution from the third and fourth lines on the 
right hand side of \refb{etotaln} take the form
\ben
&& {3\over 4} \left\{ \ve^{\mu\tau} \, (p_{1\tau}+p_{2\tau})  +
\ve^{\mu\tau} \, p_{3\tau}\, {k\cdot p_-\over k\cdot p_3} \right\} \eps_{3,\mu\nu\rho\sigma}
k^\nu \, p_-^\rho\, p_-^\sigma \nonumber \\
&& -{3\over 8} \left\{ k\cdot (p_1+p_2) + {(k\cdot p_-)^2 \over k\cdot p_3}\right\}
\eps_{3,\mu\nu\rho\sigma} \, \ve^{\mu\nu}\, p_-^\rho\, p_-^\sigma \nonumber \\
&& -{3\over 8} \left\{ {\ve_{\mu\nu} p_1^\mu p_1^\nu\over k\cdot p_1}
+ {\ve_{\mu\nu} p_2^\mu p_2^\nu\over k\cdot p_2}
+ {\ve_{\mu\nu} p_-^\mu p_-^\nu\over k\cdot p_3}\right\}
\eps_{3,\mu\nu\rho\sigma} \, k^\mu \, k^\nu \, p_-^\rho\, p_-^\sigma\, .
\een
However the actual amplitude computed from eq.~(70) of \cite{1512.00803}
has some additional terms. These  are given by
\be
-{1\over k \cdot p_3} \, R_{\mu\rho\nu\sigma} \left[{1\over 2} \, p_1^\rho \, p_2^\sigma\, 
\eps_3^{\mu\nu a b} \, p_{-a} \, p_{-b} -{1\over 4} \, p_3^\mu \, p_3^\nu \, p_-^\rho \,
\eps_3^{\sigma a b c} \, p_{-a} \, p_{-b} \, p_{-c}
\right]\, ,
\ee
where $\RR_{\mu\rho\nu\sigma}$ is given in \refb{edefR}. This form of the correction terms
is consistent with the general form of the corrections to the sub-subleading soft graviton theorem
given in \refb{edelalphan}, and can be traced to a non-minimal three point coupling between a
soft graviton, a massive rank four symmetric tensor field and another massive 
field at the same mass level. The  relevant diagram has the same structure as Fig.~\ref{fstr2} 
with the
finite energy external graviton replaced by the massive symmetric rank four tensor field
and the internal line representing either a massive rank four symmetric tensor or another
field at the same mass level.

\sectiono{Infrared  divergences} \label{sir}

In four space-time dimensions the loop amplitudes suffer from infrared divergences 
that have to
be removed by either summing over final states and averaging over 
initial states\cite{kinoshita,lee,bloch,weinberg1,weinberg2}, or
by changing the description of the scattering 
states\cite{kulish-faddeev,1308.6285,1705.04311}. 
Hence in four dimensions the structure of soft theorems for loop amplitudes is 
sensitive to the divergent infra-red effects\cite{1405.1015,1405.3413}. 
For this reason for loop 
amplitudes we focus on space-time 
dimensions $D\ge 5$ for which the S-matrix elements are finite -- at least before
taking the soft limit. Our goal in this section will be to explore if our analysis of soft
theorem in
section \ref{s1} based on 1PI effective action, that includes loop amplitudes as
well, could be affected by infrared issues
in $D\ge 5$ even though there are no divergences before taking the soft limit.
We shall first consider the possible effects of soft divergences and then briefly
discuss the effect of collinear divergences that can arise when some of the finite
energy  external 
states are  massless.

\subsection{Soft divergences} \label{sinfra}

\begin{figure}
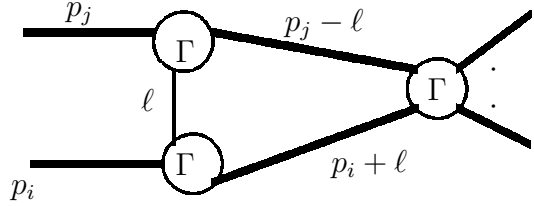


\begin{center}

\figsubtleb

\caption{Potentially infrared divergent contribution to the amplitude.
\label{fs1}} 

\end{center}

\end{figure}

Soft divergences refer to divergences that arise from regions of loop momentum
integration in which all components of the loop momentum becomes small.
The absence of soft divergences in $D\ge 5$ for 
amplitudes without soft external lines
has been
illustrated in Fig.~\ref{fs1}. Here $\Gamma$'s represent amputated Green's
functions
and the thin internal line carrying momentum 
$\ell$ represents a massless soft line, i.e. we consider the limit $\ell_\mu\to 0$.
In this limit, if we pick the internal states carrying momenta $p_j-\ell$ and
$p_i+\ell$ to be of the same mass as the external states carrying momentum
$p_j$ and $p_i$ respectively, then in the $\ell_\mu\to 0$ limit the integrand of the
Feynman diagram goes as
\be \label{ediv1}
\II = \{ \ell^2 \, (-2p_j\cdot \ell + p_j^2 + M_j^2) \, (2p_i \cdot \ell
+p_i^2 + M_i^2) )\}^{-1} \times \hbox{finite}
\ee 
where the $\ell^2$ factor in the denominator comes from the propagator
carrying momentum $\ell$ and the $(-2p_j\cdot\ell+ p_j^2 + M_j^2)$ and 
$(2p_i\cdot\ell+p_i^2 + M_i^2)$
factors arise from the propagators carrying momenta $p_j-\ell$ and 
$p_i+\ell$ respectively. The on-shell condition for the external
states carrying momenta $p_i$ and $p_j$ sets $p_i^2+M_i^2$ and 
$p_j^2+M_j^2$ to zero. Even 
though the integrand $\II$ has four powers of $\ell_\mu$ in the denominator and 
therefore
diverges in the $\ell_\mu\to 0$ limit, the integral $\int d^D\ell\, \II$ 
is convergent for $D\ge 5$. Similar power counting\cite{sterman} 
shows that there are
no collinear divergences -- divergences arising from regions of loop
momenta when one or more internal momenta of a massless state 
becomes collinear to the external momentum of a massless state.
This will be discussed in section \ref{scollinear}.
Furthermore, adding more loops containing soft or collinear 
momenta does not lead  to any new divergence.

Now the right hand side of \refb{etotaln} contains not just the amplitudes
without soft external legs, but their derivatives with respect to external
momenta, and absence of infrared divergence in the original amplitude does
not necessarily imply absence of infrared divergence in its derivatives.
To see this let us take a derivative of \refb{ediv1} with respect to $p_{i\mu}$
and then use the on-shell condition $p_i^2+M_i^2=0$, $p_j^2+M_j^2=0$.
This generates an expression of the form
\be
{\p \II\over \p p_{i\mu}} 
= \{ \ell^2 \, (-2p_j\cdot \ell) \, (2p_i \cdot \ell)^2 \}^{-1} \times \hbox{finite}\times 
(-2 p_i^\mu) + \hbox{less divergent terms}\, .
\ee 
Now in the small $\ell_\mu$ limit the integrand has 5 powers of $\ell_\mu$ in the
denominator and therefore the integral has a logarithmic divergence in five
dimensions. Similarly if we take two derivatives of $\II$ and then use the on-shell
condition, then the leading and subleading
divergent pieces are given by
\ben
{\p^2 \II\over \p p_{i\mu} \p p_{i\nu}} 
&=& \{ \ell^2 \, (-2p_j\cdot \ell) \, (2p_i \cdot \ell)^3 )\}^{-1} \times \hbox{finite}\times 
(8 p_i^\mu\, p_i^\nu) \nonumber \\ && 
+  \{ \ell^2 \, (-2p_j\cdot \ell) \, (2p_i \cdot \ell)^2 )\}^{-1} \times \hbox{finite}\times 
(-2\, \eta^{\mu\nu}) 
\nonumber \\ &&
+ 
\{ \ell^2 \, (-2p_j\cdot \ell) \, (2p_i \cdot \ell)^2 )\}^{-1} \times \hbox{finite}\times 
(-2 p_i^\mu) \nonumber \\ && 
+ \{ \ell^2 \, (-2p_j\cdot \ell) \, (2p_i \cdot \ell)^2 )\}^{-1} \times \hbox{finite}\times 
(-2 p_i^\nu) \nonumber \\ &&
+
\hbox{less divergent terms}\, .
\een
The first term on the right hand side 
has six powers of $\ell_\mu$ in the denominator in the small $\ell_\mu$
limit. Therefore the integral is logarithmically divergent in six dimensions and
linearly divergent in five dimensions. The contribution to the integral 
from the second, third and fourth terms are free from
divergence in six dimensions and are logarithmically divergent in five dimensions.
It follows from this analysis that for $D=5,6$
the divergent parts of the derivatives
of $\Gamma_{(i)}^\gamma$ are of the form:
\ben \label{eir1}
D=5 &: \displaystyle
{\p \Gamma_{(i)}^\gamma\over \p p_{ia}} = p_{i}^{a}\, \check \Gamma_{(i)}^\alpha
+\hbox{finite},
\qquad & {\p^2 \Gamma_{(i)}^\alpha\over \p p_{ia} \p p_{ib}} 
= \eta^{ab}\, \check \Gamma_{(i)}^\alpha +
p_{i}^{a} \, \check \Gamma_{(i)}^{\alpha\, b}+ 
p_{i}^{b} \, \check \Gamma_{(i)}^{\alpha \, a}+\hbox{finite} \nonumber \\
D=6 &: \displaystyle{\p \Gamma_{(i)}^\gamma\over \p p_{ia}} = 
\hbox{finite},
\qquad &{\p^2 \Gamma_{(i)}^\alpha\over \p p_{ia} \p p_{ib}} 
= p_{i}^{a} \, p_i^b \, \check \Gamma_{(i)}^{\prime\alpha}+\hbox{finite} 
\een
for some functions $\check \Gamma_{(i)}^\alpha$, 
$\check \Gamma_{(i)}^{\alpha b}$ and $\check \Gamma_{(i)}^{\prime\alpha}$. 

We shall now argue that these divergences do not make the right hand side of
\refb{etotaln} diverge.  Since the divergences are more severe in $D=5$ let us consider the
$D=5$ case -- this will automatically extend to the $D=6$ case.
The potential sources of divergence are the terms involving
derivatives of $\Gamma_{(i)}^\alpha$ in the second, third and fourth lines on the
right hand side of \refb{etotaln}. Now using the first equation in \refb{eir1} we see that
the potentially divergent 
term on the second line is proportional to $p_{ia} p_{ib} - p_{ib} p_{ia}$ and
therefore vanishes. The same argument shows that the cross terms in the third and
the fourth lines of \refb{etotaln}
involving one factor of $J$ are also free from divergences. The 
remaining potentially divergent term in the third and the fourth line is proportional to
\ben
&& \left\{ p_{i}^{b} {\p\over \p p_{ia}}
- p_{i}^a {\p\over \p p_{ib}} \right\} \, \left\{ p_{i}^{d} {\p\over \p p_{ic}}
- p_{i}^c {\p\over \p p_{id}} \right\} \Gamma_{(i)}^{\alpha}(p_i) \nonumber \\
&=&\left\{ 
p_i^b \, p_i^d \, {\p^2 \Gamma_{(i)}^{\alpha}(p_i) \over \p p_{ia} \, \p p_{ic}}
+ \eta^{ad} \, p_i^b \, {\p \Gamma_{(i)}^{\alpha}(p_i) \over \p p_{ic}}\right\}
- \{a\leftrightarrow b\}  - \{c\leftrightarrow d\} + \{a\leftrightarrow b, \, 
c\leftrightarrow d\} \nonumber \\ 
&=& \left\{  p_i^b\, p_i^d \, p_{i}^{a} \, \check \Gamma_{(i)}^{\alpha c}+ 
p_i^b \, p_i^d \, p_i^c \, \check \Gamma_{(i)}^{\alpha a} 
+ p_i^b \, p_i^d \, \eta^{ac}  \, \check \Gamma_{(i)}^\alpha
+ \eta^{ad} \, p_i^b 
\, p_i^c 
\, \check \Gamma_{(i)}^\alpha
\right\} \nonumber \\  &&
- \{a\leftrightarrow b\}  - \{c\leftrightarrow d\} + \{a\leftrightarrow b, \, 
c\leftrightarrow d\} \nonumber \\
&=& 0\, .
\een
Therefore we see that the right hand side of \refb{etotaln} does
not have any infrared divergence from the terms involving derivatives of
$\Gamma_{(i)}^\alpha$ for $D\ge 5$. One might worry that since the individual terms
are divergent, one needs a regularization before claiming that they cancel. This
can be done by keeping the external momenta slightly off-shell while computing the
right hand side of \refb{etotaln}.
This is in any case needed to define derivatives with respect to $p_{i\mu}$ for which we
need to treat all components of $p_i$ as independent.

\begin{figure}
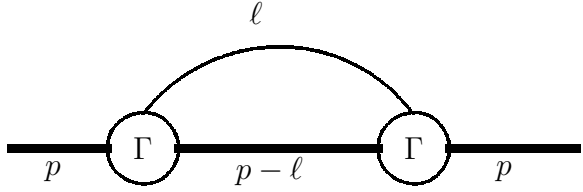


\begin{center}
\figirloop
\vskip -.7in

\caption{Infrared divergences in self-energy graphs. As usual the thin line denotes
a particle carrying soft momentum.
\label{f5}}

\end{center}

\end{figure}

Another potential 
source of infrared divergence on the right hand side of \refb{etotaln},
\refb{edelalphan}
is the derivative of the
self energy contribution proportional to
$\KK$ (and its inverse proportional to 
$\Xi$). Consider for example the contribution to $\KK^{\alpha\beta}$
from a diagram of the form shown in Fig.~\ref{f5} with the thin line denoting a
massless particle carrying soft momentum $\ell$. When the momentum $\ell$  
is small, the integrand is proportional to
\be
\II' = \{\ell^2 \, (-2 p\cdot \ell+p^2+M^2)\}^{-1} \times \hbox{finite} \, .
\ee
In this case $\int d^D\ell \, \II'$ has no divergence from the small $\ell_\mu$
region for $D\ge 4$. However since for $p^2+M^2=0$,
\be 
{\p^2 \II' \over \p p_\mu \p p_\nu} = \{\ell^2 \, (-2 p\cdot \ell)^3\}^{-1} \times
8\, p^\mu \,
p^\nu  \times \hbox{finite} + \hbox{less divergent terms}\, ,
\ee
and 
\be 
{\p^3 \II' \over \p p_\mu \p p_\nu \p p_\rho} =- 
\{\ell^2 \, (-2 p\cdot \ell)^4\}^{-1} \times 48\, p^\mu \,
p^\nu \, p^\rho\,  \times \hbox{finite} + \hbox{less divergent terms}\, ,
\ee
${\p^2 \II' / \p p_\mu \p p_\nu}$ diverges logarithmically for $D=5$ and
${\p^3 \II' / \p p_\mu \p p_\nu \p p_\rho}$ diverges linearly for $D=5$ and logarithmically for
$D=6$. It follows from this that the the first derivative of $\KK^{\alpha\beta}$ has no
divergence for $D\ge 5$, but for $D=5, 6$ 
the second and third derivatives of $\KK^{\alpha\beta}$ 
can have
divergent pieces of the form
\ben
D=5 &:& {\p^2 \KK^{\alpha\beta}(-p_i)\over \p p_{i\mu} \p p_{i\nu}}
= p_i^\mu \, p_i^\nu\, \check\KK^{\alpha\beta}(-p_i) +\hbox{finite}, \nonumber \\
&& {\p^3 \KK^{\alpha\beta}(-p_i)\over \p p_{i\mu} \p p_{i\nu} \p p_{i\rho}}
=  \left(\eta^{\mu\rho} p_i^\nu + \eta^{\mu\nu} p_i^\rho + \eta^{\nu\rho} p_i^\mu
\right) \, \check\KK^{\alpha\beta}(-p_i) 
\nonumber \\ && \hskip 1.2in +
p_i^\mu \, p_i^\nu\, \check\KK^{\alpha\beta\rho}(-p_i) +
 p_i^\mu \, p_i^\rho\, \check\KK^{\alpha\beta\nu}(-p_i) +
  p_i^\rho \, p_i^\nu\, \check\KK^{\alpha\beta\mu}(-p_i) +\hbox{finite},
\nonumber \\
D=6 &:& {\p^2 \KK^{\alpha\beta}(-p_i)\over \p p_{i\mu} \p p_{i\nu}}
= \hbox{finite}, \nonumber \\
&& {\p^3 \KK^{\alpha\beta}(-p_i)\over \p p_{i\mu} \p p_{i\nu} \p p_{i\rho}}
= p_i^\mu \, p_i^\nu\, p_i^\rho \, \check\KK^{\prime\alpha\beta}(-p_i) 
+ \hbox{finite},
\een
for some functions $\check\KK^{\alpha\beta}(-p_i)$,
$\check\KK^{\alpha\beta\mu}(-p_i)$ and
$\check\KK^{\prime\alpha\beta}(-p_i)$. Similar result holds for the derivatives of
$\Xi^i$. 

It is easy to check that these divergences also do not affect 
the final expression for the sub-subleading soft theorem given in
\refb{etotaln}. Via \refb{edelalphan}
this contains second derivative of $\KK$ and $\Xi^i$ with respect to
momenta and therefore has logarithmic divergence in $D=5$. However the divergent
piece in $\p^2 \KK/\p p_\mu \p  p_\nu$ (and $\p^2 \Xi/\p p_\mu \p  p_\nu$)
is proportional to $p^\mu p^\nu$. Substituting this into \refb{edelalphan} 
and using \refb{edefR} one can easily
verify that the corresponding contributions vanish and therefore the final expression 
for the sub-subleading soft
theorem is free from infrared divergences.

To summarize, we have shown that the right hand side of the sub-subleading soft
theorem given in \refb{etotaln} is free from infrared divergences.
Nevertheless since at the intermediate stages
of the analysis one encounters derivatives of $\Gamma_{(i)}^\alpha$,
$\KK^{\alpha\beta}$ and $\Xi^i_{\alpha\beta}$ that are infrared divergent, 
one could worry
whether all the terms have been properly accounted for. To this end we note that 
the original amplitude involving the soft graviton is manifestly free from
infrared divergences for any finite value of the soft momentum. Therefore any difference
between the original amplitude and \refb{etotaln} must be finite for any
finite value of $k$.  We shall now analyze
whether there can be such finite pieces that are left over in the difference between the
actual amplitude and the one given in \refb{etotaln}.

\begin{figure}
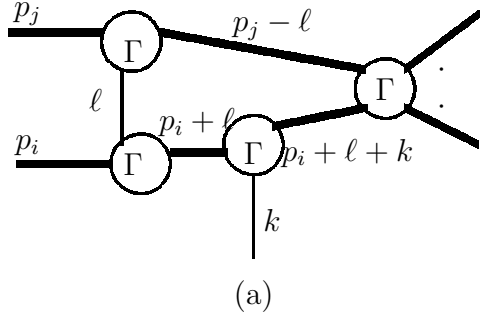


\begin{center}
\figsubtle
\end{center}

\caption{An apparently infrared divergent contribution to Fig.~\ref{f2}. \label{fir1}
}

\end{figure}

Before proceeding further, it will be useful to get some insight into the origin of the apparent
infrared divergences arising in the soft limit. 
Let us consider for example the diagram shown in Fig.~\ref{fir1}
representing a possible contribution to Fig.~\ref{f2}. As long as $k$ is finite, this represents
an infrared finite contribution in $D\ge 5$
since in the limit when $\ell$ becomes small there are at most
four powers of $\ell$ in the denominator --  one each from the propagators carrying momentum
$p_i+\ell$ and $p_j-\ell$,  and two from the propagator carrying momentum $\ell$. However
if we take the $k\to 0$ limit then the propagator carrying momentum $p_i+k+\ell$
supplies another factor of $\ell$ in the denominator, causing the integral to diverge 
logarithmically in $D=5$. 
In $D\ge 6$ this still represents a finite integral, but if we attempt
to expand the integrand in Taylor series expansion in $k$, as is needed for computing the
sub-subleading contribution, the next term in the Taylor series expansion will diverge 
logarithmically in $D=6$ and linearly in $D=5$.

These divergences explain the origin of the infrared divergences appearing in the
naive Taylor series expansion \refb{e8}
in powers of the soft momentum. For
example in $D=5$, 
the contribution of Fig.~\ref{fir1} can diverge as $\ln (p_i\cdot k)$ as $k^\mu\to 0$,
and this shows up as
logarithmic divergence in the $k$-independent terms in the
naive Taylor series expansion \refb{e8}.
On the other hand in $D=6$, the contribution from Fig.~\ref{fir1} is finite in the
$k\to 0$ limit, but has a subleading contribution proportional to $p_i\cdot k 
\ln (p_i\cdot k)$.
This shows up as a logarithmic divergence in the coefficient of the order $k^\mu$ terms 
in the naive Taylor series expansion \refb{e8}. A similar analysis can be carried out for
the diagrams contributing to Fig.~\ref{f1}.

We now try to determine the tensor structures of the singular terms by analyzing the
divergences in individual terms arising during the analysis in section 
\ref{s1}.\footnote{The reason that we can do this is due to the fact that the general 
formul\ae\ 
\refb{e8}, \refb{e4.8n} which express the amplitudes with a soft external state to ones without
it, are valid for off-shell  momenta of the finite energy external lines. Since 
for these there are no infrared divergences,
the presence of infrared divergences in the Taylor series expansion
of the original amplitudes in powers of the soft momentum $k$ can be inferred 
from possible infrared divergences that arise in the  Taylor series expansion of
\refb{e8}, \refb{e4.8n} about on-shell external momenta. \label{fo2}}
We shall illustrate
this with an example.
In expression \refb{e8} for Fig.~\ref{f2}, the $k^\mu$ independent contribution (which
represents a contribution to the subleading soft graviton amplitude)
involving a single derivative with respect to $p_{i\mu}$ is expected to be logarithmically
divergent. According to \refb{eir1} the divergent term in $\p \Gamma_{(i)}^\alpha
/\p p_{i\mu}$  is expected to be proportional to
$p_i^\mu$. Substituting this into \refb{e8} we see that the divergent part of this term is
proportional to $\ve_{\mu\nu} p_i^\mu p_i^\nu$. This can also be seen directly from
Fig.~\ref{fir1},  but analyzing the divergent
part of \refb{e8} yields the result in simpler fashion. Therefore we conclude that the
possible error in the analysis of the $k$ independent term in \refb{e8} in $D=5$
is proportional to $\ve_{\mu\nu} p_i^\mu p_i^\nu$.

With this insight we shall now try to  determine the  tensor structures of the
terms that could possibly diverge in the $k\to 0$ limit.
First let us consider the subleading soft graviton theorem. In this case intermediate steps
of the analysis involve at most one derivative of $\Gamma_{(i)}^\alpha$ 
and two derivatives of $\KK$ and $\Xi^i$ with respect to the
external momenta. 
These are free from divergences for $D\ge 6$, so we have to analyze
the possible logarithmically divergent contributions for $D= 5$. 
The potentially infrared divergent terms from derivative of $\Gamma_{(i)}^\delta$ are
\be \label{ediv2}
\eps_{i,\alpha}  
(p_i\cdot k)^{-1} \, \ve_{\mu b}\,  p_i^\mu  k_a p_{i}^{b} {\p\over \p p_{ia}} \, 
\Gamma_{(i)}^{\alpha}(p_i) \qquad  \hbox{and} \qquad 
- \eps_{i,\alpha} \,  \ve_{\mu b}\,  p_i^\mu  \, {\p\over \p p_{ib}}  
\Gamma_{(i)}^{\alpha}(p_i) 
\ee
coming from Figs.~\ref{f1} and \ref{f2} respectively. The divergent piece of 
${\p
\Gamma_{(i)}^{\alpha}(p_i) / \p p_{i\mu}}$ is proportional to $p_i^\mu$. Therefore
the divergent pieces in both terms in \refb{ediv2} are proportional to
\be \label{ediv3}
\ve_{\mu \nu}\,  p_i^\mu \, p_{i}^{\nu} \, ,
\ee
and come with opposite coefficients. However one may worry that after the cancellation
of the divergent pieces one may be left with an extra finite piece proportional to
\refb{ediv3}. This will generate an extra term of the form
\be \label{ediv4}
\sum_i \eps_{i,\alpha}  \, \ve_{\mu \nu}\,  p_i^\mu \, p_{i}^{\nu}  \, 
\wh \Gamma_{(i)}^{\prime\alpha}
\ee
for some amplitude $\wh \Gamma^{\prime\alpha}_{(i)}$. 

Another potential source of
logarithmic divergence in $D=5$ are the terms in \refb{e4.8n}
involving second derivative of $\KK$, obtained
after Taylor series expansion in soft momentum $k$ up to subleading order.
After using the fact that the divergent part of
$\p^2 \KK(-p)/\p p_\mu \p  p_\nu$
is proportional to $p^\mu p^\nu$, and carefully examining all the
terms proportional to $\p^2 \KK(-p)/\p p_\mu \p  p_\nu$ appearing in the
intermediate steps, one can see that the possible correction 
takes the form
\be \label{ediv9}
\sum_i \, \eps_{i,\alpha} \, \ve_{\mu\nu} \, p_i^\mu \, p_i^\nu\, \, 
\wh\Gamma_{(i)}^{\prime\prime\alpha}\, ,
\ee
for some amplitude $\Gamma_{(i)}^{\prime\prime\alpha}$.

Since the possible ambiguities from both sources  are proportional to
$\ve_{\mu \nu}\,  p_i^\mu \, p_{i}^{\nu}$,
they can be clubbed together. 
Therefore the net ambiguity in the subleading soft theorem takes the
form of an additive term of the form
\be \label{ediv45}
\sum_i \eps_{i,\alpha}  \, \ve_{\mu \nu}\,  p_i^\mu \, p_{i}^{\nu}  \, 
\wh \Gamma_{(i)}^{\alpha}
\ee
for some amplitude $\wh \Gamma^{\alpha}_{(i)}$. $\wh \Gamma^{\alpha}_{(i)}$ has at
most logarithmic divergence in the $k^\mu\to 0$ limit.

We shall now argue that an additional 
term of the form \refb{ediv45} in the subleading soft graviton
amplitude is
inconsistent with gauge invariance\footnote{Note that due to the way we have 
described the coupling of soft gravitons -- by covariantizing the vertices without the
soft graviton --  possible corrections to the soft graviton
amplitude should be invariant 
under gauge transformation of the soft graviton 
without using on-shell condition for other external states.} 
and therefore must vanish. For this let us consider 
shifting $\ve_{\mu\nu}$ by a pure gauge term
\be \label{egtrs}
(k_\mu \xi_\nu + k_\nu \xi_\mu)
\ee
for any vector $\xi$ satisfying
$k\cdot \xi=0$. Then \refb{ediv45} changes by
\be
2\, \sum_i \eps_{i,\alpha}  \, \xi\cdot p_i \, k\cdot p_i \, \wh \Gamma_{(i)}^\alpha \, .
\ee
We now see that this does not vanish for general $\xi$ and $k$ unless 
$\sum_i \eps_{i,\alpha} p_{i\mu} p_{i\nu}  \wh \Gamma_{(i)}^\alpha$ is
proportional to $\eta_{\mu\nu}$. In the latter
case \refb{ediv45} itself vanishes.
 Therefore \refb{ediv45} is not
gauge invariant, and the amplitude cannot have an additional contribution of the form
given in \refb{ediv45}. 
This shows that the subleading soft theorem is unaffected by infrared
divergences for $D\ge 5$.

Next we consider sub-subleading soft graviton theorem. In this case the intermediate stages of 
the analysis involve at most two derivatives of $\Gamma_{(i)}^\alpha$,
and at most three derivatives of $\KK$  with respect to the external
momenta. These are free from divergences for $D\ge 7$, so we need to analyze the cases
$D=5$ and $D=6$. Let us first consider the case $D=6$. In this case terms with single derivatives of 
$\Gamma_{(i)}^\alpha$ are free from infrared divergences; so we need to analyze the terns 
with two derivatives of $\Gamma_{(i)}^\alpha$. Such terms arise from two sources. First there
is a contribution from Fig.~\ref{f2} given by the last term in \refb{e8}. Since the divergent part
of ${\p^2
\Gamma_{(i)}^{\alpha}(p_i) / \p p_{i\mu} \p p_{i\nu}}$ is proportional to $p_i^\mu p_i^\nu$ in
$D=6$, the divergent contribution to the last term in \refb{e8} is proportional to 
\be \label{ediv6}
\ve_{\mu\nu} \, p_i^\mu \, p_i^\nu \, p_i\cdot k\, .
\ee
The other contribution involving two derivatives of $\Gamma_{(i)}^\alpha$ comes from Fig.~\ref{f1} and
is given by the Taylor series expansion of the $\Gamma_{(i)}^\delta$ factor
in \refb{e4.8n}. This is proportional to
\be \label{ediv7}
(p_i\cdot k)^{-1} \, \ve_{\mu\nu} \, p_i^\mu \, p_i^\nu \, (p_i\cdot k)^2\, ,
\ee
which is the same as \refb{ediv6}. Therefore in $D=6$, the amplitude may have
a potentially ambiguous contribution
proportional to
\be \label{ediv8}
\sum_i \, \eps_{i,\alpha} \ve_{\mu\nu} p_i^\mu p_i^\nu\, (p_i\cdot k)\, 
\bar\Gamma_{(i)}^{\prime\alpha}\, ,
\ee
for some amplitude $\bar\Gamma_{(i)}^{\prime\alpha}$. 

The potentially divergent self energy contributions in $D=6$
come from terms involving three
derivatives of $\KK$ in \refb{e4.8n}. Using the fact that the 
divergent part of $\p^3 \KK/\p p_\mu \p  p_\nu \p p_\rho$
is proportional to $p^\mu p^\nu p^\rho$ 
one can see that the possible correction is proportional to
\be \label{ediv8a}
\sum_i \, \eps_{i,\alpha} \ve_{\mu\nu} p_i^\mu p_i^\nu\, (p_i\cdot k)\, 
\bar\Gamma_{(i)}^{\prime\prime\alpha}\, ,
\ee
for some amplitude $\bar\Gamma_{(i)}^{\prime\prime\alpha}$. This has the same form
as \refb{ediv8} and can be clubbed with it. Therefore the net ambiguous term in the
sub-subleading soft graviton theorem in $D=6$ is an additive term of the form
\be \label{ediv85}
\sum_i \, \eps_{i,\alpha} \ve_{\mu\nu} p_i^\mu p_i^\nu\, (p_i\cdot k)\, 
\bar\Gamma_{(i)}^{\alpha}\, ,
\ee
for some amplitude $\bar\Gamma_{(i)}^{\alpha}$. In the $k^\mu\to 0$ limit
$\bar\Gamma_{(i)}^{\alpha}$ can have logarithmic divergence.

Now under a gauge transformation of
$\ve$ given in \refb{egtrs}, \refb{ediv85} changes by
\be 
2\, \sum_i \, \eps_{i,\alpha} \, \xi\cdot p_i \, (k\cdot p_i)^2 \bar\Gamma_{(i)}^\alpha\, .
\ee
This does not vanish for general $\xi$ and $k$ satisfying $k^2=0$, $\xi\cdot k=0$
unless
$\sum_i \eps_{i,\alpha} p_{i\mu} p_{i\nu} p_{i\rho}  \bar \Gamma_{(i)}^\alpha$ is
proportional to
\be
\eta_{\mu\nu} A_\rho + \eta_{\mu\rho} A_\nu + \eta_{\nu\rho} A_\mu\, ,
\ee 
for some function $A_\mu$.
In this case \refb{ediv85} itself vanishes.
Therefore adding a term of the form
\refb{ediv85} to the amplitude is inconsistent with gauge invariance. This in turn
proves that sub-subleading soft graviton theorem is unaffected by infrared
divergences for $D= 6$. 

One can  carry out a similar analysis for sub-subleading soft graviton theorem
in $D=5$. 
In this case there are many types of terms that can have infrared
divergences during the intermediate stages of the analysis, and therefore the possible
ambiguity is given by the sum of all such terms. 
One finds that all such possibly divergent terms can be clubbed into 
the form\footnote{The
divergent term \refb{ediv85} 
in $D=6$ is s special case of this where we choose $A^{\mu\nu\rho}$
to be $\sum_i p_i^\mu p_i^\nu p_i^\rho \eps_{i,\alpha} \bar\Gamma_{(i)}^{\alpha}\ $.}
\be \label{eresidue}
\ve_{\mu\nu} \, k_\rho \, A^{\mu\nu\rho}
\ee
for some amplitude $A^{\mu\nu\rho}$ which has at most logarithmic divergence as
$k^\mu\to 0$. In particular the $(p_i\cdot k)$ terms in the denominator are always cancelled. 
Without loss of generality we can take $A^{\mu\nu\rho}$ to be symmetric
in the indices $\mu,\nu$.
The requirement of gauge invariance now imposes
the constraint
\be
\xi_\mu \, k_\nu\, k_\rho \, A^{\mu\nu\rho} = 0\, .
\ee 
This can be satisfied for general $\xi$ and $k$ satisfying $\xi\cdot k=0$, $k^2=0$
if in the $k\to 0$ limit
\be \label{eform10}
A^{\mu\nu\rho} = P^\mu \, \eta^{\nu\rho} + P^\nu \, \eta^{\mu\rho} + Q^\rho \, \eta^{\mu\nu} 
+ B^{\mu\nu\rho}\, ,
\ee
for some function $P^\mu,Q^\mu,B^{\mu\nu\rho}$ with $B^{\mu\nu\rho}$ symmetric
under $\mu\leftrightarrow\nu$ and satisfying
\be
B^{\mu\nu\rho} + B^{\mu\rho\nu} = 0\, ,
\ee
for all $\mu, \nu, \rho$. It is easy to see that this, together with the relation
$B^{\mu\nu\rho}=B^{\nu\mu\rho}$, gives
\be
B^{\mu\nu\rho}=0\, .
\ee
Therefore we are left with the contribution to \refb{eform10} from the terms proportional to
the vectors $P$ and $Q$. However using \refb{eepscond} one can check that their
contribution to the amplitude  
\refb{eresidue} vanishes. Therefore even in five dimensions
the sub-subleading soft graviton 
theorem does not have any correction from the infrared
divergent terms. 

\subsection{Collinear divergences} \label{scollinear}

When some of the finite energy external states are massless, we can also have collinear
divergences. Again as mentioned in footnote \ref{fo2}, we can analyze their effect by 
examining the presence of these divergences in \refb{e8}, \refb{e4.8n} and their 
derivatives in the on-shell limit.

Let us for example consider Fig.~\ref{fs1} representing a possible contribution to
the $\Gamma_{(i)}^\alpha$ factor appearing in \refb{e8}. Potential collinear divergences arise
when one of the external states $i$ or $j$ represent massless particle. Let the $i$-th particle
be massless. Without loss of generality we can choose a frame in which this particle
moves along $x^{D-1}$ so that the only nonzero component of momenta are 
$p_i^0$ and $p_i^{D-1}$. For any momentum $p$ we now define $p^\pm = p^0\pm p^{D-1}$
and $\vec p_\perp=(p^1,\cdots p^{D-2})$ so that $p^2 = -p^+p^- + \vec p_\perp^2$.
In this language collinear region will correspond to region of loop momentum integration
where 
\be \label{ecoll}
\ell^+\sim 1, \qquad \vec l_\perp\sim \lambda, \qquad\ell^-\sim \lambda^2\, ,
\ee 
for
some small $\lambda$.   Therefore the small denominator factors of the integrand in 
Fig.~\ref{fs1} take the
form
\be \label{ecoll1}
\II|_{p_i^2=0} \sim (-\ell^+ \ell^- + \vec \ell_\perp^2-i\eps)^{-1} \, \{-(p_i^++\ell^+) \ell^-  +\vec\ell_\perp^2  -i\eps\}^{-1}\, .
\ee
The collinear region is $-p_i^+\le \ell^+\le 0$ since using the $i\eps$ prescription one
can easily verify that outside this region the $\ell^-$ integration contour
can be deformed away from the singularities\cite{sterman}. Note that we have not
included the denominator factor of the line carrying momentum $p_j-\ell$ since 
this remains finite in the limit \refb{ecoll}. In this limit  expression
\refb{ecoll1} goes as $\lambda^{-4}$ for $p_i^2=0$, whereas the 
$d\ell^- d^{D-2}\ell_\perp$ goes as $\lambda^D$. Therefore for $D\ge 5$ there are no
divergences.

However now consider taking derivatives with respect to $p_{i\mu}$ by first keeping $p_i$ off-shell and 
and setting
$p_i^2=0$ {\it after taking the derivative}. We get
\be \label{ecoll2}
\left. {\p \II\over \p p_{i\mu}}\right|_{p_i^2=0} \sim (-\ell^+ \ell^- + \vec \ell_\perp^2-i\eps)^{-1} \, \{-(p_i^++\ell^+) \ell^-  +\vec\ell_\perp^2 
 -i\eps\}^{-2} (-2) (p_i^\mu + \ell^\mu)\, ,
\ee
\be \label{ecoll3}
\left. {\p^2 \II\over \p p_{i\mu} \p p_{i\nu}}\right|_{p_i^2=0}
 \sim (-\ell^+ \ell^- + \vec \ell_\perp^2-i\eps)^{-1} \, \{-(p_i^++\ell^+) \ell^-  +\vec\ell_\perp^2 
 -i\eps\}^{-3} (8) (p_i^\mu + \ell^\mu) (p_i^\nu+\ell^\nu)\, .
\ee
In order to analyze these let us define $\tilde\ell^\mu$ via
\be \label{ellmu}
\ell^\mu = {\ell^+\over p_i^+} \, p_i^\mu + \tilde \ell_i^\mu
\ee
$\tilde\ell^+$ vanishes, and we have $\tilde\ell^-=\ell^-$ and $\tilde\ell_\perp=\ell_\perp$.
When we substitute \refb{ellmu} into \refb{ecoll2}, \refb{ecoll3} the terms proportional to $p_i^\mu$
are divergent since we now have six factors of $\lambda$ in the denominator
of \refb{ecoll2} and eight factors of $\lambda$ in the denominator of \refb{ecoll3}.
However our analysis
of \S\ref{sinfra} shows that divergent terms proportional to $p_i^\mu$ or
$p_i^\nu$ do not cause any problem.  If we choose
the terms proportional to $\tilde \ell^\mu$ and/or $\tilde \ell^\nu$ then for
$\mu,\nu=-$ it is easy to see that the degrees of divergence of \refb{ecoll2} and
\refb{ecoll3} remain the same as \refb{ecoll1} and therefore there is no divergence.
However there is a potential problem if we choose $\mu,\nu=\perp$ in \refb{ecoll3} since
now the integrand goes as $\lambda^{-6}$ and the integration measure goes as 
$\lambda^D$. Therefore the integral is divergent in $D=5,6$.

We must however remember that we also have to take into account possible numerator
factors from the vertices. If the internal graviton with 
momentum $\ell$ had been a physical graviton
then it would always carry polarization transverse to $\ell$. This would couple to
momentum components of $p_i$ transverse to $\ell$, giving a result proportional to 
$\vec\ell_\perp^2$ and killing the divergence for $D\ge 5$. This would be the case if we
work in a physical gauge where only the transverse components of the graviton 
propagate.\footnote{There is no conflict between choosing a physical gauge for the 
internal graviton and a covariant gauge for the 1PI action. We can compute the
1PI action using physical gauge, then subtract the gauge fixing term to get the gauge
invariant 1PI action and then gauge fix it using covariant gauge condition.
We can follow the same procedure if the internal particle had been a 
massless vector particle instead of a graviton. In this case we
would only get a single factor of $\vec \ell_\perp$ from the vertex. Naive power counting
then shows that \refb{ecoll3} 
is logarithmically divergent for $D=5$. However since the numerator
will have three powers of $\vec\ell_\perp$, the apparently divergent term would vanish
by $\vec\ell_\perp\to -\vec\ell_\perp$ symmetry.}
Alternatively
if we use de Donder gauge where longitudinal modes of the graviton also propagate, then the
divergent contributions will vanish after summing over different Feynman 
diagrams\cite{1109.0270}.

A similar analysis can be carried out for Fig.~\ref{f5} to show that there is no 
collinear divergence in the derivatives of this up to the desired order.

\bigskip

{\bf Acknowledgement:}
A.L is indebted to Miguel Campiglia for never ending discussions  on soft theorems and related topics. A.L. is grateful to Prahar Mitra and Madhusudan Raman for a number of discussions and Prahar Mitra for  detailed conversations pertaining to \cite{1611.07534}. We  would
like to thank Massimo Bianchi, Henriette Elvang and Andrea Guerrieri  
for useful communications.
Work of A.L is supported by Ramanujan Fellowship.
The work of A.S. was
supported in part by the 
J. C. Bose fellowship of 
the Department of Science and Technology, India.


\begin{thebibliography}{99}


\bibitem{weinberg1} 
  S.~Weinberg,
  ``Photons and Gravitons in s Matrix Theory: 
  Derivation of Charge Conservation and Equality of Gravitational and Inertial Mass,''
  Phys.\ Rev.\  {\bf 135}, B1049 (1964).
  doi:10.1103/PhysRev.135.B1049

\bibitem{weinberg2} 
  S.~Weinberg,
  ``Infrared photons and gravitons,''
  Phys.\ Rev.\  {\bf 140}, B516 (1965).
  doi:10.1103/PhysRev.140.B516

\bibitem{jackiw1} 
  D.~J.~Gross and R.~Jackiw,
  ``Low-Energy Theorem for Graviton Scattering,''
  Phys.\ Rev.\  {\bf 166}, 1287 (1968).
  doi:10.1103/PhysRev.166.1287
  
\bibitem{jackiw2} 
  R.~Jackiw,
  ``Low-Energy Theorems for Massless Bosons: Photons and Gravitons,''
  Phys.\ Rev.\  {\bf 168}, 1623 (1968).
  doi:10.1103/PhysRev.168.1623

\bibitem{1103.2981} 
  C.~D.~White,
  ``Factorization Properties of Soft Graviton Amplitudes,''
  JHEP {\bf 1105}, 060 (2011)
  doi:10.1007/JHEP05(2011)060
  [arXiv:1103.2981 [hep-th]].

\bibitem{1404.4091} 
  F.~Cachazo and A.~Strominger,
  ``Evidence for a New Soft Graviton Theorem,''
  arXiv:1404.4091 [hep-th].


\bibitem{1404.7749} 
  B.~U.~W.~Schwab and A.~Volovich,
  ``Subleading Soft Theorem in Arbitrary Dimensions from Scattering Equations,''
  Phys.\ Rev.\ Lett.\  {\bf 113}, no. 10, 101601 (2014)
  doi:10.1103/PhysRevLett.113.101601
  [arXiv:1404.7749 [hep-th]].

\bibitem{1405.1410} 
  S.~He, Y.~t.~Huang and C.~Wen,
  ``Loop Corrections to Soft Theorems in Gauge Theories and Gravity,''
  JHEP {\bf 1412}, 115 (2014)
  doi:10.1007/JHEP12(2014)115
  [arXiv:1405.1410 [hep-th]].

\bibitem{1405.2346} 
  A.~J.~Larkoski,
  ``Conformal Invariance of the Subleading Soft Theorem in Gauge Theory,''
  Phys.\ Rev.\ D {\bf 90}, no. 8, 087701 (2014)
  doi:10.1103/PhysRevD.90.087701
  [arXiv:1405.2346 [hep-th]].

\bibitem{1405.3413} 
  F.~Cachazo and E.~Y.~Yuan,
  ``Are Soft Theorems Renormalized?,''
  arXiv:1405.3413 [hep-th].

\bibitem{1405.3533} 
  N.~Afkhami-Jeddi,
  ``Soft Graviton Theorem in Arbitrary Dimensions,''
  arXiv:1405.3533 [hep-th].

\bibitem{1406.6574} 
  J.~Broedel, M.~de Leeuw, J.~Plefka and M.~Rosso,
  ``Constraining subleading soft gluon and graviton theorems,''
  Phys.\ Rev.\ D {\bf 90}, no. 6, 065024 (2014)
  doi:10.1103/PhysRevD.90.065024
  [arXiv:1406.6574 [hep-th]].

\bibitem{1406.6987} 
  Z.~Bern, S.~Davies, P.~Di Vecchia and J.~Nohle,
  ``Low-Energy Behavior of Gluons and Gravitons from Gauge Invariance,''
  Phys.\ Rev.\ D {\bf 90}, no. 8, 084035 (2014)
  doi:10.1103/PhysRevD.90.084035
  [arXiv:1406.6987 [hep-th]].
  
\bibitem{1406.7184} 
  C.~D.~White,
  ``Diagrammatic insights into next-to-soft corrections,''
  Phys.\ Lett.\ B {\bf 737}, 216 (2014)
  doi:10.1016/j.physletb.2014.08.041
  [arXiv:1406.7184 [hep-th]].
  
\bibitem{1407.5936} 
  M.~Zlotnikov,
  ``Sub-sub-leading soft-graviton theorem in arbitrary dimension,''
  JHEP {\bf 1410}, 148 (2014)
  doi:10.1007/JHEP10(2014)148
  [arXiv:1407.5936 [hep-th]].

\bibitem{1407.5982} 
  C.~Kalousios and F.~Rojas,
  ``Next to subleading soft-graviton theorem in arbitrary dimensions,''
  JHEP {\bf 1501}, 107 (2015)
  doi:10.1007/JHEP01(2015)107
  [arXiv:1407.5982 [hep-th]].

\bibitem{1408.4179} 
  Y.~J.~Du, B.~Feng, C.~H.~Fu and Y.~Wang,
  ``Note on Soft Graviton theorem by KLT Relation,''
  JHEP {\bf 1411}, 090 (2014)
  doi:10.1007/JHEP11(2014)090
  [arXiv:1408.4179 [hep-th]].

\bibitem{1410.6406} 
  D.~Bonocore, E.~Laenen, L.~Magnea, L.~Vernazza and C.~D.~White,
  ``The method of regions and next-to-soft corrections in DrellÐYan production,''
  Phys.\ Lett.\ B {\bf 742}, 375 (2015)
  doi:10.1016/j.physletb.2015.02.008
  [arXiv:1410.6406 [hep-ph]].

\bibitem{1412.3699} 
  A.~Sabio Vera and M.~A.~Vazquez-Mozo,
  ``The Double Copy Structure of Soft Gravitons,''
  JHEP {\bf 1503}, 070 (2015)
  doi:10.1007/JHEP03(2015)070
  [arXiv:1412.3699 [hep-th]].

\bibitem{1503.04816} 
  F.~Cachazo, S.~He and E.~Y.~Yuan,
  ``New Double Soft Emission Theorems,''
  Phys.\ Rev.\ D {\bf 92}, no. 6, 065030 (2015)
  doi:10.1103/PhysRevD.92.065030
  [arXiv:1503.04816 [hep-th]].

\bibitem{1504.01364} 
  A.~E.~Lipstein,
  ``Soft Theorems from Conformal Field Theory,''
  JHEP {\bf 1506}, 166 (2015)
  doi:10.1007/JHEP06(2015)166
  [arXiv:1504.01364 [hep-th]].

\bibitem{1507.08882} 
  S.~D.~Alston, D.~C.~Dunbar and W.~B.~Perkins,
  ``$n$-point amplitudes with a single negative-helicity graviton,''
  Phys.\ Rev.\ D {\bf 92}, no. 6, 065024 (2015)
  doi:10.1103/PhysRevD.92.065024
  [arXiv:1507.08882 [hep-th]].

\bibitem{1509.07840} 
  Y.~t.~Huang and C.~Wen,
  ``Soft theorems from anomalous symmetries,''
  JHEP {\bf 1512}, 143 (2015)
  doi:10.1007/JHEP12(2015)143
  [arXiv:1509.07840 [hep-th]].


\bibitem{1604.00650} 
  J.~Rao and B.~Feng,
  ``Note on Identities Inspired by New Soft Theorems,''
  JHEP {\bf 1604}, 173 (2016)
  doi:10.1007/JHEP04(2016)173
  [arXiv:1604.00650 [hep-th]].

\bibitem{1604.03893} 
  F.~Cachazo, P.~Cha and S.~Mizera,
  ``Extensions of Theories from Soft Limits,''
  JHEP {\bf 1606}, 170 (2016)
  doi:10.1007/JHEP06(2016)170
  [arXiv:1604.03893 [hep-th]].

\bibitem{1607.02700} 
  A.~P.~Saha,
  ``Double Soft Theorem for Perturbative Gravity,''
  JHEP {\bf 1609}, 165 (2016)
  doi:10.1007/JHEP09(2016)165
  [arXiv:1607.02700 [hep-th]].

\bibitem{1611.02172} 
  A.~Luna, S.~Melville, S.~G.~Naculich and C.~D.~White,
  ``Next-to-soft corrections to high energy scattering in QCD and gravity,''
  JHEP {\bf 1701}, 052 (2017)
  doi:10.1007/JHEP01(2017)052
  [arXiv:1611.02172 [hep-th]].

\bibitem{1611.03137} 
  C.~Cheung, K.~Kampf, J.~Novotny, C.~H.~Shen and J.~Trnka,
  ``A Periodic Table of Effective Field Theories,''
  arXiv:1611.03137 [hep-th].

\bibitem{1702.02350} 
  A.~P.~Saha,
  ``Double Soft Theorem for Perturbative Gravity II: Some Details on CHY Soft Limits,''
  arXiv:1702.02350 [hep-th].


\bibitem{1312.2229} 
  A.~Strominger,
  ``On BMS Invariance of Gravitational Scattering,''
  JHEP {\bf 1407}, 152 (2014)
  doi:10.1007/JHEP07(2014)152
  [arXiv:1312.2229 [hep-th]].

\bibitem{1401.7026} 
  T.~He, V.~Lysov, P.~Mitra and A.~Strominger,
  ``BMS supertranslations and WeinbergÕs soft graviton theorem,''
  JHEP {\bf 1505}, 151 (2015)
  doi:10.1007/JHEP05(2015)151
  [arXiv:1401.7026 [hep-th]].

\bibitem{1411.5745} 
  A.~Strominger and A.~Zhiboedov,
  ``Gravitational Memory, BMS Supertranslations and Soft Theorems,''
  JHEP {\bf 1601}, 086 (2016)
  doi:10.1007/JHEP01(2016)086
  [arXiv:1411.5745 [hep-th]].


\bibitem{1506.05789} 
  S.~G.~Avery and B.~U.~W.~Schwab,
  ``Burg-Metzner-Sachs symmetry, string theory, and soft theorems,''
  Phys.\ Rev.\ D {\bf 93}, 026003 (2016)
  doi:10.1103/PhysRevD.93.026003
  [arXiv:1506.05789 [hep-th]].

\bibitem{1509.01406} 
  M.~Campiglia and A.~Laddha,
  ``Asymptotic symmetries of gravity and soft theorems for massive particles,''
  JHEP {\bf 1512}, 094 (2015)
  doi:10.1007/JHEP12(2015)094
  [arXiv:1509.01406 [hep-th]].

\bibitem{1605.09094} 
  M.~Campiglia and A.~Laddha,
  ``Sub-subleading soft gravitons: New symmetries of quantum gravity?,''
  Phys.\ Lett.\ B {\bf 764}, 218 (2017)
  doi:10.1016/j.physletb.2016.11.046
  [arXiv:1605.09094 [gr-qc]].

\bibitem{1608.00685} 
  M.~Campiglia and A.~Laddha,
  ``Sub-subleading soft gravitons and large diffeomorphisms,''
  JHEP {\bf 1701}, 036 (2017)
  doi:10.1007/JHEP01(2017)036
  [arXiv:1608.00685 [gr-qc]].

\bibitem{1612.08294} 
  E.~Conde and P.~Mao,
  ``BMS Supertranslations and Not So Soft Gravitons,''
  arXiv:1612.08294 [hep-th].

\bibitem{1701.00496} 
  T.~He, D.~Kapec, A.~M.~Raclariu and A.~Strominger,
  ``Loop-Corrected Virasoro Symmetry of 4D Quantum Gravity,''
  arXiv:1701.00496 [hep-th].

\bibitem{1612.05886} 
  M.~Asorey, A.~P.~Balachandran, F.~Lizzi and G.~Marmo,
  ``Equations of Motion as Constraints: Superselection Rules, Ward Identities,''
  arXiv:1612.05886 [hep-th].

\bibitem{1703.05448} 
  A.~Strominger,
  ``Lectures on the Infrared Structure of Gravity and Gauge Theory,''
  arXiv:1703.05448 [hep-th].



\bibitem{ademollo} 
  M.~Ademollo, A.~D'Adda, R.~D'Auria, F.~Gliozzi, E.~Napolitano, S.~Sciuto and P.~Di Vecchia,
  ``Soft Dilations and Scale Renormalization in Dual Theories,''
  Nucl.\ Phys.\ B {\bf 94}, 221 (1975).
  doi:10.1016/0550-3213(75)90491-5

\bibitem{shapiro} 
  J.~A.~Shapiro,
  ``On the Renormalization of Dual Models,''
  Phys.\ Rev.\ D {\bf 11}, 2937 (1975).
  doi:10.1103/PhysRevD.11.2937

\bibitem{1406.4172} 
  B.~U.~W.~Schwab,
  ``Subleading Soft Factor for String Disk Amplitudes,''
  JHEP {\bf 1408}, 062 (2014)
  doi:10.1007/JHEP08(2014)062
  [arXiv:1406.4172 [hep-th]].

\bibitem{1406.5155} 
  M.~Bianchi, S.~He, Y.~t.~Huang and C.~Wen,
  ``More on Soft Theorems: Trees, Loops and Strings,''
  Phys.\ Rev.\ D {\bf 92}, no. 6, 065022 (2015)
  doi:10.1103/PhysRevD.92.065022
  [arXiv:1406.5155 [hep-th]].

\bibitem{1411.6661} 
  B.~U.~W.~Schwab,
  ``A Note on Soft Factors for Closed String Scattering,''
  JHEP {\bf 1503}, 140 (2015)
  doi:10.1007/JHEP03(2015)140
  [arXiv:1411.6661 [hep-th]].

\bibitem{1502.05258}
 P.~Di Vecchia, R.~Marotta and M.~Mojaza,
  ``Soft theorem for the graviton, dilaton and the Kalb-Ramond field in the bosonic string,''
  JHEP {\bf 1505}, 137 (2015)
  doi:10.1007/JHEP05(2015)137
  [arXiv:1502.05258 [hep-th]].

\bibitem{1505.05854} 
  M.~Bianchi and A.~L.~Guerrieri,
  ``On the soft limit of open string disk amplitudes with massive states,''
  JHEP {\bf 1509}, 164 (2015)
  doi:10.1007/JHEP09(2015)164
  [arXiv:1505.05854 [hep-th]].

\bibitem{1507.08829} 
  A.~L.~Guerrieri,
  ``Soft behavior of string amplitudes with external massive states,''
  Nuovo Cim.\ C {\bf 39}, no. 1, 221 (2016)
  doi:10.1393/ncc/i2016-16221-2
  [arXiv:1507.08829 [hep-th]].

\bibitem{1511.04921} 
  P.~Di Vecchia, R.~Marotta and M.~Mojaza,
  ``Soft Theorems from String Theory,''
  Fortsch.\ Phys.\  {\bf 64}, 389 (2016)
  doi:10.1002/prop.201500068
  [arXiv:1511.04921 [hep-th]].
  
  \bibitem{1512.00803} 
  M.~Bianchi and A.~L.~Guerrieri,
  ``On the soft limit of closed string amplitudes with massive states,''
  Nucl.\ Phys.\ B {\bf 905}, 188 (2016)
  doi:10.1016/j.nuclphysb.2016.02.005
  [arXiv:1512.00803 [hep-th]].

\bibitem{1601.03457} 
  M.~Bianchi and A.~L.~Guerrieri,
  ``On the soft limit of tree-level string amplitudes,''
  arXiv:1601.03457 [hep-th].

\bibitem{1604.03355} 
  P.~Di Vecchia, R.~Marotta and M.~Mojaza,
  ``Subsubleading soft theorems of gravitons and dilatons in the bosonic string,''
  JHEP {\bf 1606}, 054 (2016)
  doi:10.1007/JHEP06(2016)054
  [arXiv:1604.03355 [hep-th]].

\bibitem{1610.03481} 
  P.~Di Vecchia, R.~Marotta and M.~Mojaza,
  ``Soft behavior of a closed massless state in superstring and universality in the soft behavior of the dilaton,''
  JHEP {\bf 1612}, 020 (2016)
  doi:10.1007/JHEP12(2016)020
  [arXiv:1610.03481 [hep-th]].


\bibitem{1702.03934} 
  A.~Sen,
  ``Soft Theorems in Superstring Theory,''
  arXiv:1702.03934 [hep-th].

\bibitem{1703.00024} 
  A.~Sen,
  ``Subleading Soft Graviton Theorem for Loop Amplitudes,''
  arXiv:1703.00024 [hep-th].

\bibitem{1405.1015} 
  Z.~Bern, S.~Davies and J.~Nohle,
  ``On Loop Corrections to Subleading Soft Behavior of Gluons and Gravitons,''
  Phys.\ Rev.\ D {\bf 90}, no. 8, 085015 (2014)
  doi:10.1103/PhysRevD.90.085015
  [arXiv:1405.1015 [hep-th]].

\bibitem{1611.07534} 
  H.~Elvang, C.~R.~T.~Jones and S.~G.~Naculich,
  ``Soft Photon and Graviton Theorems in Effective Field Theory,''
  arXiv:1611.07534 [hep-th].

\bibitem{1605.09677} 
  M.~Campiglia and A.~Laddha,
  ``Subleading soft photons and large gauge transformations,''
  JHEP {\bf 1611}, 012 (2016)
  doi:10.1007/JHEP11(2016)012
  [arXiv:1605.09677 [hep-th]].


\bibitem{wald} 
  R.~M.~Wald,
  ``General Relativity,''
  Chicago, Usa: Univ. Pr. ( 1984) 491p
  doi:10.7208/chicago/9780226870373.001.0001

\bibitem{9411092} 
  S.~Y.~Choi, J.~S.~Shim and H.~S.~Song,
  ``Factorization and polarization in linearized gravity,''
  Phys.\ Rev.\ D {\bf 51}, 2751 (1995)
  doi:10.1103/PhysRevD.51.2751
  [hep-th/9411092].

\bibitem{1308.1697} 
  H.~Elvang and Y.~t.~Huang,
  ``Scattering Amplitudes,''
  arXiv:1308.1697 [hep-th].

\bibitem{1310.5353} 
  L.~J.~Dixon,
  ``A brief introduction to modern amplitude methods,''
  doi:10.5170/CERN-2014-008.31
  arXiv:1310.5353 [hep-ph].

\bibitem{kinoshita} 
  T.~Kinoshita,
  ``Mass singularities of Feynman amplitudes,''
  J.\ Math.\ Phys.\  {\bf 3}, 650 (1962).
  doi:10.1063/1.1724268

\bibitem{lee} 
  T.~D.~Lee and M.~Nauenberg,
  ``Degenerate Systems and Mass Singularities,''
  Phys.\ Rev.\  {\bf 133}, B1549 (1964).
  doi:10.1103/PhysRev.133.B1549

\bibitem{bloch} 
  F.~Bloch and A.~Nordsieck,
  ``Note on the Radiation Field of the electron,''
  Phys.\ Rev.\  {\bf 52}, 54 (1937).
  doi:10.1103/PhysRev.52.54

\bibitem{kulish-faddeev} 
  P.~P.~Kulish and L.~D.~Faddeev,
  ``Asymptotic conditions and infrared divergences in quantum electrodynamics,''
  Theor.\ Math.\ Phys.\  {\bf 4}, 745 (1970)
  [Teor.\ Mat.\ Fiz.\  {\bf 4}, 153 (1970)].
  doi:10.1007/BF01066485
  
\bibitem{1308.6285} 
  J.~Ware, R.~Saotome and R.~Akhoury,
  ``Construction of an asymptotic S matrix for perturbative quantum gravity,''
  JHEP {\bf 1310}, 159 (2013)
  doi:10.1007/JHEP10(2013)159
  [arXiv:1308.6285 [hep-th]].


\bibitem{1705.04311} 
  D.~Kapec, M.~Perry, A.~M.~Raclariu and A.~Strominger,
  ``Infrared Divergences in QED, Revisited,''
  arXiv:1705.04311 [hep-th].


\bibitem{sterman} 
  G.~F.~Sterman,
  ``An Introduction to quantum field theory,'' Cambridge University Press (1993).

\bibitem{1109.0270} 
  R.~Akhoury, R.~Saotome and G.~Sterman,
  ``Collinear and Soft Divergences in Perturbative Quantum Gravity,''
  Phys.\ Rev.\ D {\bf 84}, 104040 (2011)
  doi:10.1103/PhysRevD.84.104040
  [arXiv:1109.0270 [hep-th]].

\end{thebibliography}
\end{document}